\documentclass[twocolumn, prx, reprint, amsfonts, amsmath, amssymb, superscriptaddress, longbibliography, floatfix, aps]{revtex4-2}
\usepackage[T1]{fontenc}
\usepackage{lmodern}
\usepackage{graphicx}
\usepackage[dvipsnames]{xcolor}
\usepackage[colorlinks,linkcolor=Blue,citecolor=Blue]{hyperref}
\usepackage{braket}
\usepackage{mathtools}
\usepackage{enumitem}
\usepackage{microtype}

\begin{document}

\title{Ultra-critical Floquet Non-Fermi Liquid}

\author{Li-kun Shi}
\thanks{These authors contributed equally to this work.}
\affiliation{Institut f{\"u}r Theoretische Physik, Universit{\"a}t Leipzig, Br{\"u}derstra{\ss}e 16, 04103, Leipzig, Germany}
\affiliation{Center for Quantum Matter, School of Physics, Zhejiang University, Hangzhou 310058, China}

\author{Oles Matsyshyn}
\thanks{These authors contributed equally to this work.}
\affiliation{Division of Physics and Applied Physics, School of Physical and Mathematical Sciences, Nanyang Technological University, Singapore 637371}

\author{Justin C. W. Song}
\affiliation{Division of Physics and Applied Physics, School of Physical and Mathematical Sciences, Nanyang Technological University, Singapore 637371}

\author{Inti Sodemann Villadiego}
\affiliation{Institut f{\"u}r Theoretische Physik, Universit{\"a}t Leipzig, Br{\"u}derstra{\ss}e 16, 04103, Leipzig, Germany}

\date{\today}

\begin{abstract}
We demonstrate that periodically driven Fermions coupled to simple bosonic baths have steady state occupations of Floquet Bloch bands that generically display non-analyticties at certain momenta which resemble the Fermi surfaces of equilibrium non-Fermi liquids. Remarkably these non-equilibrium Fermi surfaces remain sharp even when the bath is at finite temperature, leading to critical power-law decaying correlations at finite temperature, a phenomenon with no analogue in equilibrium. We also show that generically there is in-gap current rectification for clean metals lacking inversion symmetry, and explain why this occurs universally regardless of the details of collisions.
\end{abstract}

\maketitle

\begin{figure}
\includegraphics[width=0.48\textwidth]{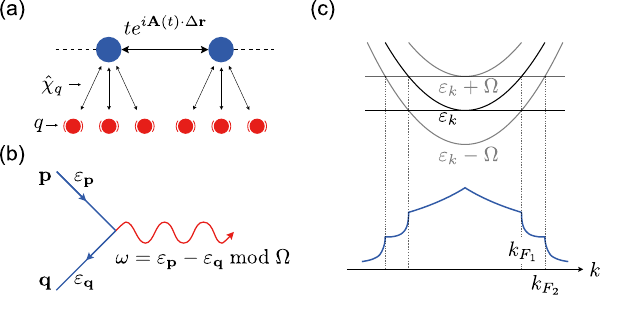}
\caption{(a) Schematic of a periodically driven tight-binding model (blue) coupled to boson modes (red) [see Eq.~\eqref{Full-Hamiltonian}], enabling (b) Floquet-Umklapp processes [see Eq.~\eqref{general-scattering}] and resulting (c) Floquet Non-Fermi liquid Fermi surfaces $k_{F_n}$ at intersections of band edges and its Floquet copies.}
\label{fig1}
\end{figure}

\textit{\color{blue}Introduction.} The transitivity of thermal equilibrium or zeroth law of thermodynamics~\cite{kardar2007statistical},  encodes a remarkable universality of thermal baths in equilibrium, namely, a system coupled to a bath always thermalizes towards the same macroscopic state regardless of the details of the bath. However, for non-equilibrium settings the details of the bath matter. In this work we will demonstrate a dramatic example of this by showing that periodically driven fermions coupled to a boson bath have a steady state occupation function with non-analyticities at certain special momenta that resemble the Fermi surfaces of a non-Fermi liquid state (see Fig.~\ref{fig1}(c) and Refs.~\cite{giamarchi2003quantum,senthil2008critical,varma2002singular}). This sharply contrasts with the case when they are coupled to a fermion bath, which displays a Fermi-Dirac staircase occupation with multiple jumps, and behaves like a Fermi liquid, as we have recently demonstrated in Refs.~\cite{matsyshyn2023fermi,shi2024floquet} (for related studies see also Refs.~\cite{seetharam2015controlled,nazarov2009quantum,tien1963multiphoton,kumari2024josephson,asmar2022impurity,le2024inverse}).

We will also show that, remarkably, these non-analyticities remain sharp even when the boson bath is at finite temperature, and thus these Floquet non-Fermi liquid Fermi surfaces do not suffer from thermal smearing, leading to correlations that decay like power laws at long distances even at finite temperature. This behavior, which we refer to as ``ultra-critical'', has no analogue in equilibrium, where Fermi surfaces of Fermi and Non-Fermi liquids invariably smear at finite temperature and correlations decay exponentially at long distances.

\smallskip

\textit{\color{blue}Floquet-Boltzmann equation.} We consider a fermionic system ($S$) coupled to a bosonic bath ($B$), described by the Hamiltonian [see Fig.~\ref{fig1}(a)]:
\begin{equation}
\hat{H}(t) = \hat{H}_S(t) +\hat{H}_B +\hat{H}_{SB} + \text{h.c.},
\label{Full-Hamiltonian}
\end{equation}
where $\hat{H}_S(t) = \sum_{\alpha \beta} \bra{\alpha} \hat{h}_{t} \ket{\beta}  \hat{a}_\alpha^\dagger \hat{a}_\beta$, $\hat{H}_B = \sum_q \hbar \omega_q \hat{b}^\dagger_q \hat{b}_q$, and $\hat{H}_{SB} = \sum_{q,\nu \eta} \bra{\nu} \hat{\chi}_{q} \ket{\eta} \hat{b}_q \hat{a}_\nu^\dagger \hat{a}_\eta$. $\hat{h}_{t}$ and $\hat{\chi}_{q}$ are  the the system single particle Hamiltonian matrix and its coupling matrix to the mode $q$ of the bath, $\hat{a}_\alpha^\dagger$, $\hat{a}_\alpha$ ($\hat{b}^\dagger_q$, $\hat{b}_q$) are fermionic (bosonic) creation and annihilation operators for state $\ket{\alpha}$ (mode $q$). The bath is in thermal equilibrium with 
Bose-Einstein occupation $N_{q}=1/(e^{\beta \hbar\omega_q}-1)$ and temperature $k_B T_0 = 1 /  {\beta}$, but the system can be driven out of equilibrium via the time-dependence of $\hat{h}_{t}$ (hence the subscript $t$). Starting from the full density matrix for the system and bath, $\hat{\eta}_t$,  and following the analysis of Raichev and Basko \cite{VaskoRaichev}  for weak system-bath coupling (see Appendix~\ref{appendixA1}), we arrive at a quantum kinetic equation for the system one-body density matrix,
$\rho_{\gamma\delta} \equiv {\rm Tr}[\hat{\eta}_t \hat a_\delta^\dagger \hat a_\gamma]$, 
\begin{equation}
\hbar\partial_t \hat{\rho}_t + i[\hat{h}_{t},\hat{\rho}_t]= \hat{I}^e_t + \hat{I}^a_t,
\label{EoMfull}
\end{equation}
where $\hat{I}^e_t$ and $\hat{I}^a_t$ are respectively the emission and absorption collision operators, which are given by:
\begin{align}
\begin{aligned}
& \hat{I}^e_t = \frac{1}{\hbar}\sum_{q} (N_{q}+1)  \int_{-\infty}^t dt' 
e^{i\omega_q (t-t')}
\big[ \hat{U} (t,t') 
\\
& \big(  \hat{\rho}_{t'} \hat{\chi}_{q} (1-\hat{\rho}_{t'})  
+ \hat{\rho}_{t'} {\rm tr} 
\big[ \hat{\rho}_{t'} \hat{\chi}_{q} \big] \big)
\hat{U}(t',t), \hat \chi_{q}^{\dagger}\big]+\text{h.c.},
\label{eq:Je}
\end{aligned}
\end{align}
where $\hat{U}$ is the system unitary evolution matrix in the absence of coupling to the bath.
The absorption operator $\hat{I}^a_t$ can be obtained from $\hat{I}^e_t$ by substituting $N_{q}+1\to N_{q},~ \omega_q \to-\omega_q,~\hat\chi\to\hat\chi^\dagger$ \cite{VaskoRaichev}. We will apply this general formalism to investigate the steady states of Floquet Bloch bands. For simplicity we consider tight binding models with one site per unit cell (i.e. no Berry phases) periodically driven by a spatially uniform electric field, with an associated vector potential $\mathbf{A}(t) = \mathbf{A}(t + T)$. Therefore, the system Hamiltonian is $\hat{h}_{t}=\sum_{\bf k} \epsilon({\bf k} -{\bf A}(t)) \ket{{\bf k}} \bra{{\bf k}}$. We couple each site, ${\bf r}$, of the tight biding model to its own collection of local boson modes, namely, the label for boson modes can be written as $q=({\bf r},\lambda)$, where $\lambda$ labels the different boson modes coupled to each site [see Fig.~\ref{fig1}(a)]. The frequency of the boson modes is site independent and only depends on $\lambda$, namely, $\omega_q=\omega_\lambda$. We consider the limit where $\lambda$ becomes dense and the density of states (DOS) of the bath, $\nu_B (\omega)=\sum_\lambda \delta(\omega-\omega_\lambda)$, approaches a continuous function which we will specify in the coming sections. Moreover, we take the boson modes to be coupled to the on-site fermion density, namely the coupling matrix is simply the projector onto a system site $\hat{\chi}_{q}=\chi_0\ket{{\bf r}} \bra{{\bf r}}$, and $\chi_0$ is the same for all boson modes.

Starting from Eq.~(\ref{EoMfull}),  one can show that in the limit of weak coupling to the bath ($\chi_0\to 0$) the one-body density matrix of the fermionic system approaches the steady state (see Appendix~\ref{appendixA1}):
\begin{align}
\begin{aligned}
\rho_t = \sum_{\bf k} f_{\bf k}\ket{{\bf k}} \bra{{\bf k}},
\end{aligned}\label{rhot}
\end{align}
\noindent where the occupations, $f_{\bf k}$, are determined by the solutions of the time-independent Floquet-Boltzmann equation:
\begin{align}
& \sum_{{\bf q}}
( 
f_{\bf q} W_{{\bf q} \to {\bf p}} \bar{f}_{\bf p}
-
f_{\bf p} W_{{\bf p}\to {\bf q} }\bar{f}_{\bf q}
)=0,
\quad
\bar{f}_{\bf p} \equiv 1 - f_{\bf p},
\label{Floquet-Boltzmann}
\end{align}
where the scattering rates are given by:
\begin{align}
\begin{aligned}
& W_{{\bf q} \to {\bf p} } = 2\pi |\chi_0|^2
\sum\nolimits_{l} 
\Phi^{(l)}_{{\bf q},{\bf p}}
\, S(\varepsilon_{\bf q} - \varepsilon_{\bf p} + l \Omega) ,
\\
& \Phi^{(l)}_{{\bf q},{\bf p}} = 
\sum\nolimits_{l'}
|\varphi_{l',{\bf p}} \varphi_{l'+l, {\bf q}}^*|^2 ,
\label{general-scattering}
\end{aligned}
\end{align}

\noindent where $\varepsilon_{\bf k}$ and $\varphi_{l,\bf k}$ are respectively the Floquet band energy and $l$-th harmonic of the wave-function, i.e., the solution of the Schr{\"o}dinger equation for $h_t$, is $\ket{\psi_{\bf k}(t)}=\sum_l e^{-i(\varepsilon_{\bf k}+ l\Omega) t}\varphi_{l,{\bf k}} \ket{{\bf k}}$. The function $S$, which encapsulates both emission and absorption processes, is given by:  
\begin{align}
\begin{aligned}
S(\omega) = [N_b (|\omega|)+\Theta(\omega)] \nu_B(|\omega|),
\label{S-function}
\end{aligned}
\end{align}
and plays a crucial role on the properties of the solutions of this Floquet-Boltzmann equation. The argument of $S$ can be interpreted as the energy of a boson that is emitted (absorbed) to (from) the bath when it is positive (negative), and therefore we see that the scattering behaves as if there was energy conservation modulo $\Omega$ (Floquet-Umklapp) [see Fig.~\ref{fig1}(b)].
Similar Floquet-Boltzmann equations have been employed in previous studies~\cite{seetharam2015controlled,genske2015floquet,esin2018quantized,seetharam2019steady}. For concreteness we will focus on the case of a parabolic band dispersion in one- and two-dimensions: 
\begin{align}
\epsilon_{\bf k} (t) & = 
[ {\bf k} - {\bf A}(t)]^2/2m,
\label{ParabolicModels}
\end{align} 
where $A (t) = A_0 \sin (\Omega t + \phi_0) $ in 1D and 
in 2D we will restrict to circularly polarized light, ${\bf A}(t) = A_0 [ \sin(\Omega t + \phi_0), \sin(\Omega t + \phi_0 \pm \pi/2) ]$, so that the steady state has rotational symmetry in order to simplify the numerical solutions (but our main conclusions remain qualitatively valid for arbitrary polarization). The Floquet energies and wavefunctions for this problem are shown in Appendix~\ref{Dimensionless-Models}. 

Notice that while the density matrix in Eq.~\eqref{rhot} is time-independent in canonical momentum basis, ${\bf k}$, physical quantities will generically oscillate periodically since they depend on the mechanical momentum ${\bf k}-{\bf A}(t)$. This steady state is an example of a periodic Gibbs ensemble (PGE)~\cite{lazarides2014periodic,lazarides2014equilibrium,lazarides2015fate,khemani2016phase}, which does not arise from many-body self-thermalization but rather from the coupling to a bath. We have recently shown that the steady state  when the system is coupled to a fermion bath is another PGE with a Fermi-Dirac staircase occupation with multiple jumps that behaves like a Fermi-liquid at low temperatures~\cite{matsyshyn2023fermi,shi2024floquet}.
As we will show, however, the solutions of Eq.~\eqref{Floquet-Boltzmann} lead to a dramatically different 
PGE where the occupation $f_{\bf k}$ does not feature jumps, but instead displays higher order non-analyticities. These effectively behave as the Fermi surfaces of a non-Fermi liquid, and we refer to this state ``Floquet Non-Fermi liquid''.
Remarkably, these non-analyticities remain sharp even when the bath is at finite temperature, a property we will call ``ultra-critical''. 
Our findings can be viewed as establishing that these non-analyticities appear for an infinitesimally weak electron-boson interaction ($\chi_0\to 0$), in contrast to the equilibrium, where non-Fermi liquid behavior typically emerges from strong interactions~\cite{giamarchi2003quantum,senthil2008critical,varma2002singular}. In Appendix~\ref{Persistence} we provide a rigorous demonstration that these non-analyticities remain robust after including electron-electron collisions (as derived in Ref.~\cite{genske2015floquet}) in Eq.~\eqref{Floquet-Boltzmann}. 
While this is reminiscent to setting in equilibrium where critical boson modes lead non-Fermi liquid behavior, e.g. within the Hertz-Millis framework~\cite{hertz1976quantum,millis1993effect}, we caution that our states should not be confused with equilibrium non-Fermi liquids, since non-equilibrium conditions are crucial in our setting.

\smallskip

\begin{figure}
\includegraphics[width=0.48\textwidth]{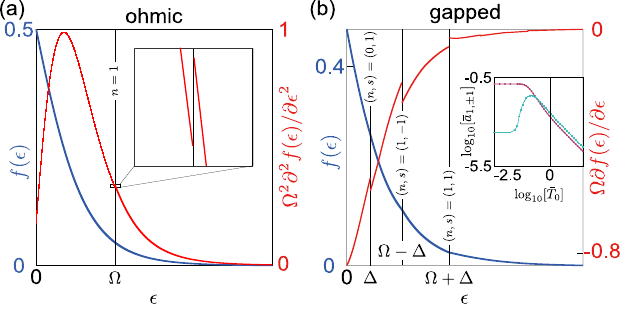}
\caption{Steady-state occupation and FFSs for 2D parabolic bands coupled to (a) Ohmic and (b) gapped baths at finite temperature.
The occupation (blue) and derivatives (red) are plotted as functions of $\epsilon = k^2 / 2m$. 
Inset (b) shows temperature $\bar{T}_0 = k_B T_0 /  \Omega$ dependence of the amplitude of non-analyticitiy $\bar{a}_{n=1, s=1} = \Omega a_{n=1, s=1}$ (green) and $\bar{a}_{n=1, s=-1} = \Omega a_{n=1, s=-1}$ (purple). 
Parameters used: $c_1 / c_2 \Omega =3 $, $ \Delta / \Omega = 3/10$, $A_0/\sqrt{6m\Omega} = 1/5$, particle density $n_0 / (2m\Omega) = 3/13$, bath temperatures $\bar{T}_0 = 3/10$.}
\label{fig2}
\end{figure}

\textit{\color{blue}Floquet non-Fermi liquid with ohmic bath.} We start by considering an Ohmic bosonic bath characterized by a density of states that vanishes linearly at low frequency:
\begin{align}
\nu_B^{\rm ohm} (\omega) = (c_1 \omega + c_2 \omega^2 +\cdots) \Theta(\omega).
\end{align}
We solve the Floquet-Boltzmann Eq.~\eqref{Floquet-Boltzmann} numerically~\cite{pal2024nonlinearsolve} by keeping the scattering matrix to the
second order of $A_0^2$. We have found that $f_{\bf k}$ has non-analyticities at $|{\bf k}|= k_{Fn}$ given by [see Fig.~\ref{fig2}(a)]:  
\begin{align}
k_{Fn}^2/2m =n \Omega, \ n=1,2,...
\label{FFs-ohm}
\end{align}
which are the ``Floquet Fermi surfaces'' (FFSs) \footnote{We also see a non-analiticity ${\bf k}=0$ associated with $n=0$, but here we focus on those with finite radius}. The size of FFSs is independent of the fermion density and thus generally different from the size of the equilibrium Fermi surface. The origin of these non-analyticities can be traced back to three crucial ingredients: (a) the absence of detailed balance in the scattering rates from Eq.~\eqref{general-scattering}, (b) the existence of a non-analyticity in DOS of the Floquet energies, which for a parabolic band is the DOS edge from the band bottom at ${\bf k}=0$, (c) the existence of a non-analytictiy of the $S$ function from Eq.~\eqref{S-function}, which for the Ohmic bath occurs at $\omega=0$. The idea is that states near a FFSs ($|{\bf k}|\approx k_{Fn}$), can scatter into or from those near the bottom of the band (${\bf k}\approx 0$) while absorbing or emitting bosons from the bath with negligibly small energy $\omega \approx 0$ [see Figs.~\ref{fig1}(b),(c)], which becomes allowed in the Floquet setting because energy is conserved only modulo $\Omega$ (Floquet-Umklapp). We have found that when the driving amplitude is small ($ |{\bf E}| \ll \Omega^2 /v_F $), the non-analyticity with $n=1$ is the strongest one (see Fig.~\ref{fig2}).

The degree of non-analyticity depends on space dimensionality ($d=1,2$).
In Appendix~\ref{SAnalysis}, we prove that for the Ohmic bath the second derivative of $f(k)$ exhibits non-analyticities of the form: $d^2 f/dk^2 = (d^2 f/dk^2)_{\text{reg}} + \sum_n b_n (k - k_{F_n})^{(d-2)/2} \, \Theta(k - k_{F_n})$. Here  $b_n$ is a constant, and $(d^2 f/dk^2)_{\text{reg}}$ includes the analytic part and weaker non-analyticities. Namely, these correspond to a square-root kink in $d f/dk$ in 1D, and a jump in $d^2 f/dk^2$ in 2D as illustrated in Fig.~\ref{fig2}(a).

\smallskip

\textit{\color{blue}Floquet Non-Fermi liquid with gapped bath.}
In order to illustrate that the non-analyticities of the $S$ function from Eq.~\eqref{S-function} affect the location of the FFSs, we now consider a bosonic bath with a gap ($\Delta$) in its DOS, given by:
\begin{align}
\nu_B^{\rm gap} (\omega) = \Theta(\omega - \Delta),
\quad
\Delta > 0.
\label{gap-bath}
\end{align}
We have verified that the behavior remains qualitatively similar if the above Theta function is replaced by other similar functions with a single discontinuity at $\omega=\Delta$, but we will focus on the above case Eq.~\eqref{gap-bath} for simplicity. Interestingly, in this case the non-analyticities of $f_{\bf k}$ are found at $|{\bf k}|= k_{Fn,s}$ given by [see Fig.~\ref{fig2}(b)]:
\begin{align}
k_{Fn,s}^2/2m=
n \Omega+s\Delta,
\label{FFs-gap}
\end{align}
\noindent where $n,s$ can be any integers as long as the right hand side is positive. We have observed that for weak  amplitudes the strongest non-analyticities are those with indices $(n,s)\in\{(1,1),(1,-1),(0,1)\}$. We will focus here in characterizing primarily those with $(n,s)=(1,\pm1)$, which involve non-trivial Floquet-Umklapp scattering. These non-analyticities are more pronounced in the gapped bath compared to the Ohmic case,  and appear in the first derivative of the occupation function [see Fig.~\ref{fig2}(b)]:
$
d f/dk = (d f/dk)_\text{reg} + \sum_{n,s} a_{n,s}
(k - k_{F_{n,s}})^{(d-2)/2}
\Theta(k - k_{F_{n,s}})$,
where $d$ is the dimensionality, and $a_{n,s}$ is the strength of the non-analyticity (see Appendix~\ref{SAnalysis} for proof).
Fig.~\ref{fig2}(b) shows that $a_{n=1,s=\pm1}$ decays as $1/T_0$  with bath temperature. 
Furthermore, for weak drivings, $a_{n=1,s=-1}$ scales as $A_0^{2}$, while $a_{n=1,s=+1}$ as $A_0^{4}$ (see Appendix~\ref{Floquet-Boltzmann-in-1Dand2D}).

\smallskip

\textit{\color{blue} Friedel oscillations and particle-hole continuum.}
To further substantiate the analogy between these non-analyticities and Fermi surfaces of a non-Fermi liquid, we will show that they give rise to phenomena that are often seen as fingerprints of the presence of Fermi surfaces. We begin by illustrating that density correlations display power-law decaying oscillations analogous to Friedel oscillations. The pair-correlation function of the fluid, $g({\bf r}_1, {\bf r}_2, t)$, measures the probability at time $t$ of finding a particle at location ${\bf r_2}$ given that another one is at ${\bf r}_1$.
We find that (see details in Appendix~\ref{PairCorrelation}) translational invariance and time independence of the one-body density matrix in canonical momentum basis [see Eq.~\eqref{EoMfull}], lead to $g({\bf r}_1, {\bf r}_2, t)=g(0, {\bf r}_2-{\bf r}_1,0)\equiv g({\bf r}_2-{\bf r}_1)$, and that this function is related to the real space Fourier transform, $\tilde{f}({\bf r})$, of the momentum occupation, $f({\bf k})$, as follows $g({\bf r})=1 - (\tilde{f}({\bf r}) / \tilde{f}(0))^2$. Moreover, in 2D the isotropy of $f_{\bf k}$ under circularly polarized light, leads to a pair correlation that only depends on the distance $|{\bf r}_2-{\bf r}_1|$. For the gapped bath we find the asymptotic behavior ($|{\bf r}| \to \infty$): 
\begin{align}
\begin{aligned}
\hspace{-7px}
\tilde{f}({\bf r})
\approx 
- \sum_{n,s}
\frac{(\pi a_{n,s}) (k_{F_{n,s}} / 2)^{\frac{d-1}{2}}}{ (\pi |{\bf r}|)^{ \frac{2d+1}{2} }}
\sin\Bigl( k_{F_{n,s}} |{\bf r}| + \frac{\pi}{4} \Bigr) ,
\label{fr-asymptotic}
\end{aligned}
\end{align}
where $n,s$ labels different Fermi surfaces.
Therefore, each Floquet Fermi surface contributes to a Friedel-like oscillation of $g({\bf r})$ with period $2k_{Fn,s}$ and decaying with power $1/|{\bf r}|^{2 d+1}$, as we illustrate in Appendix~\ref{PairCorrelation}. Remarkably, the above  power-law correlations remain even when the bath is at finite temperature, because the Floquet Fermi surface remains sharp, which is the behavior that we call ultra-critical.  This contrasts with the Friedel oscillations in equilibrium whose amplitude decays as $1/|{\bf r}|^{d+1}$ at zero temperature but exponentially at finite temperatures~\cite{giuliani2008quantum}, due to the thermal smearing of the Fermi surface.

Another manifestation of a sharp Fermi surface is the existence of non-analyticities of dynamical correlations at frequencies, $\omega$, and wave-vectors, ${\bf q}$, that match energies and wave-vectors of particle-hole excitations arbitrarily close to the fermi-surface, i.e. at the edge of the particle-hole continuum. In equilibrium the boundary of this particle-hole continuum smears out at finite temperatures due to the thermal smearing of the Fermi surface~\cite{giuliani2008quantum}. To investigate the presence of a particle-hole continuum in our setting, we consider the unequal-time density correlation function $C({\bf r}_1, t_1 ; {\bf r}_2, t_2)=
\braket{ n({\bf r}_1, t_1) n({\bf r}_2, t_2) }
-
\braket{ n({\bf r}_1, t_1) }
\braket{ n({\bf r}_2, t_2) }$. Translational invariance leads to $C({\bf r}_1, t_1 ; {\bf r}_2, t_2)=C({\bf r}_1-{\bf r}_2, t_1 ;0 , t_2)$. Because of the periodic driving this function depends not only on $t=t_1-t_2$, but also periodically on $\bar{t}=(t_1+t_2)/2$. Therefore, for simplicity, we will focus on the correlation averaged over one period, defined as  $\bar{C}({\bf r}, t)=\int_0^T C({\bf r}, t+t_2 ; 0, t_2) dt_2/T$. Its Fourier transform is (see Appendix~\ref{Appendix-DensityNoise}):
\begin{align}\begin{aligned}
\hspace{-7pt}
\bar{C} ({\bf q}, \omega)
= \frac{1}{V}
& \sum_{{\bf k},l}
f_{ {\bf k} + {\bf q} } \bar{f}_{\bf k}
\Phi^{(l)}_{{\bf k},{\bf k}+{\bf q}}
\delta ( \varepsilon_{\bf k} - \varepsilon_{ {\bf k} + {\bf q} } - l \omega ).
\label{Eq-DensityNoise}
\end{aligned}\end{align}
\begin{figure}
\includegraphics[width=0.48\textwidth]{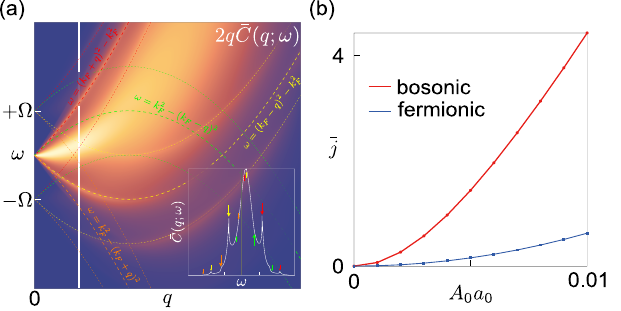}
\caption{
(a) Correlator $2 q \bar{C}(q;\omega)$ from Eq.~\eqref{Eq-DensityNoise} for  1D parabolic model coupled to a gapped bath with $T_0 / \Omega =1/5$ and $\Delta/ \Omega=1/100$.
Thick (thin) dashed lines are for $l=0$ ($l=\pm1$) non-analyticities [see Eq.~\eqref{ph-boundaries}].
The inset shows the cut (white) at $q / \sqrt{m \Omega} = 2/3$, with big (small) arrows being crossings between the thick (thin) dashed lines and the cut.
Parameters used:
$A_0/\sqrt{m \Omega} = 4/5$, and the particle density $n_0 / \sqrt{m \Omega} = 1/2$.
(b) Rectification current density $\bar{j} = j / 10^{-4} t_1$ as a function of the  driving amplitude $A_0$ for a 1D Bloch band coupled to an ohmic bosonic bath or an ideal fermionic bath using Eqs.~\eqref{Floquet-Boltzmann} and \eqref{CurrentDensity}.
Parameters:
$t_2/t_1 = 1/2$, $\Omega / t_1 = 2/3$, $k_B T_0 / t_1 = 1/100$ and particle density $n_0 a_0 = \sqrt{1/3}$.
}
\label{fig3}
\end{figure}

\noindent For an FFS with radius $k_F$ we have found that this function displays non-analyticites at finite bath temperatures for [see Fig.~\ref{fig3}(a)]:
\begin{align}
\omega= \pm k_F |{\bf q}| / m  
\pm |{\bf q}|^2/ 2m+l\Omega, 
\label{ph-boundaries}
\end{align}
\noindent where $l=0$ are the same values expected for a parabolic Fermi surface of radius $k_F$ in equilibrium~\cite{giuliani2008quantum}, and $l\neq 0$ are Floquet copies. The angular dependence of  $\bar{C} ({\bf q}, \omega)$ in 2D is non-trivial and is discussed in Appendix~\ref{Appendix-DensityNoise}.

\smallskip

\textit{\color{blue}Current rectification.}
Recently, we demonstrated that the steady state of a metal coupled to a fermion bath sustains a non-zero rectified electric current even when it is driven by monochromatic light whose frequency lies in its optical gap \cite{shi2023berry,matsyshyn2023fermi}, offering a counter-example to claims that such current should generally vanish~\cite{belinicher1986transient,onishi2022effects,golub2022raman,ivchenko1988magneto,pershoguba2022direct}. Here we will argue that this conclusion is  general and that in particular remains valid for boson baths. The steady state current for a single band model averaged over one period is \cite{matsyshyn2023fermi}:
\begin{align}
{\bf j}= \int d^d k f_{\bf k}\partial_{\bf k}  \varepsilon_{\bf k}/(2 \pi)^d.
\label{CurrentDensity}
\end{align}
\noindent where $\varepsilon_{\bf k}=\int_0^T \epsilon({\bf k} -{\bf A}(t)) dt/T$ is the Floquet energy. Crucially, in the  Floquet setting the steady state occupation $f_{\bf k}$ cannot be expressed as a function of the Floquet energy, $\varepsilon_{\bf k}$, alone. If that was the case (i.e. in equilibrium), then ${\bf j}$ could be expressed as the integral of a total derivative over the Brillouin zone and would always vanish. However, the lack of detailed balance combined with the fact that the scattering rates in Eq.~\eqref{Floquet-Boltzmann} depend explicitly on Floquet wave-functions and scattering matrix elements prohibit expressing $f_{\bf k}$ as a local function of $\varepsilon_{\bf k}$ (in contrast to equilibrium). Therefore, one expects that generically, as long as there is no ${\bf k} \to -{\bf k}$ symmetries, the steady state would have a non-zero rectified current, regardless of details of the collisions or the precise nature of the bath. In particular, the the non-analyticities that we have found in the current work, or the multiple steps found in our previous study with fermion baths~\cite{shi2024floquet}, are not crucial for the existence of the in-gap rectified current.

To illustrate this explicitly, we consider a 1D Bloch band without $k \to -k$ symmetry with dispersion $\epsilon_k= - t_1 \cos (a_0 k) - t_2 \sin (2 a_0 k)$, driven by $A(t) = A_0 \sin (\Omega t)$. Fig.~\ref{fig3}(b) shows the current in the steady state when the system is coupled to the bosonic bath and also the fermionic bath from Refs.~\cite{matsyshyn2023fermi,shi2024floquet}. We see that in both cases the rectified current is non-zero.

\smallskip

\textit{\color{blue} Generalizations and discussion.} We have observed the appearance of non-analyticities in the fermion occupation of Floquet bands coupled to boson baths which resemble Fermi surfaces in non-Fermi liquids. While we have illustrated this in detail for a parabolic band model our results are quite general. In fact, we can now conjecture a general criterion for the  location of the FFSs. Suppose that the Floquet spectrum has a non-analyticity in its DOS at a Floquet energy $\varepsilon_*$, e.g. arising from a local band minimum. Suppose that the DOS of the boson bath has a non-analyticity at some energy $\omega_*$. Then we expect the presence of a collection of Floquet Fermi Surfaces associated with these non-analyticities parameterized by two integers $(n,s)$ at the following momenta: 
\begin{align}
\varepsilon_{\bf k}= \varepsilon_*+n \Omega +s \omega_*.
\end{align}
\noindent where $\varepsilon_{\bf k}$ is the Floquet dispersion and  $\Omega$ is the driving frequency. We have shown that these non-analyticities remain sharp even when the boson bath is at finite temperature, leading to finite temperature Friedel-like oscillations of the density correlations, and sharp particle-hole continua edges, which are often viewed as characteristic phenomena of sharp Fermi surfaces.  

We have also demonstrated that the steady state occupation of a Floquet band coupled to a boson bath generally has a rectified current. This is a general conclusion independent of the details of collisions, and provides a general demonstration that there will be finite current rectification in ideal inversion-breaking metals illuminated by monochromatic light even when the frequency lies within its optical gap, as we have previously argued~\cite{matsyshyn2023fermi,shi2023berry,shi2024floquet}.

Our work illustrates how the steady states of open and driven quantum fluids can retain intriguing quantum characteristics. Let us summarize the key experimental conditions to realize such true Floquet steady states of electrons in materials. The visibility of Floquet effects is controlled by the amplitude of Floquet harmonics, which typically scale as $| \varphi_{l=\pm 1} \big|^2 \sim (ev_F |{\bf E}|/\hbar \Omega^2)^2$ (see Appendix~\ref{Dimensionless-Models}). Therefore, in the clean limit it is advantageous to have as low frequency as possible, so that a small electric field amplitude is sufficient to reach sizable values of these harmonics and consequently there is less heating from the continuous irradiation. But this frequency should also be larger than the inverse electron lifetime; otherwise, Floquet effects are also smeared. For example, in a clean material with a lifetime of tens of picoseconds, radiations with frequencies on the order of $100$~ GHz (i.e. microwaves) would be desirable.

The above conditions are similar to those used to observe a variety of beautiful phenomena in high-mobility 2DEGs~\cite{dmitriev2012nonequilibrium}, such as microwave-induced resistance oscillations (MIRO). MIRO are reminiscent of quantum oscillations under magnetic fields, but with a period consistent with a Fermi surface of area $\Omega$, such as the one we have found for $n=1$ from Eq.~(\ref{FFs-ohm}). Therefore, there is a tantalizing possibility that an ultracritical Floquet non-Fermi liquid state is realized in these materials. However, we note that we have also shown that a Floquet Fermi liquid could also give rise to MIRO over the same period~\cite{shi2024floquet}, and other mechanisms for MIRO have been proposed~\cite{dmitriev2012nonequilibrium}. This allows us to reiterate that for out-of-equilibrium settings, the details of the bath and relaxation mechanisms matter much more than in equilibrium. Thus, further work is needed to fully clarify the precise nature of steady states in these systems and to solve open puzzles of MIRO (see, e.g., Ref.~\cite{shi2015shubnikov}).

Although there have been several interesting experimental studies of Floquet physics with mid-infrared ultra-fast optical techniques~\cite{wang2013observation,mciver2020light,shan2021giant,aeschlimann2021survival,zhou2023pseudospin,zhang2024light,bielinski2025floquet,merboldt2024observation,choi2024direct}, we believe that realizing true Floquet steady states in clean materials with non-trivial electronic structures remains a largely open and remarkably fertile ground for discovering new physics. As we have emphasized, a strategy to reach such non-trivial quantum true Floquet steady-states is to drive with much lower frequencies. In this regard, we would like to highlight that the experimental investigation of steady states under microwaves in other clean materials with non-trivial electronic structures, such as graphene, remains an open research frontier (see, however, Ref.~\cite{mönch2020observation}).

\textit{\color{blue}Acknowledgements.} 
We would like to thank Matthias Thamm, Achim Rosch, Mark Rudner, Takashi Oka, Aris Alexandradinata, and Sebastian Diehl for stimulating and helpful discussions.
We acknowledge
support by the Deutsche Forschungsgemeinschaft (DFG) through research grant project numbers 542614019; 518372354; 555335098 (IS) and Singapore Ministry of Education under its Academic Research Fund Tier 2 Grant No. MOE-T2EP50222-0011 (JCWS) and Tier 3 Grant No. MOE- MOET32023-0003 Quantum Geometric Advantage (JCWS).
\bibliography{floquet-non-fermi-liquid}

\begin{thebibliography}{43}%
\makeatletter
\providecommand \@ifxundefined [1]{%
 \@ifx{#1\undefined}
}%
\providecommand \@ifnum [1]{%
 \ifnum #1\expandafter \@firstoftwo
 \else \expandafter \@secondoftwo
 \fi
}%
\providecommand \@ifx [1]{%
 \ifx #1\expandafter \@firstoftwo
 \else \expandafter \@secondoftwo
 \fi
}%
\providecommand \natexlab [1]{#1}%
\providecommand \enquote  [1]{``#1''}%
\providecommand \bibnamefont  [1]{#1}%
\providecommand \bibfnamefont [1]{#1}%
\providecommand \citenamefont [1]{#1}%
\providecommand \href@noop [0]{\@secondoftwo}%
\providecommand \href [0]{\begingroup \@sanitize@url \@href}%
\providecommand \@href[1]{\@@startlink{#1}\@@href}%
\providecommand \@@href[1]{\endgroup#1\@@endlink}%
\providecommand \@sanitize@url [0]{\catcode `\\12\catcode `\$12\catcode `\&12\catcode `\#12\catcode `\^12\catcode `\_12\catcode `\%12\relax}%
\providecommand \@@startlink[1]{}%
\providecommand \@@endlink[0]{}%
\providecommand \url  [0]{\begingroup\@sanitize@url \@url }%
\providecommand \@url [1]{\endgroup\@href {#1}{\urlprefix }}%
\providecommand \urlprefix  [0]{URL }%
\providecommand \Eprint [0]{\href }%
\providecommand \doibase [0]{https://doi.org/}%
\providecommand \selectlanguage [0]{\@gobble}%
\providecommand \bibinfo  [0]{\@secondoftwo}%
\providecommand \bibfield  [0]{\@secondoftwo}%
\providecommand \translation [1]{[#1]}%
\providecommand \BibitemOpen [0]{}%
\providecommand \bibitemStop [0]{}%
\providecommand \bibitemNoStop [0]{.\EOS\space}%
\providecommand \EOS [0]{\spacefactor3000\relax}%
\providecommand \BibitemShut  [1]{\csname bibitem#1\endcsname}%
\let\auto@bib@innerbib\@empty
\bibitem [{\citenamefont {Kardar}(2007)}]{kardar2007statistical}%
  \BibitemOpen
  \bibfield  {author} {\bibinfo {author} {\bibfnamefont {M.}~\bibnamefont {Kardar}},\ }\href@noop {} {\emph {\bibinfo {title} {Statistical physics of particles}}}\ (\bibinfo  {publisher} {Cambridge University Press},\ \bibinfo {year} {2007})\BibitemShut {NoStop}%
\bibitem [{\citenamefont {Giamarchi}(2003)}]{giamarchi2003quantum}%
  \BibitemOpen
  \bibfield  {author} {\bibinfo {author} {\bibfnamefont {T.}~\bibnamefont {Giamarchi}},\ }\href@noop {} {\emph {\bibinfo {title} {Quantum physics in one dimension}}},\ Vol.\ \bibinfo {volume} {121}\ (\bibinfo  {publisher} {Clarendon press},\ \bibinfo {year} {2003})\BibitemShut {NoStop}%
\bibitem [{\citenamefont {Senthil}(2008)}]{senthil2008critical}%
  \BibitemOpen
  \bibfield  {author} {\bibinfo {author} {\bibfnamefont {T.}~\bibnamefont {Senthil}},\ }\bibfield  {title} {\bibinfo {title} {Critical fermi surfaces and non-fermi liquid metals},\ }\href@noop {} {\bibfield  {journal} {\bibinfo  {journal} {Physical Review B}\ }\textbf {\bibinfo {volume} {78}},\ \bibinfo {pages} {035103} (\bibinfo {year} {2008})}\BibitemShut {NoStop}%
\bibitem [{\citenamefont {Varma}\ \emph {et~al.}(2002)\citenamefont {Varma}, \citenamefont {Nussinov},\ and\ \citenamefont {Van~Saarloos}}]{varma2002singular}%
  \BibitemOpen
  \bibfield  {author} {\bibinfo {author} {\bibfnamefont {C.~M.}\ \bibnamefont {Varma}}, \bibinfo {author} {\bibfnamefont {Z.}~\bibnamefont {Nussinov}},\ and\ \bibinfo {author} {\bibfnamefont {W.}~\bibnamefont {Van~Saarloos}},\ }\bibfield  {title} {\bibinfo {title} {Singular or non-fermi liquids},\ }\href@noop {} {\bibfield  {journal} {\bibinfo  {journal} {Physics Reports}\ }\textbf {\bibinfo {volume} {361}},\ \bibinfo {pages} {267} (\bibinfo {year} {2002})}\BibitemShut {NoStop}%
\bibitem [{\citenamefont {Matsyshyn}\ \emph {et~al.}(2023)\citenamefont {Matsyshyn}, \citenamefont {Song}, \citenamefont {Villadiego},\ and\ \citenamefont {Shi}}]{matsyshyn2023fermi}%
  \BibitemOpen
  \bibfield  {author} {\bibinfo {author} {\bibfnamefont {O.}~\bibnamefont {Matsyshyn}}, \bibinfo {author} {\bibfnamefont {J.~C.}\ \bibnamefont {Song}}, \bibinfo {author} {\bibfnamefont {I.~S.}\ \bibnamefont {Villadiego}},\ and\ \bibinfo {author} {\bibfnamefont {L.-k.}\ \bibnamefont {Shi}},\ }\bibfield  {title} {\bibinfo {title} {Fermi-dirac staircase occupation of floquet bands and current rectification inside the optical gap of metals: An exact approach},\ }\href@noop {} {\bibfield  {journal} {\bibinfo  {journal} {Physical Review B}\ }\textbf {\bibinfo {volume} {107}},\ \bibinfo {pages} {195135} (\bibinfo {year} {2023})}\BibitemShut {NoStop}%
\bibitem [{\citenamefont {Shi}\ \emph {et~al.}(2024)\citenamefont {Shi}, \citenamefont {Matsyshyn}, \citenamefont {Song},\ and\ \citenamefont {Villadiego}}]{shi2024floquet}%
  \BibitemOpen
  \bibfield  {author} {\bibinfo {author} {\bibfnamefont {L.-k.}\ \bibnamefont {Shi}}, \bibinfo {author} {\bibfnamefont {O.}~\bibnamefont {Matsyshyn}}, \bibinfo {author} {\bibfnamefont {J.~C.}\ \bibnamefont {Song}},\ and\ \bibinfo {author} {\bibfnamefont {I.~S.}\ \bibnamefont {Villadiego}},\ }\bibfield  {title} {\bibinfo {title} {Floquet fermi liquid},\ }\href@noop {} {\bibfield  {journal} {\bibinfo  {journal} {Physical Review Letters}\ }\textbf {\bibinfo {volume} {132}},\ \bibinfo {pages} {146402} (\bibinfo {year} {2024})}\BibitemShut {NoStop}%
\bibitem [{\citenamefont {Seetharam}\ \emph {et~al.}(2015)\citenamefont {Seetharam}, \citenamefont {Bardyn}, \citenamefont {Lindner}, \citenamefont {Rudner},\ and\ \citenamefont {Refael}}]{seetharam2015controlled}%
  \BibitemOpen
  \bibfield  {author} {\bibinfo {author} {\bibfnamefont {K.~I.}\ \bibnamefont {Seetharam}}, \bibinfo {author} {\bibfnamefont {C.-E.}\ \bibnamefont {Bardyn}}, \bibinfo {author} {\bibfnamefont {N.~H.}\ \bibnamefont {Lindner}}, \bibinfo {author} {\bibfnamefont {M.~S.}\ \bibnamefont {Rudner}},\ and\ \bibinfo {author} {\bibfnamefont {G.}~\bibnamefont {Refael}},\ }\bibfield  {title} {\bibinfo {title} {Controlled population of floquet-bloch states via coupling to bose and fermi baths},\ }\href@noop {} {\bibfield  {journal} {\bibinfo  {journal} {Physical Review X}\ }\textbf {\bibinfo {volume} {5}},\ \bibinfo {pages} {041050} (\bibinfo {year} {2015})}\BibitemShut {NoStop}%
\bibitem [{\citenamefont {Nazarov}\ and\ \citenamefont {Blanter}(2009)}]{nazarov2009quantum}%
  \BibitemOpen
  \bibfield  {author} {\bibinfo {author} {\bibfnamefont {Y.~V.}\ \bibnamefont {Nazarov}}\ and\ \bibinfo {author} {\bibfnamefont {Y.~M.}\ \bibnamefont {Blanter}},\ }\href@noop {} {\emph {\bibinfo {title} {Quantum transport: introduction to nanoscience}}}\ (\bibinfo  {publisher} {Cambridge university press},\ \bibinfo {year} {2009})\BibitemShut {NoStop}%
\bibitem [{\citenamefont {Tien}\ and\ \citenamefont {Gordon}(1963)}]{tien1963multiphoton}%
  \BibitemOpen
  \bibfield  {author} {\bibinfo {author} {\bibfnamefont {P.}~\bibnamefont {Tien}}\ and\ \bibinfo {author} {\bibfnamefont {J.}~\bibnamefont {Gordon}},\ }\bibfield  {title} {\bibinfo {title} {Multiphoton process observed in the interaction of microwave fields with the tunneling between superconductor films},\ }\href@noop {} {\bibfield  {journal} {\bibinfo  {journal} {Physical Review}\ }\textbf {\bibinfo {volume} {129}},\ \bibinfo {pages} {647} (\bibinfo {year} {1963})}\BibitemShut {NoStop}%
\bibitem [{\citenamefont {Kumari}\ \emph {et~al.}(2024)\citenamefont {Kumari}, \citenamefont {Seradjeh},\ and\ \citenamefont {Kundu}}]{kumari2024josephson}%
  \BibitemOpen
  \bibfield  {author} {\bibinfo {author} {\bibfnamefont {R.}~\bibnamefont {Kumari}}, \bibinfo {author} {\bibfnamefont {B.}~\bibnamefont {Seradjeh}},\ and\ \bibinfo {author} {\bibfnamefont {A.}~\bibnamefont {Kundu}},\ }\bibfield  {title} {\bibinfo {title} {Josephson-current signatures of unpaired floquet majorana fermions},\ }\href@noop {} {\bibfield  {journal} {\bibinfo  {journal} {Physical Review Letters}\ }\textbf {\bibinfo {volume} {133}},\ \bibinfo {pages} {196601} (\bibinfo {year} {2024})}\BibitemShut {NoStop}%
\bibitem [{\citenamefont {Asmar}\ and\ \citenamefont {Tse}(2022)}]{asmar2022impurity}%
  \BibitemOpen
  \bibfield  {author} {\bibinfo {author} {\bibfnamefont {M.~M.}\ \bibnamefont {Asmar}}\ and\ \bibinfo {author} {\bibfnamefont {W.-K.}\ \bibnamefont {Tse}},\ }\bibfield  {title} {\bibinfo {title} {Impurity screening and friedel oscillations in floquet-driven two-dimensional metals},\ }\href@noop {} {\bibfield  {journal} {\bibinfo  {journal} {Journal of Physics: Condensed Matter}\ }\textbf {\bibinfo {volume} {34}},\ \bibinfo {pages} {315602} (\bibinfo {year} {2022})}\BibitemShut {NoStop}%
\bibitem [{\citenamefont {Le}\ \emph {et~al.}(2024)\citenamefont {Le}, \citenamefont {Jiang}, \citenamefont {Tu}, \citenamefont {Bian}, \citenamefont {Ma}, \citenamefont {Shi}, \citenamefont {Jia}, \citenamefont {Li}, \citenamefont {Lyu}, \citenamefont {Cao} \emph {et~al.}}]{le2024inverse}%
  \BibitemOpen
  \bibfield  {author} {\bibinfo {author} {\bibfnamefont {T.}~\bibnamefont {Le}}, \bibinfo {author} {\bibfnamefont {R.}~\bibnamefont {Jiang}}, \bibinfo {author} {\bibfnamefont {L.}~\bibnamefont {Tu}}, \bibinfo {author} {\bibfnamefont {R.}~\bibnamefont {Bian}}, \bibinfo {author} {\bibfnamefont {Y.}~\bibnamefont {Ma}}, \bibinfo {author} {\bibfnamefont {Y.}~\bibnamefont {Shi}}, \bibinfo {author} {\bibfnamefont {K.}~\bibnamefont {Jia}}, \bibinfo {author} {\bibfnamefont {Z.}~\bibnamefont {Li}}, \bibinfo {author} {\bibfnamefont {Z.}~\bibnamefont {Lyu}}, \bibinfo {author} {\bibfnamefont {X.}~\bibnamefont {Cao}}, \emph {et~al.},\ }\bibfield  {title} {\bibinfo {title} {Inverse-current quantum electro-oscillations in a charge density wave insulator},\ }\href@noop {} {\bibfield  {journal} {\bibinfo  {journal} {Physical Review B}\ }\textbf {\bibinfo {volume} {109}},\ \bibinfo {pages} {245123} (\bibinfo {year} {2024})}\BibitemShut {NoStop}%
\bibitem [{\citenamefont {Vasko}\ and\ \citenamefont {Raichev}(2005)}]{VaskoRaichev}%
  \BibitemOpen
  \bibfield  {author} {\bibinfo {author} {\bibfnamefont {F.~T.}\ \bibnamefont {Vasko}}\ and\ \bibinfo {author} {\bibfnamefont {O.~E.}\ \bibnamefont {Raichev}},\ }\href@noop {} {\emph {\bibinfo {title} {Quantum kinetic theory and applications}}}\ (\bibinfo  {publisher} {Springer},\ \bibinfo {year} {2005})\BibitemShut {NoStop}%
\bibitem [{\citenamefont {Genske}\ and\ \citenamefont {Rosch}(2015)}]{genske2015floquet}%
  \BibitemOpen
  \bibfield  {author} {\bibinfo {author} {\bibfnamefont {M.}~\bibnamefont {Genske}}\ and\ \bibinfo {author} {\bibfnamefont {A.}~\bibnamefont {Rosch}},\ }\bibfield  {title} {\bibinfo {title} {Floquet-boltzmann equation for periodically driven fermi systems},\ }\href@noop {} {\bibfield  {journal} {\bibinfo  {journal} {Physical Review A}\ }\textbf {\bibinfo {volume} {92}},\ \bibinfo {pages} {062108} (\bibinfo {year} {2015})}\BibitemShut {NoStop}%
\bibitem [{\citenamefont {Esin}\ \emph {et~al.}(2018)\citenamefont {Esin}, \citenamefont {Rudner}, \citenamefont {Refael},\ and\ \citenamefont {Lindner}}]{esin2018quantized}%
  \BibitemOpen
  \bibfield  {author} {\bibinfo {author} {\bibfnamefont {I.}~\bibnamefont {Esin}}, \bibinfo {author} {\bibfnamefont {M.~S.}\ \bibnamefont {Rudner}}, \bibinfo {author} {\bibfnamefont {G.}~\bibnamefont {Refael}},\ and\ \bibinfo {author} {\bibfnamefont {N.~H.}\ \bibnamefont {Lindner}},\ }\bibfield  {title} {\bibinfo {title} {Quantized transport and steady states of floquet topological insulators},\ }\href@noop {} {\bibfield  {journal} {\bibinfo  {journal} {Physical Review B}\ }\textbf {\bibinfo {volume} {97}},\ \bibinfo {pages} {245401} (\bibinfo {year} {2018})}\BibitemShut {NoStop}%
\bibitem [{\citenamefont {Seetharam}\ \emph {et~al.}(2019)\citenamefont {Seetharam}, \citenamefont {Bardyn}, \citenamefont {Lindner}, \citenamefont {Rudner},\ and\ \citenamefont {Refael}}]{seetharam2019steady}%
  \BibitemOpen
  \bibfield  {author} {\bibinfo {author} {\bibfnamefont {K.~I.}\ \bibnamefont {Seetharam}}, \bibinfo {author} {\bibfnamefont {C.-E.}\ \bibnamefont {Bardyn}}, \bibinfo {author} {\bibfnamefont {N.~H.}\ \bibnamefont {Lindner}}, \bibinfo {author} {\bibfnamefont {M.~S.}\ \bibnamefont {Rudner}},\ and\ \bibinfo {author} {\bibfnamefont {G.}~\bibnamefont {Refael}},\ }\bibfield  {title} {\bibinfo {title} {Steady states of interacting floquet insulators},\ }\href@noop {} {\bibfield  {journal} {\bibinfo  {journal} {Physical Review B}\ }\textbf {\bibinfo {volume} {99}},\ \bibinfo {pages} {014307} (\bibinfo {year} {2019})}\BibitemShut {NoStop}%
\bibitem [{\citenamefont {Lazarides}\ \emph {et~al.}(2014{\natexlab{a}})\citenamefont {Lazarides}, \citenamefont {Das},\ and\ \citenamefont {Moessner}}]{lazarides2014periodic}%
  \BibitemOpen
  \bibfield  {author} {\bibinfo {author} {\bibfnamefont {A.}~\bibnamefont {Lazarides}}, \bibinfo {author} {\bibfnamefont {A.}~\bibnamefont {Das}},\ and\ \bibinfo {author} {\bibfnamefont {R.}~\bibnamefont {Moessner}},\ }\bibfield  {title} {\bibinfo {title} {Periodic thermodynamics of isolated quantum systems},\ }\href@noop {} {\bibfield  {journal} {\bibinfo  {journal} {Physical review letters}\ }\textbf {\bibinfo {volume} {112}},\ \bibinfo {pages} {150401} (\bibinfo {year} {2014}{\natexlab{a}})}\BibitemShut {NoStop}%
\bibitem [{\citenamefont {Lazarides}\ \emph {et~al.}(2014{\natexlab{b}})\citenamefont {Lazarides}, \citenamefont {Das},\ and\ \citenamefont {Moessner}}]{lazarides2014equilibrium}%
  \BibitemOpen
  \bibfield  {author} {\bibinfo {author} {\bibfnamefont {A.}~\bibnamefont {Lazarides}}, \bibinfo {author} {\bibfnamefont {A.}~\bibnamefont {Das}},\ and\ \bibinfo {author} {\bibfnamefont {R.}~\bibnamefont {Moessner}},\ }\bibfield  {title} {\bibinfo {title} {Equilibrium states of generic quantum systems subject to periodic driving},\ }\href@noop {} {\bibfield  {journal} {\bibinfo  {journal} {Physical Review E}\ }\textbf {\bibinfo {volume} {90}},\ \bibinfo {pages} {012110} (\bibinfo {year} {2014}{\natexlab{b}})}\BibitemShut {NoStop}%
\bibitem [{\citenamefont {Lazarides}\ \emph {et~al.}(2015)\citenamefont {Lazarides}, \citenamefont {Das},\ and\ \citenamefont {Moessner}}]{lazarides2015fate}%
  \BibitemOpen
  \bibfield  {author} {\bibinfo {author} {\bibfnamefont {A.}~\bibnamefont {Lazarides}}, \bibinfo {author} {\bibfnamefont {A.}~\bibnamefont {Das}},\ and\ \bibinfo {author} {\bibfnamefont {R.}~\bibnamefont {Moessner}},\ }\bibfield  {title} {\bibinfo {title} {Fate of many-body localization under periodic driving},\ }\href@noop {} {\bibfield  {journal} {\bibinfo  {journal} {Physical review letters}\ }\textbf {\bibinfo {volume} {115}},\ \bibinfo {pages} {030402} (\bibinfo {year} {2015})}\BibitemShut {NoStop}%
\bibitem [{\citenamefont {Khemani}\ \emph {et~al.}(2016)\citenamefont {Khemani}, \citenamefont {Lazarides}, \citenamefont {Moessner},\ and\ \citenamefont {Sondhi}}]{khemani2016phase}%
  \BibitemOpen
  \bibfield  {author} {\bibinfo {author} {\bibfnamefont {V.}~\bibnamefont {Khemani}}, \bibinfo {author} {\bibfnamefont {A.}~\bibnamefont {Lazarides}}, \bibinfo {author} {\bibfnamefont {R.}~\bibnamefont {Moessner}},\ and\ \bibinfo {author} {\bibfnamefont {S.~L.}\ \bibnamefont {Sondhi}},\ }\bibfield  {title} {\bibinfo {title} {Phase structure of driven quantum systems},\ }\href@noop {} {\bibfield  {journal} {\bibinfo  {journal} {Physical review letters}\ }\textbf {\bibinfo {volume} {116}},\ \bibinfo {pages} {250401} (\bibinfo {year} {2016})}\BibitemShut {NoStop}%
\bibitem [{\citenamefont {Hertz}(1976)}]{hertz1976quantum}%
  \BibitemOpen
  \bibfield  {author} {\bibinfo {author} {\bibfnamefont {J.~A.}\ \bibnamefont {Hertz}},\ }\bibfield  {title} {\bibinfo {title} {Quantum critical phenomena},\ }\href@noop {} {\bibfield  {journal} {\bibinfo  {journal} {Physical Review B}\ }\textbf {\bibinfo {volume} {14}},\ \bibinfo {pages} {1165} (\bibinfo {year} {1976})}\BibitemShut {NoStop}%
\bibitem [{\citenamefont {Millis}(1993)}]{millis1993effect}%
  \BibitemOpen
  \bibfield  {author} {\bibinfo {author} {\bibfnamefont {A.}~\bibnamefont {Millis}},\ }\bibfield  {title} {\bibinfo {title} {Effect of a nonzero temperature on quantum critical points in itinerant fermion systems},\ }\href@noop {} {\bibfield  {journal} {\bibinfo  {journal} {Physical Review B}\ }\textbf {\bibinfo {volume} {48}},\ \bibinfo {pages} {7183} (\bibinfo {year} {1993})}\BibitemShut {NoStop}%
\bibitem [{\citenamefont {Pal}\ \emph {et~al.}(2024)\citenamefont {Pal}, \citenamefont {Holtorf}, \citenamefont {Larsson}, \citenamefont {Loman}, \citenamefont {Schaefer}, \citenamefont {Qu}, \citenamefont {Edelman}, \citenamefont {Rackauckas} \emph {et~al.}}]{pal2024nonlinearsolve}%
  \BibitemOpen
  \bibfield  {author} {\bibinfo {author} {\bibfnamefont {A.}~\bibnamefont {Pal}}, \bibinfo {author} {\bibfnamefont {F.}~\bibnamefont {Holtorf}}, \bibinfo {author} {\bibfnamefont {A.}~\bibnamefont {Larsson}}, \bibinfo {author} {\bibfnamefont {T.}~\bibnamefont {Loman}}, \bibinfo {author} {\bibfnamefont {F.}~\bibnamefont {Schaefer}}, \bibinfo {author} {\bibfnamefont {Q.}~\bibnamefont {Qu}}, \bibinfo {author} {\bibfnamefont {A.}~\bibnamefont {Edelman}}, \bibinfo {author} {\bibfnamefont {C.}~\bibnamefont {Rackauckas}}, \emph {et~al.},\ }\bibfield  {title} {\bibinfo {title} {Nonlinearsolve. jl: High-performance and robust solvers for systems of nonlinear equations in julia},\ }\href@noop {} {\bibfield  {journal} {\bibinfo  {journal} {arXiv preprint arXiv:2403.16341}\ } (\bibinfo {year} {2024})}\BibitemShut {NoStop}%
\bibitem [{Note1()}]{Note1}%
  \BibitemOpen
  \bibinfo {note} {We also see a non-analiticity ${\protect \bf k}=0$ associated with $n=0$, but here we focus on those with finite radius}\BibitemShut {NoStop}%
\bibitem [{\citenamefont {Giuliani}\ and\ \citenamefont {Vignale}(2008)}]{giuliani2008quantum}%
  \BibitemOpen
  \bibfield  {author} {\bibinfo {author} {\bibfnamefont {G.}~\bibnamefont {Giuliani}}\ and\ \bibinfo {author} {\bibfnamefont {G.}~\bibnamefont {Vignale}},\ }\href@noop {} {\emph {\bibinfo {title} {Quantum theory of the electron liquid}}}\ (\bibinfo  {publisher} {Cambridge university press},\ \bibinfo {year} {2008})\BibitemShut {NoStop}%
\bibitem [{\citenamefont {Shi}\ \emph {et~al.}(2023)\citenamefont {Shi}, \citenamefont {Matsyshyn}, \citenamefont {Song},\ and\ \citenamefont {Villadiego}}]{shi2023berry}%
  \BibitemOpen
  \bibfield  {author} {\bibinfo {author} {\bibfnamefont {L.-k.}\ \bibnamefont {Shi}}, \bibinfo {author} {\bibfnamefont {O.}~\bibnamefont {Matsyshyn}}, \bibinfo {author} {\bibfnamefont {J.~C.}\ \bibnamefont {Song}},\ and\ \bibinfo {author} {\bibfnamefont {I.~S.}\ \bibnamefont {Villadiego}},\ }\bibfield  {title} {\bibinfo {title} {Berry-dipole photovoltaic demon and the thermodynamics of photocurrent generation within the optical gap of metals},\ }\href@noop {} {\bibfield  {journal} {\bibinfo  {journal} {Physical Review B}\ }\textbf {\bibinfo {volume} {107}},\ \bibinfo {pages} {125151} (\bibinfo {year} {2023})}\BibitemShut {NoStop}%
\bibitem [{\citenamefont {Belinicher}\ \emph {et~al.}(1986)\citenamefont {Belinicher}, \citenamefont {Ivchenko},\ and\ \citenamefont {Pikus}}]{belinicher1986transient}%
  \BibitemOpen
  \bibfield  {author} {\bibinfo {author} {\bibfnamefont {V.}~\bibnamefont {Belinicher}}, \bibinfo {author} {\bibfnamefont {E.}~\bibnamefont {Ivchenko}},\ and\ \bibinfo {author} {\bibfnamefont {G.}~\bibnamefont {Pikus}},\ }\bibfield  {title} {\bibinfo {title} {Transient photocurrent in gyrotropic crystals},\ }\href@noop {} {\bibfield  {journal} {\bibinfo  {journal} {Soviet Physics Semiconductors-Ussr}\ }\textbf {\bibinfo {volume} {20}},\ \bibinfo {pages} {558} (\bibinfo {year} {1986})}\BibitemShut {NoStop}%
\bibitem [{\citenamefont {Onishi}\ \emph {et~al.}(2022)\citenamefont {Onishi}, \citenamefont {Watanabe}, \citenamefont {Morimoto},\ and\ \citenamefont {Nagaosa}}]{onishi2022effects}%
  \BibitemOpen
  \bibfield  {author} {\bibinfo {author} {\bibfnamefont {Y.}~\bibnamefont {Onishi}}, \bibinfo {author} {\bibfnamefont {H.}~\bibnamefont {Watanabe}}, \bibinfo {author} {\bibfnamefont {T.}~\bibnamefont {Morimoto}},\ and\ \bibinfo {author} {\bibfnamefont {N.}~\bibnamefont {Nagaosa}},\ }\bibfield  {title} {\bibinfo {title} {Effects of relaxation on the photovoltaic effect and possibility for photocurrent within the transparent region},\ }\href@noop {} {\bibfield  {journal} {\bibinfo  {journal} {Physical Review B}\ }\textbf {\bibinfo {volume} {106}},\ \bibinfo {pages} {235110} (\bibinfo {year} {2022})}\BibitemShut {NoStop}%
\bibitem [{\citenamefont {Golub}\ and\ \citenamefont {Glazov}(2022)}]{golub2022raman}%
  \BibitemOpen
  \bibfield  {author} {\bibinfo {author} {\bibfnamefont {L.}~\bibnamefont {Golub}}\ and\ \bibinfo {author} {\bibfnamefont {M.}~\bibnamefont {Glazov}},\ }\bibfield  {title} {\bibinfo {title} {Raman photogalvanic effect: Photocurrent at inelastic light scattering},\ }\href@noop {} {\bibfield  {journal} {\bibinfo  {journal} {Physical Review B}\ }\textbf {\bibinfo {volume} {106}},\ \bibinfo {pages} {205205} (\bibinfo {year} {2022})}\BibitemShut {NoStop}%
\bibitem [{\citenamefont {Ivchenko}\ \emph {et~al.}(1988)\citenamefont {Ivchenko}, \citenamefont {Lyanda-Geller},\ and\ \citenamefont {Pikus}}]{ivchenko1988magneto}%
  \BibitemOpen
  \bibfield  {author} {\bibinfo {author} {\bibfnamefont {E.}~\bibnamefont {Ivchenko}}, \bibinfo {author} {\bibfnamefont {Y.~B.}\ \bibnamefont {Lyanda-Geller}},\ and\ \bibinfo {author} {\bibfnamefont {G.}~\bibnamefont {Pikus}},\ }\bibfield  {title} {\bibinfo {title} {Magneto-photogalvanic effects in noncentrosymmetric crystals},\ }\href@noop {} {\bibfield  {journal} {\bibinfo  {journal} {Ferroelectrics}\ }\textbf {\bibinfo {volume} {83}},\ \bibinfo {pages} {19} (\bibinfo {year} {1988})}\BibitemShut {NoStop}%
\bibitem [{\citenamefont {Pershoguba}\ and\ \citenamefont {Yakovenko}(2022)}]{pershoguba2022direct}%
  \BibitemOpen
  \bibfield  {author} {\bibinfo {author} {\bibfnamefont {S.~S.}\ \bibnamefont {Pershoguba}}\ and\ \bibinfo {author} {\bibfnamefont {V.~M.}\ \bibnamefont {Yakovenko}},\ }\bibfield  {title} {\bibinfo {title} {Direct current in a stirred optical lattice},\ }\href@noop {} {\bibfield  {journal} {\bibinfo  {journal} {Annals of Physics}\ }\textbf {\bibinfo {volume} {447}},\ \bibinfo {pages} {169075} (\bibinfo {year} {2022})}\BibitemShut {NoStop}%
\bibitem [{\citenamefont {Dmitriev}\ \emph {et~al.}(2012)\citenamefont {Dmitriev}, \citenamefont {Mirlin}, \citenamefont {Polyakov},\ and\ \citenamefont {Zudov}}]{dmitriev2012nonequilibrium}%
  \BibitemOpen
  \bibfield  {author} {\bibinfo {author} {\bibfnamefont {I.}~\bibnamefont {Dmitriev}}, \bibinfo {author} {\bibfnamefont {A.}~\bibnamefont {Mirlin}}, \bibinfo {author} {\bibfnamefont {D.}~\bibnamefont {Polyakov}},\ and\ \bibinfo {author} {\bibfnamefont {M.}~\bibnamefont {Zudov}},\ }\bibfield  {title} {\bibinfo {title} {Nonequilibrium phenomena in high landau levels},\ }\href@noop {} {\bibfield  {journal} {\bibinfo  {journal} {Reviews of Modern Physics}\ }\textbf {\bibinfo {volume} {84}},\ \bibinfo {pages} {1709} (\bibinfo {year} {2012})}\BibitemShut {NoStop}%
\bibitem [{\citenamefont {Shi}\ \emph {et~al.}(2015)\citenamefont {Shi}, \citenamefont {Martin}, \citenamefont {Hatke}, \citenamefont {Zudov}, \citenamefont {Watson}, \citenamefont {Gardner}, \citenamefont {Manfra}, \citenamefont {Pfeiffer},\ and\ \citenamefont {West}}]{shi2015shubnikov}%
  \BibitemOpen
  \bibfield  {author} {\bibinfo {author} {\bibfnamefont {Q.}~\bibnamefont {Shi}}, \bibinfo {author} {\bibfnamefont {P.}~\bibnamefont {Martin}}, \bibinfo {author} {\bibfnamefont {A.}~\bibnamefont {Hatke}}, \bibinfo {author} {\bibfnamefont {M.}~\bibnamefont {Zudov}}, \bibinfo {author} {\bibfnamefont {J.}~\bibnamefont {Watson}}, \bibinfo {author} {\bibfnamefont {G.}~\bibnamefont {Gardner}}, \bibinfo {author} {\bibfnamefont {M.}~\bibnamefont {Manfra}}, \bibinfo {author} {\bibfnamefont {L.}~\bibnamefont {Pfeiffer}},\ and\ \bibinfo {author} {\bibfnamefont {K.}~\bibnamefont {West}},\ }\bibfield  {title} {\bibinfo {title} {Shubnikov--de haas oscillations in a two-dimensional electron gas under subterahertz radiation},\ }\href@noop {} {\bibfield  {journal} {\bibinfo  {journal} {Physical Review B}\ }\textbf {\bibinfo {volume} {92}},\ \bibinfo {pages} {081405} (\bibinfo {year} {2015})}\BibitemShut {NoStop}%
\bibitem [{\citenamefont {Wang}\ \emph {et~al.}(2013)\citenamefont {Wang}, \citenamefont {Steinberg}, \citenamefont {Jarillo-Herrero},\ and\ \citenamefont {Gedik}}]{wang2013observation}%
  \BibitemOpen
  \bibfield  {author} {\bibinfo {author} {\bibfnamefont {Y.}~\bibnamefont {Wang}}, \bibinfo {author} {\bibfnamefont {H.}~\bibnamefont {Steinberg}}, \bibinfo {author} {\bibfnamefont {P.}~\bibnamefont {Jarillo-Herrero}},\ and\ \bibinfo {author} {\bibfnamefont {N.}~\bibnamefont {Gedik}},\ }\bibfield  {title} {\bibinfo {title} {Observation of floquet-bloch states on the surface of a topological insulator},\ }\href@noop {} {\bibfield  {journal} {\bibinfo  {journal} {Science}\ }\textbf {\bibinfo {volume} {342}},\ \bibinfo {pages} {453} (\bibinfo {year} {2013})}\BibitemShut {NoStop}%
\bibitem [{\citenamefont {McIver}\ \emph {et~al.}(2020)\citenamefont {McIver}, \citenamefont {Schulte}, \citenamefont {Stein}, \citenamefont {Matsuyama}, \citenamefont {Jotzu}, \citenamefont {Meier},\ and\ \citenamefont {Cavalleri}}]{mciver2020light}%
  \BibitemOpen
  \bibfield  {author} {\bibinfo {author} {\bibfnamefont {J.~W.}\ \bibnamefont {McIver}}, \bibinfo {author} {\bibfnamefont {B.}~\bibnamefont {Schulte}}, \bibinfo {author} {\bibfnamefont {F.-U.}\ \bibnamefont {Stein}}, \bibinfo {author} {\bibfnamefont {T.}~\bibnamefont {Matsuyama}}, \bibinfo {author} {\bibfnamefont {G.}~\bibnamefont {Jotzu}}, \bibinfo {author} {\bibfnamefont {G.}~\bibnamefont {Meier}},\ and\ \bibinfo {author} {\bibfnamefont {A.}~\bibnamefont {Cavalleri}},\ }\bibfield  {title} {\bibinfo {title} {Light-induced anomalous hall effect in graphene},\ }\href@noop {} {\bibfield  {journal} {\bibinfo  {journal} {Nature physics}\ }\textbf {\bibinfo {volume} {16}},\ \bibinfo {pages} {38} (\bibinfo {year} {2020})}\BibitemShut {NoStop}%
\bibitem [{\citenamefont {Shan}\ \emph {et~al.}(2021)\citenamefont {Shan}, \citenamefont {Ye}, \citenamefont {Chu}, \citenamefont {Lee}, \citenamefont {Park}, \citenamefont {Balents},\ and\ \citenamefont {Hsieh}}]{shan2021giant}%
  \BibitemOpen
  \bibfield  {author} {\bibinfo {author} {\bibfnamefont {J.-Y.}\ \bibnamefont {Shan}}, \bibinfo {author} {\bibfnamefont {M.}~\bibnamefont {Ye}}, \bibinfo {author} {\bibfnamefont {H.}~\bibnamefont {Chu}}, \bibinfo {author} {\bibfnamefont {S.}~\bibnamefont {Lee}}, \bibinfo {author} {\bibfnamefont {J.-G.}\ \bibnamefont {Park}}, \bibinfo {author} {\bibfnamefont {L.}~\bibnamefont {Balents}},\ and\ \bibinfo {author} {\bibfnamefont {D.}~\bibnamefont {Hsieh}},\ }\bibfield  {title} {\bibinfo {title} {Giant modulation of optical nonlinearity by floquet engineering},\ }\href@noop {} {\bibfield  {journal} {\bibinfo  {journal} {Nature}\ }\textbf {\bibinfo {volume} {600}},\ \bibinfo {pages} {235} (\bibinfo {year} {2021})}\BibitemShut {NoStop}%
\bibitem [{\citenamefont {Aeschlimann}\ \emph {et~al.}(2021)\citenamefont {Aeschlimann}, \citenamefont {Sato}, \citenamefont {Krause}, \citenamefont {Ch{\'a}vez-Cervantes}, \citenamefont {De~Giovannini}, \citenamefont {Hübener}, \citenamefont {Forti}, \citenamefont {Coletti}, \citenamefont {Hanff}, \citenamefont {Rossnagel} \emph {et~al.}}]{aeschlimann2021survival}%
  \BibitemOpen
  \bibfield  {author} {\bibinfo {author} {\bibfnamefont {S.}~\bibnamefont {Aeschlimann}}, \bibinfo {author} {\bibfnamefont {S.~A.}\ \bibnamefont {Sato}}, \bibinfo {author} {\bibfnamefont {R.}~\bibnamefont {Krause}}, \bibinfo {author} {\bibfnamefont {M.}~\bibnamefont {Ch{\'a}vez-Cervantes}}, \bibinfo {author} {\bibfnamefont {U.}~\bibnamefont {De~Giovannini}}, \bibinfo {author} {\bibfnamefont {H.}~\bibnamefont {Hübener}}, \bibinfo {author} {\bibfnamefont {S.}~\bibnamefont {Forti}}, \bibinfo {author} {\bibfnamefont {C.}~\bibnamefont {Coletti}}, \bibinfo {author} {\bibfnamefont {K.}~\bibnamefont {Hanff}}, \bibinfo {author} {\bibfnamefont {K.}~\bibnamefont {Rossnagel}}, \emph {et~al.},\ }\bibfield  {title} {\bibinfo {title} {Survival of floquet--bloch states in the presence of scattering},\ }\href@noop {} {\bibfield  {journal} {\bibinfo  {journal} {Nano letters}\ }\textbf {\bibinfo {volume} {21}},\ \bibinfo {pages} {5028} (\bibinfo {year} {2021})}\BibitemShut {NoStop}%
\bibitem [{\citenamefont {Zhou}\ \emph {et~al.}(2023)\citenamefont {Zhou}, \citenamefont {Bao}, \citenamefont {Fan}, \citenamefont {Zhou}, \citenamefont {Gao}, \citenamefont {Zhong}, \citenamefont {Lin}, \citenamefont {Liu}, \citenamefont {Yu}, \citenamefont {Tang} \emph {et~al.}}]{zhou2023pseudospin}%
  \BibitemOpen
  \bibfield  {author} {\bibinfo {author} {\bibfnamefont {S.}~\bibnamefont {Zhou}}, \bibinfo {author} {\bibfnamefont {C.}~\bibnamefont {Bao}}, \bibinfo {author} {\bibfnamefont {B.}~\bibnamefont {Fan}}, \bibinfo {author} {\bibfnamefont {H.}~\bibnamefont {Zhou}}, \bibinfo {author} {\bibfnamefont {Q.}~\bibnamefont {Gao}}, \bibinfo {author} {\bibfnamefont {H.}~\bibnamefont {Zhong}}, \bibinfo {author} {\bibfnamefont {T.}~\bibnamefont {Lin}}, \bibinfo {author} {\bibfnamefont {H.}~\bibnamefont {Liu}}, \bibinfo {author} {\bibfnamefont {P.}~\bibnamefont {Yu}}, \bibinfo {author} {\bibfnamefont {P.}~\bibnamefont {Tang}}, \emph {et~al.},\ }\bibfield  {title} {\bibinfo {title} {Pseudospin-selective floquet band engineering in black phosphorus},\ }\href@noop {} {\bibfield  {journal} {\bibinfo  {journal} {Nature}\ }\textbf {\bibinfo {volume} {614}},\ \bibinfo {pages} {75} (\bibinfo {year} {2023})}\BibitemShut {NoStop}%
\bibitem [{\citenamefont {Zhang}\ \emph {et~al.}(2024)\citenamefont {Zhang}, \citenamefont {Carbin}, \citenamefont {Culver}, \citenamefont {Du}, \citenamefont {Wang}, \citenamefont {Cheong}, \citenamefont {Roy},\ and\ \citenamefont {Kogar}}]{zhang2024light}%
  \BibitemOpen
  \bibfield  {author} {\bibinfo {author} {\bibfnamefont {X.}~\bibnamefont {Zhang}}, \bibinfo {author} {\bibfnamefont {T.}~\bibnamefont {Carbin}}, \bibinfo {author} {\bibfnamefont {A.~B.}\ \bibnamefont {Culver}}, \bibinfo {author} {\bibfnamefont {K.}~\bibnamefont {Du}}, \bibinfo {author} {\bibfnamefont {K.}~\bibnamefont {Wang}}, \bibinfo {author} {\bibfnamefont {S.-W.}\ \bibnamefont {Cheong}}, \bibinfo {author} {\bibfnamefont {R.}~\bibnamefont {Roy}},\ and\ \bibinfo {author} {\bibfnamefont {A.}~\bibnamefont {Kogar}},\ }\bibfield  {title} {\bibinfo {title} {Light-induced electronic polarization in antiferromagnetic cr2o3},\ }\href@noop {} {\bibfield  {journal} {\bibinfo  {journal} {Nature materials}\ }\textbf {\bibinfo {volume} {23}},\ \bibinfo {pages} {790} (\bibinfo {year} {2024})}\BibitemShut {NoStop}%
\bibitem [{\citenamefont {Bielinski}\ \emph {et~al.}(2025)\citenamefont {Bielinski}, \citenamefont {Chari}, \citenamefont {May-Mann}, \citenamefont {Kim}, \citenamefont {Zwettler}, \citenamefont {Deng}, \citenamefont {Aishwarya}, \citenamefont {Roychowdhury}, \citenamefont {Shekhar}, \citenamefont {Hashimoto} \emph {et~al.}}]{bielinski2025floquet}%
  \BibitemOpen
  \bibfield  {author} {\bibinfo {author} {\bibfnamefont {N.}~\bibnamefont {Bielinski}}, \bibinfo {author} {\bibfnamefont {R.}~\bibnamefont {Chari}}, \bibinfo {author} {\bibfnamefont {J.}~\bibnamefont {May-Mann}}, \bibinfo {author} {\bibfnamefont {S.}~\bibnamefont {Kim}}, \bibinfo {author} {\bibfnamefont {J.}~\bibnamefont {Zwettler}}, \bibinfo {author} {\bibfnamefont {Y.}~\bibnamefont {Deng}}, \bibinfo {author} {\bibfnamefont {A.}~\bibnamefont {Aishwarya}}, \bibinfo {author} {\bibfnamefont {S.}~\bibnamefont {Roychowdhury}}, \bibinfo {author} {\bibfnamefont {C.}~\bibnamefont {Shekhar}}, \bibinfo {author} {\bibfnamefont {M.}~\bibnamefont {Hashimoto}}, \emph {et~al.},\ }\bibfield  {title} {\bibinfo {title} {Floquet--bloch manipulation of the dirac gap in a topological antiferromagnet},\ }\href@noop {} {\bibfield  {journal} {\bibinfo  {journal} {Nature Physics}\ ,\ \bibinfo {pages} {1}} (\bibinfo {year} {2025})}\BibitemShut {NoStop}%
\bibitem [{\citenamefont {Merboldt}\ \emph {et~al.}(2024)\citenamefont {Merboldt}, \citenamefont {Sch{\"u}ler}, \citenamefont {Schmitt}, \citenamefont {Bange}, \citenamefont {Bennecke}, \citenamefont {Gadge}, \citenamefont {Pierz}, \citenamefont {Schumacher}, \citenamefont {Momeni}, \citenamefont {Steil} \emph {et~al.}}]{merboldt2024observation}%
  \BibitemOpen
  \bibfield  {author} {\bibinfo {author} {\bibfnamefont {M.}~\bibnamefont {Merboldt}}, \bibinfo {author} {\bibfnamefont {M.}~\bibnamefont {Sch{\"u}ler}}, \bibinfo {author} {\bibfnamefont {D.}~\bibnamefont {Schmitt}}, \bibinfo {author} {\bibfnamefont {J.~P.}\ \bibnamefont {Bange}}, \bibinfo {author} {\bibfnamefont {W.}~\bibnamefont {Bennecke}}, \bibinfo {author} {\bibfnamefont {K.}~\bibnamefont {Gadge}}, \bibinfo {author} {\bibfnamefont {K.}~\bibnamefont {Pierz}}, \bibinfo {author} {\bibfnamefont {H.~W.}\ \bibnamefont {Schumacher}}, \bibinfo {author} {\bibfnamefont {D.}~\bibnamefont {Momeni}}, \bibinfo {author} {\bibfnamefont {D.}~\bibnamefont {Steil}}, \emph {et~al.},\ }\bibfield  {title} {\bibinfo {title} {Observation of floquet states in graphene},\ }\href@noop {} {\bibfield  {journal} {\bibinfo  {journal} {arXiv preprint arXiv:2404.12791}\ } (\bibinfo {year} {2024})}\BibitemShut {NoStop}%
\bibitem [{\citenamefont {Choi}\ \emph {et~al.}(2024)\citenamefont {Choi}, \citenamefont {Mogi}, \citenamefont {De~Giovannini}, \citenamefont {Azoury}, \citenamefont {Lv}, \citenamefont {Su}, \citenamefont {H{\"u}bener}, \citenamefont {Rubio},\ and\ \citenamefont {Gedik}}]{choi2024direct}%
  \BibitemOpen
  \bibfield  {author} {\bibinfo {author} {\bibfnamefont {D.}~\bibnamefont {Choi}}, \bibinfo {author} {\bibfnamefont {M.}~\bibnamefont {Mogi}}, \bibinfo {author} {\bibfnamefont {U.}~\bibnamefont {De~Giovannini}}, \bibinfo {author} {\bibfnamefont {D.}~\bibnamefont {Azoury}}, \bibinfo {author} {\bibfnamefont {B.}~\bibnamefont {Lv}}, \bibinfo {author} {\bibfnamefont {Y.}~\bibnamefont {Su}}, \bibinfo {author} {\bibfnamefont {H.}~\bibnamefont {H{\"u}bener}}, \bibinfo {author} {\bibfnamefont {A.}~\bibnamefont {Rubio}},\ and\ \bibinfo {author} {\bibfnamefont {N.}~\bibnamefont {Gedik}},\ }\bibfield  {title} {\bibinfo {title} {Direct observation of floquet-bloch states in monolayer graphene},\ }\href@noop {} {\bibfield  {journal} {\bibinfo  {journal} {arXiv preprint arXiv:2404.14392}\ } (\bibinfo {year} {2024})}\BibitemShut {NoStop}%
\bibitem [{\citenamefont {Mönch}\ \emph {et~al.}(2020)\citenamefont {Mönch}, \citenamefont {Bandurin}, \citenamefont {Dmitriev}, \citenamefont {Phinney}, \citenamefont {Yahniuk}, \citenamefont {Taniguchi}, \citenamefont {Watanabe}, \citenamefont {Jarillo-Herrero},\ and\ \citenamefont {Ganichev}}]{mönch2020observation}%
  \BibitemOpen
  \bibfield  {author} {\bibinfo {author} {\bibfnamefont {E.}~\bibnamefont {Mönch}}, \bibinfo {author} {\bibfnamefont {D.~A.}\ \bibnamefont {Bandurin}}, \bibinfo {author} {\bibfnamefont {I.~A.}\ \bibnamefont {Dmitriev}}, \bibinfo {author} {\bibfnamefont {I.~Y.}\ \bibnamefont {Phinney}}, \bibinfo {author} {\bibfnamefont {I.}~\bibnamefont {Yahniuk}}, \bibinfo {author} {\bibfnamefont {T.}~\bibnamefont {Taniguchi}}, \bibinfo {author} {\bibfnamefont {K.}~\bibnamefont {Watanabe}}, \bibinfo {author} {\bibfnamefont {P.}~\bibnamefont {Jarillo-Herrero}},\ and\ \bibinfo {author} {\bibfnamefont {S.~D.}\ \bibnamefont {Ganichev}},\ }\bibfield  {title} {\bibinfo {title} {Observation of terahertz-induced magnetooscillations in graphene},\ }\href@noop {} {\bibfield  {journal} {\bibinfo  {journal} {Nano Letters}\ }\textbf {\bibinfo {volume} {20}},\ \bibinfo {pages} {5943} (\bibinfo {year} {2020})}\BibitemShut {NoStop}%
\end{thebibliography}%
\clearpage

\appendix

\renewcommand{\theequation}{\thesection-\arabic{equation}}
\renewcommand{\thefigure}{\thesection-\arabic{figure}}
\renewcommand{\thetable}{\thesection-\Roman{table}}

\onecolumngrid

\section*{Supplemental Material for ``The Ultra-critical Floquet Non-Fermi Liquid''}

\section{Derivation of the electron collision integral}
\label{appendixA1}

This section presents a detailed derivation of the electron collision integral, employing a formalism that combines fermionic and bosonic degrees of freedom from Ref.\cite{VaskoRaichev}. The full system contains fermionic and bosonic degrees of freedom. Fermionic creation and annihilation operators have the standard anticommutation relations:
\begin{align}\begin{aligned}
\{\hat a_\alpha,\hat a^\dagger_\beta\} = \delta_{\alpha\beta},
\quad
\{\hat a^\dagger_\alpha,\hat a^\dagger_\beta\} = 0,
\quad
\{\hat a_\alpha,\hat a_\beta\} = 0
\end{aligned}\end{align}
Here, $\alpha$ and $\beta$ denote labels of fermionic states.
The bosonic subsystem (i.e., the bath) is comprised of a collection of independent bosonic modes, labeled by $q$. Boson creation and annihilation operators follow the standard bosonic commutation relations:
\begin{align}\begin{aligned}
\big[\hat{b}_q,\hat{b}^\dagger_{q'}\big] = \delta_{q q'},
\quad
\big[\hat{b}_q^\dagger,\hat{b}^\dagger_{q'}\big] = 0,
\quad
\big[\hat{b}_q,\hat{b}_{q'}\big] = 0
\end{aligned}\end{align}

The analysis of Ref. \cite{VaskoRaichev}, employs the concept of partial traces. The trace over the full many body Hilbert space can be split as $\rm Tr = Tr_e Tr_b$, where $\rm Tr_e$ and $\rm Tr_b$ are partial traces over fermionic and bosonic degrees of freedom, respectively. We will further need to decompose the traces over bosonic degrees of freedom as follows: $\rm Tr_b = Tr_q Tr_b^{q}$, ${\rm Tr_b^q}={\rm Tr_b^{q^\prime}}{\rm Tr_b^{q,q^\prime}}$, where $\rm Tr_b^{q_1,q_2\ldots}$ denotes the trace over bosonic modes excluding $(\rm q_1,q_2\ldots)$.

We begin by defining a one-electron density matrix (DM) $n^t_{\gamma\delta}$, which acts non-trivially as an operator on the bosonic degrees of freedom:
\begin{align}\begin{aligned}
n^t_{\gamma\delta} \equiv {\rm Tr_e}[\eta_t \hat a_\delta^\dagger \hat a_\gamma]
\end{aligned}\end{align}
The more standard one-body electronic density matrix $\rho_{\gamma\delta}$ can be obtained by tracing the above $n^t_{\gamma\delta}$ over bosonic degrees of freedom, namely:
\begin{align}\begin{aligned}
\rho_{\gamma\delta} = {\rm Tr_b} [\hat n^t_{\gamma\delta}]
\end{aligned}\end{align}
The equation of motion (EoM) for $n^t_{\gamma\delta}$ can be derived as follows (see also Ref.~\cite{VaskoRaichev}):

\begin{align}\begin{aligned}
i\hbar\partial_t \hat n^t_{\gamma\delta} &= {\rm Tr_e}\left\{[\hat{H}(t),\eta_t] \hat a^\dagger_\delta \hat a_\gamma\right\} \\
&= [\hat{H}_S(t),\hat n^t]_{\gamma\delta}+[\hat{H}_B,\hat n^t_{\gamma\delta}]+\sum_q [\chi_q \hat b_q +h.c.,\hat n^t]_{\gamma\delta} 
+ [{\rm Tr_e}(\hat a_\nu^\dagger \hat a^\dagger_\delta \hat a_\eta \hat a_\gamma \eta_t),\sum_q \hat \chi^{\nu\eta}_{q}\hat b_q +h.c.],
\label{EoMn}\end{aligned}\end{align}
where $\hat{H}(t)$ is defined in the main text. This leads to the EoM for the fermionic DM:
\begin{align}\begin{aligned}
\hbar\partial_t \rho^t_{\gamma\delta} + i[\hat{H}_S(t),\hat \rho^t]_{\gamma\delta} &= -i{\rm Tr_b}\Big(\sum_q [\hat \chi_q \hat b_q +h.c.,n^t]_{\gamma\delta}\Big) = {{\cal I}_{\gamma\delta}}[t,\hat{\rho}(t)] .
\label{rho}\end{aligned}\end{align}
We note that the term proportional to $H_B$ vanishes due to the trace permutation property. Additionally, the term arising to the last term in Eq.~\eqref{EoMn} vanishes since $b_q$ has no matrix structure in the fermionic subspace. We recognize the right-hand side of Eq.~\eqref{rho} as the collision integral. Our goal is to eliminate $n^t$ from Eq.~\eqref{rho} using Eq.~\eqref{EoMn}, in order to obtain a self-consistent equation that involves only the standard one-body electronic density matrix $\rho_{\gamma\delta}$.

To proceed with the elimination of $n^t$ from Eq.~\eqref{rho}, we begin by decomposing the trace over bosonic modes. We introduce the quantity $\hat n^{t,(q)} ={\rm Tr_b^q}[\hat n^{t}]$, which serves as a fundamental component for constructing the full bosonic trace in the collision integral from Eq.~\eqref{rho}. The equation of motion for matrix elements of $\hat n^{t,(q)}$ can be derived as follows (see Ref.~\cite{VaskoRaichev}):
\begin{align}\begin{aligned}
i\hbar\partial_t \hat n^{t,(q)}_{\gamma\delta} &= [\hat{H}_S(t)+\hbar\omega_q\hat b^\dagger_q \hat b_q,\hat n^{t,(q)}]_{\gamma\delta}+ [\hat \chi_q \hat b_q +h.c.,\hat n^{t,(q)}]_{\gamma\delta} \\
&\quad +\sum_{q\neq q^\prime}{\rm Tr_b^{q^\prime}}[\hat \chi_{q^\prime} \hat b_{q^\prime} +h.c.,\hat n^{t,(q,q^\prime)}]_{\gamma\delta} + \big[{\rm Tr_b^qTr_e}(\hat a_\nu^\dagger \hat a^\dagger_\delta \hat a_\eta \hat a_\gamma \hat \eta_t), \hat \chi^{\nu\eta}_{q}\hat b_q +h.c.\big] .
\label{EoMnq}\end{aligned}\end{align}
We now employ the assumption of weak coupling between fermions and the bosonic bath. As we can see from Eq.~\eqref{rho}, to evaluate the collision integral up to the order $\hat\chi^2$, $\hat n^{t,(q)}$ has to be solved only in the linear order in $\hat\chi$. Thus, it is useful to split $\hat n^{t}$ into terms that are order zero in $\hat\chi$ and terms that vanish when $\hat\chi\rightarrow 0$, by introducing $\hat\kappa^{(q)}_{\gamma\delta} = \hat n^{(q)}_{\gamma\delta}-\hat{\bar n}^{(q)}_{\gamma\delta}$, where $\hat{\bar n}^{(q)}_{\gamma\delta}$ is the solution for $\hat n^{t}$ at zero coupling $\hat\chi=0$. The scaling with $\hat\chi$ of $\hat n^{(q)}$ is expected to have the following structure:

\begin{align}\begin{aligned}
\hat n^{(q)} &= \begin{pmatrix}
\hat \rho(t) &\sim \hat \chi_q  \\
\sim \hat \chi_q^\dagger& \text{Thermal Bosons} 
\end{pmatrix} 
= \underbrace{\begin{pmatrix}
\hat \rho(t) &0 \\
0& \text{Thermal Bosons}
\end{pmatrix}}_{\hat {\bar n}^{(q)}}+\underbrace{\begin{pmatrix}
0 &\sim \hat \chi_q  \\
\sim \hat \chi_q^\dagger& 0
\end{pmatrix}}_{\hat \kappa_q}
\label{kappaintro}\end{aligned}\end{align}
It can be shown that the coupling-less component of $\hat n^{t}$ obeys the following EoMs:
\begin{align}\begin{aligned}
& i\hbar\partial_t \hat{\bar{n}}_{\gamma\delta}^{t,(q)} = [\hat{H}_S(t)+\hbar\omega_q \hat b^\dagger_q \hat b_q,\hat n^{t,(q)}]_{\gamma\delta} ,
\\
& i\hbar\partial_t \hat{\bar{n}}_{\gamma\delta}^{t,(q,q^\prime)} = [\hat{H}_S(t)+\hbar\omega_q \hat b^\dagger_q \hat b_q+\hbar\omega_{q^\prime} \hat b^\dagger_{q^\prime} \hat b_{q^\prime},\hat n^{t,(q,q^\prime)}]_{\gamma\delta}
\label{nqbar}\end{aligned}\end{align}
Note that $\hat{n}^{t,(q,q^\prime)} ={\rm Tr_b^{q,q'}}[\hat n^{t}]$ has a diagonal structure in both $q$ and $q'$ modes, implying that the commutator proportional to $\sum_{\rm q\neq q'}[\ldots]$ in Eq.~\eqref{EoMnq} vanishes (is of higher order in $\hat\chi_q$). Thus we find the effective EoM for $n^{t,(q)}_{\gamma\delta}$ as:
\begin{align}\begin{aligned}
i\hbar\partial_t \hat n^{t,(q)}_{\gamma\delta} &= [\hat{H}_S(t)+\hbar\omega_q\hat b^\dagger_q \hat b_q,\hat n^{t,(q)}]_{\gamma\delta}+ [\hat \chi_q \hat b_q +h.c.,\hat n^{t,(q)}]_{\gamma\delta} \\
&\quad + \big[{\rm Tr_b^qTr_e}(\hat a_\nu^\dagger \hat a^\dagger_\delta \hat a_\eta \hat a_\gamma \hat \eta_t), \hat \chi^{\nu\eta}_{q}\hat b_q +h.c.\big]+{\cal O}(\hat\chi^{2})
\end{aligned}\end{align}
This equation forms the basis for further analysis of the electron collision integral, taking into account the weak coupling approximation and the structure of the bosonic and fermionic subspaces.

\subsection{Approximation of the four-fermion trace}

This subsection presents an approximation for the four-fermion trace $\hat g_{\eta\delta,(q)}^{\gamma\nu,t}\equiv{\rm Tr_b^qTr_e}(\hat a_\nu^\dagger \hat a^\dagger_\delta \hat a_\eta \hat a_\gamma \eta_t)$ to the zeroth order in $\hat\chi$. The fermionic commutation relations imply the following index permutation properties for $\hat g$:
\begin{align}\begin{aligned}
\hat g_{\eta\delta,(q)}^{\gamma\nu,t} &= \hat g_{\gamma\nu,(q)}^{\eta\delta,t} = -\hat g_{\gamma\delta,(q)}^{\eta\nu,t} = -\hat g_{\eta\nu,(q)}^{\gamma\delta,t} .
\end{aligned}\end{align}
We employ an approximation proposed by Vasko and Raichev \cite{VaskoRaichev}, which is similar to a time-dependent Hartree-Fock decomposition:

\begin{align}\begin{aligned}
\hat g_{\eta\delta,(q)}^{\gamma\nu,t} &\equiv {\rm Tr_b^qTr_e}(\hat a_\nu^\dagger \hat a^\dagger_\delta \hat a_\eta \hat a_\gamma \hat{\eta}_t)
\approx 
\frac{\hat {\bar n}^{t,(q)}_{\gamma\nu}\rho^t_{\eta\delta} - \hat{\bar n}^{t,(q)}_{\gamma\delta}\rho^t_{\eta\nu} + [(\gamma,\nu)\leftrightarrow(\eta,\delta)]}{2},
\label{gapp}\end{aligned}\end{align}

This approximation is justified by the fact that both the exact quantity and the right-hand side of Eq.~\eqref{gapp} obey the same equations of motion (EoMs) in the limit $\hat \chi_q\to 0$ \cite{VaskoRaichev}. To illustrate the self-consistency of this approximation, we compare the exact evolution of $\hat g$:
\begin{align}\begin{aligned}
i\hbar\partial_t \hat g_{\eta\delta,(q)}^{\gamma\nu,t} &= {\rm Tr_b^qTr_e}(\hat a_\nu^\dagger \hat a^\dagger_\delta \hat a_\eta \hat a_\gamma [\hat{H}(t),\hat \eta_t]) = {\rm Tr_b^qTr_e}
\big(\hat a_\nu^\dagger \hat a^\dagger_\delta \hat a_\eta \hat a_\gamma \big[\hat{H}_S(t)+\sum_{q^\prime}\hbar\omega_{q^\prime}\hat b^\dagger_{q^\prime}\hat b_{q^\prime},\hat \eta_t\big]
\big) \\
&= H_S^{\alpha\beta}(t)
\big(
\delta_{\gamma\alpha}\hat g_{\eta\delta,(q)}^{\beta\nu,t} -\delta_{\nu\beta}\hat g_{\eta\delta,(q)}^{\gamma\alpha,t} +\delta_{\eta\alpha}\hat g_{\beta\delta,(q)}^{\gamma\nu,t} -\delta_{\beta\delta} \hat g_{\eta\alpha,(q)}^{\gamma\nu,t}
\big)
+
\hbar\omega_q
\big[\hat b^\dagger_{q}\hat b_{q},\hat g_{\eta\delta,(q)}^{\gamma\nu,t}\big];
\label{exactEoM}\end{aligned}\end{align}
with the approximated EoM:
\begin{align}\begin{aligned}
i\hbar\partial_t \hat g_{\eta\delta,(q)}^{\gamma\nu,t} &=  (i\hbar\partial_t \hat {\bar n}^{t,(q)}_{\gamma\nu})\rho^t_{\eta\delta} +\hat {\bar n}^{t,(q)}_{\gamma\nu}(i\hbar\partial_t\rho^t_{\eta\delta}) - (i\hbar\partial_t\hat {\bar n}^{t,(q)}_{\gamma\delta})\rho^t_{\eta\nu} - \hat{\bar n}^{t,(q)}_{\gamma\delta}(i\hbar\partial_t\rho_{\eta\nu}) + [(\gamma,\nu)\leftrightarrow(\eta,\delta)] \\
&= [\hat{H}_S(t),\hat {\bar n}^{t,(q)}]_{\gamma\nu}\rho^t_{\eta\delta}- [\hat{H}_S(t),\hat {\bar n}^{t,(q)}]_{\gamma\delta}\rho^t_{\eta\nu} +\hat {\bar n}^{t,(q)}_{\gamma\nu}[\hat{H}_S(t),\hat \rho^t]_{\eta\delta} \\
&\quad -\hat { \bar n}^{t,(q)}_{\gamma\delta}[\hat{H}_S(t),\hat \rho^t]_{\eta\nu}+\hbar\omega_q[\hat b^\dagger_{q}\hat b_{q},\hat g_{\eta\delta,(q)}^{\gamma\nu,t}]+ [(\gamma,\nu)\leftrightarrow(\eta,\delta)] \\
&= H_S^{\gamma\alpha}(t)\hat g^{\alpha\nu,t}_{\eta\delta,(q)} - H_S^{\alpha\delta}(t)\hat g^{\gamma\nu,t}_{\eta\alpha,(q)} +\hat g^{\gamma\nu,t}_{\alpha\delta,(q)}H_S^{\eta\alpha}(t)- \hat g^{\gamma\alpha,t}_{\eta\delta,(q)} H_S^{\alpha\nu}(t) + \hbar\omega_q[\hat b^\dagger_{q}\hat b_{q},\hat g_{\eta\delta,(q)}^{\gamma\nu,t}] \\
&= H_S^{\alpha\beta}(t)
\big (\delta_{\gamma\alpha}\hat g^{\beta\nu,t}_{\eta\delta,(q)} - \delta_{\beta\delta}\hat g^{\gamma\nu,t}_{\eta\alpha,(q)} +\delta_{\alpha\eta}\hat g^{\gamma\nu,t}_{\beta\delta,(q)}- \delta_{\beta\nu}\hat g^{\gamma\alpha,t}_{\eta\delta,(q)} \big) 
+
\hbar\omega_q \big[\hat b^\dagger_{q}\hat b_{q},\hat g_{\eta\delta,(q)}^{\gamma\nu,t} \big] ,
\label{apprev}\end{aligned}\end{align}
where we have utilized the EoMs from Eqs.~\eqref{nqbar} and \eqref{rho} in the limit $\hat \chi_q\to 0$. The evolution of the approximation is consistent with the exact evolution in Eq.~\eqref{exactEoM}, validating the approximation's applicability in this context.

\subsection{Continuing the collision integral derivation}

This subsection starts with the evaluation of $\hat\kappa$, which is essential for computing the collision integral to the desired order. Utilizing Eqs.~\eqref{EoMnq}, \eqref{nqbar} and the definition of $\hat\kappa$ [see text above Eq.~\eqref{kappaintro}], we derive the equation governing the evolution of $\hat\kappa$:
\begin{align}\begin{aligned}
i\hbar\partial_t \hat\kappa^{t,(q)}_{\gamma\delta} &= [\hat{H}_S(t)+\hbar\omega_q\hat b^\dagger_q \hat b_q,\hat\kappa^{t,(q)}]_{\gamma\delta}+ [\hat\chi_q \hat b_q +h.c.,\hat n^{t,(q)}]_{\gamma\delta}
+
\big[\hat g_{\eta\delta,(q)}^{\gamma\nu,t}, \hat\chi^{\nu\eta}_{q}\hat b_q +h.c.\big]
\end{aligned}\end{align}
Here, $g$ is a function of the equilibrium distribution $\bar n$ and electron density matrix $\rho$ as defined in Eq.~\eqref{gapp}. The evolution of $\rho$, as shown in Eq.~\eqref{rho}, can be simplified to:
\begin{align}\begin{aligned}
i\hbar\partial_t \rho^t_{\gamma\delta} = [\hat{H}_S(t),\hat \rho^t]_{\gamma\delta}+\sum_q {\rm Tr_q}[\hat \chi_q \hat b_q +h.c.,\hat \kappa^{t,(q)}]_{\gamma\delta}
\end{aligned}\end{align}
This simplification arises from the fact that the bosonic equilibrium occupation function $\hat{\bar n} = N_b(b^\dagger b)$ is quadratic in $b$, causing the corresponding trace term to vanish.

We now arrive at a complete set of equations sufficient to determine the collision integral:
\begin{align}
    &i\hbar\partial_t \hat{\bar n}_{\gamma\delta}^{t,(q)}= [\hat{H}_S(t)+\hbar\omega_q \hat b^\dagger_q \hat b_q,\hat {\bar n}^{t,(q)}]_{\gamma\delta}+{\cal O}(\hat\chi).\label{3rd}\\
    &i\hbar\partial_t \hat \kappa^{t,(q)}_{\gamma\delta} = [\hat{H}_S(t)+\hbar\omega_q\hat b^\dagger_q \hat b_q,\hat \kappa^{t,(q)}]_{\gamma\delta}+ [\hat \chi_q \hat b_q +h.c.,\hat {\bar n}^{t,(q)}]_{\gamma\delta}+ \left[\hat g_{\eta\delta,(q)}^{\gamma\nu,t}, \hat \chi^{\nu\eta}_{q}\hat b_q +h.c.\right]+{\cal O}(\hat\chi^2),\label{2nd}\\
    &\hbar\partial_t \hat \rho^t_{\gamma\delta} +i[\hat{H}_S(t),\hat \rho^t]_{\gamma\delta}=  -i\sum_q {\rm Tr_q}[\hat \chi_q \hat b_q +h.c.,\hat \kappa^{t,(q)}]_{\gamma\delta}+{\cal O}(\hat\chi^3)={{\cal I}_{\gamma\delta}}[t,\hat{\rho}(t)]\label{1st}.
\end{align}
By employing the electronic evolution operator [see Eq.~\eqref{eq:Je} of the main text] and transitioning to the interaction picture using the unitary transformation $\hat{{\cal U}}_{t,t_0,(q)} = \hat{U}_S(t,t_0)e^{-i\omega_q \hat{b}^\dagger_q \hat{b}_q(t-t_0)}$, we can solve Eq.~\eqref{2nd} to obtain:
\begin{align}\begin{aligned}
\hat\kappa^{t,(q)}_{\alpha\beta} &= \frac{1}{i\hbar}\int_{t_0}^t dt' \left( {\hat{{\cal U}}_{t',t,(q)}^{\dagger,\alpha\gamma} [\hat\chi_q\hat b_q +h.c.,\hat{\bar n}^{t',(q)}]_{\gamma\delta}\hat{{\cal U}}^{\delta\beta}_{t',t,(q)}} + {\hat{{\cal U}}_{t',t,(q)}^{\dagger,\alpha\gamma} \left[\hat g_{\eta\delta,(q)}^{\gamma\nu,t'}, \hat\chi^{\nu\eta}_{q}\hat b_q +h.c.\right]\hat{{\cal U}}^{\delta\beta}_{t',t,(q)}}\right)
\label{kappasol}\end{aligned}\end{align}
where $\hat{\bar n}^{t,(q)} = \hat{{\cal U}}_{t,t_0,(q)}\hat{\bar  n}^{0,(q)}\hat{{\cal U}}_{t,t_0,(q)}^\dagger$, which is the solution of Eq.~\eqref{3rd}. Eq.~\eqref{kappasol} defines $\kappa$ as a function of $\rho$.

The final step in deriving the kinetic equation is to substitute Eq.~\eqref{kappasol} into Eq.~\eqref{1st} and compute the partial traces of the collision integral:
\begin{align}\begin{aligned}
{{\cal I}_{\gamma\delta}}[t,\hat{\rho}(t)] = -i\sum_q {\rm Tr_q}[\chi_q b_q +\chi^\dagger_q b^\dagger_q,\kappa^{t,(q)}]_{\gamma\delta}
= -i\sum_q  \big([\chi_q, {\rm Tr_q}\{b_q\kappa^{t,(q)}\} +[\chi^\dagger_q, {\rm Tr_q}\{ b^\dagger_q\kappa^{t,(q)}\}]\big)_{\gamma\delta} ,
\end{aligned}\end{align}
and we note that ${{\cal I}_{\gamma\delta}}[t,\hat{\rho}(t)] = {{\cal I}^{\rm VR}_{\gamma\delta}}[t,\hat{\rho}(t)] +{\cal O}(\hat\chi^3)$.
In the subsequent steps, we utilize the following relations:
\begin{align}\begin{aligned}
\hat{\bar n}^{t,(q)} &= \rho^t\otimes\bar n_b^{t,(q)}, \qquad
\bar n_b^{t,(q)} = \frac{1}{Z_q}e^{-\beta\omega_q b^\dagger_qb_q} ,
\end{aligned}\end{align}
according to the definition from Eq.~\eqref{kappaintro}, and using the fact that ${\rm Tr_q} \equiv \sum_{n_q=0}^\infty \bra{n_q}...\ket{n_q}$, we then compute the summation over $n_q$ (not $q$) using the following identities:
\begin{align}\begin{aligned}
\sum_{n_q=0}^\infty n_q\frac{e^{-\beta \omega_q n_q}}{Z_q} = \frac{1}{e^{\beta \omega_q}-1} = N_q ,
\qquad
\sum_{n_q=0}^\infty (n_q+1)\frac{e^{-\beta \omega_q n_q}}{Z_q} &= \frac{e^{\beta \omega_q}}{e^{\beta \omega_q}-1} = 
N_q +1,
\end{aligned}\end{align}
where $Z_q = {e^{\beta \omega_q}}/({e^{\beta \omega_q}-1})$ and $(N_q+1) e^{-\beta \omega_q}=N_q$.
After performing all these summations, we derive the more general version of the collision integral as ${{\cal I}^{\rm VR}}[t,\hat{\rho}(t)]={I_e}[t,\hat{\rho}(t)]+{I_a}[t,\hat{\rho}(t)]$, where the emission component is:
\begin{align}\begin{aligned}
I_e[t,\hat{\rho}(t)] = \sum_q \int_{t_0}^t dt' (N_q+1) \Bigg(e^{-i\omega_q (t-t')}\left[  U_S(t,t')[(1-\rho^{t'})\chi^{\dagger}_{q}\rho^{t'} +\rho^{t'}{\rm Tr}\{\rho^{t'}\chi^{\dagger}_{q}\}]U_S(t',t),\chi_{q}\right] -\\e^{i\omega_q (t-t')}\left[ U_S(t,t')[\rho^{t'}\chi_{q}(1-\rho^{t'}) + \rho^{t'}{\rm Tr}\{\rho^{t'}\chi_{q}\}]U_S(t',t),\chi^\dagger_{q}\right]\Bigg),
\label{Jesupp}\end{aligned}\end{align}
and the absorption component is:
\begin{align}\begin{aligned}
I_a[t,\hat{\rho}(t)] =\sum_q \int_{t_0}^t dt' N_q\Bigg( e^{i\omega_q (t-t')}\left[U_S(t,t')[(1-\rho^{t'})\chi_{q}\rho^{t'} + \rho^{t'}{\rm Tr}\{\rho^{t'}\chi_{q}\}]U_S(t',t),\chi^\dagger_{q}\right]- \\e^{-i\omega_q (t-t')}\left[U_S(t,t')[\rho^{t'}\chi^{\dagger}_{q}(1-\rho^{t'}) +\rho^{t'}{\rm Tr}\{\rho^{t'}\chi^{\dagger}_{q}\}]U_S(t',t),\chi_{q}\right]\Bigg).
\label{Jasupp}
\end{aligned}\end{align}
These equations provide a comprehensive description of the electron-boson collision integral and are identical to those derived by  Vasko and Raichev in Ref.\cite{VaskoRaichev}. 

If we further impose translation invariance, i.e.,
\begin{align}
\begin{aligned}
\langle k | \rho | k' \rangle = \delta_{k,k'} f_k, \quad
\langle k | \chi_q | k' \rangle \propto \delta_{k, k' + k_\lambda},
\end{aligned}
\end{align}
where the label for boson modes are $q=({\bf r},\lambda)$ with $\lambda$ labels the different boson modes coupled to each site, $k_\lambda$ is the momentum for $\lambda$ mode, $f_k$ represents the occupation number for momentum $k$, the trace ${\rm Tr}\{\rho \chi_q\}$ is proportional to the identity matrix in momentum space that couples to the global particle number $N$:
\begin{align}\begin{aligned}
{\rm Tr}\{\rho \chi_q\} = \sum_k \langle k | \rho \chi_q | k \rangle = \sum_k f_k \langle k | \chi_q | k \rangle
\propto \delta_{0, k_\lambda}
\sum_k f_k = \delta_{0, k_\lambda} N ,
\end{aligned}\end{align}
The collision integrals contain commutators of the following form vanishes:
\begin{align}\begin{aligned}
\left[ U_S(t,t') \rho^{t'} {\rm Tr}\{\rho^{t'} \chi_q\} U_S(t',t), \chi_q^\dagger \right]
=
\begin{cases}
0, & \text{ when } k_\lambda\neq0 \text{ since } {\rm Tr}\{\rho \chi_q\} = 0 \\ 0, & \text{ when } k_\lambda=0 \text{ since both } {\rm Tr}\{\rho \chi_q\} \text{ and } \chi_q  \propto \text{ identity matrix} 
\end{cases}
\end{aligned}\end{align}
Therefore, all terms proportional to ${\rm Tr}\{\rho \chi_q\}$ in the collision integrals in Eqs.~\eqref{Jesupp} and~\eqref{Jasupp} can be neglected. This allows us reduce the collision integrals to:
\begin{align}
&I_e[t,\hat{\rho}(t)] = \sum_q \int_{t_0}^t dt' (N_q+1)e^{-i\omega_q (t-t')}\left[  U_S(t,t')[(1-\rho^{t'})\chi^{\dagger}_{q}\rho^{t'}]U_S(t',t),\chi_{q}\right] +h.c.,
\label{Jeappr}\\
&I_a[t,\hat{\rho}(t)] =\sum_q \int_{t_0}^t dt' N_q e^{i\omega_q (t-t')}\left[  U_S(t,t')[(1-\rho^{t'})\chi_{q}\rho^{t'}]U_S(t',t),\chi^\dagger_{q}\right] +h.c.
\end{align}
In the main text we only display the emission component of the collision integral, as the absorption component can be obtained from the emission component by replacing $N_q+1\to N_q, \omega_q\to-\omega_q, \chi\to\chi^\dagger$.

\section{Electron-boson collision integral for Floquet systems}
\label{Floquet-Vasko-Raichev}

This section applies the kinetic equation derived for Floquet systems.
We adopt the following convention for the Floquet expansion and Fourier transformations:
\begin{align}\begin{aligned}
f(\omega) = \int dt f(t) e ^{i\omega t}, 
\quad 
f(t) = \int \frac{d\omega}{2\pi}e^{-i\omega t} f(\omega) ,
\quad
f(t) = f(t+T) = \sum_n f_ne^{-in\omega t}.
\end{aligned}\end{align}
This allows us to express the evolution operator using the Floquet theorem:
\begin{align}
\begin{aligned}
i \hbar\partial_t \hat{U}_S (t,t_0) = \hat{H}_S (t) \hat{U}_S (t,t_0) ,
\quad
U_S(t,t') = \sum_k \ket{\psi_k(t)}\bra{\psi_k(t')} ,
\quad
\ket{\psi_k(t)} &= \sum_l e^{-i (\epsilon^F_k + l\Omega) t}\ket{\varphi_{k,l}} ,
\label{S-FloquetDefinitions}
\end{aligned}
\end{align}
where $\Omega = 2\pi/T$ is the frequency of the drive $H_S(t)=H_S(t+T)$ and $\epsilon^F_k$ is the system's Floquet energy. Now, let us further specialize to case of the single-band system:
\begin{align}\begin{aligned}
U_S(t,t') = \sum_k \sum_{ll^\prime} e^{-i\epsilon^F_k(t-t')-il\Omega t+il^\prime \Omega t'}\varphi_{k,l}\varphi^*_{k,l^\prime}\ket{k}\bra{k} ,
\quad
\rho^t = \sum_k f_{k,t}\ket{k} \bra{k} ,
\label{UFloq}\end{aligned}\end{align}
the reduced single-band Floquet-Boltzmann equation can be obtained by computing $\partial_t f_{k,t} = \bra{k}J_e[t,\rho]\ket{k} + \bra{k}J_a[t,\rho]\ket{k}$. Using Eqs.~\eqref{Jeappr}, \eqref{UFloq}, we derive:
\begin{align}\begin{aligned}
\bra{k}I_e[t,\rho]\ket{k} &= \sum_{l_1l_2l_3l_4} \sum_{k^\prime,q} \int_{t_0}^t dt' (N_q+1)e^{-i\omega_q(t-t')-il_1\Omega t+il_2 \Omega t'}e^{-il_3\Omega t'+il_4 \Omega t} \\
&\quad \times \bigg[\varphi_{k,l_1}\varphi^*_{k_1,l_2}\varphi_{k',l_3}\varphi^*_{k',l_4}e^{-i(\epsilon^F_{k}-\epsilon^F_{k'})(t-t')}
|\bra{k'}\chi_q\ket{k}|^2(1-f_{k,t'})f_{k',t'} \\
&\quad -\varphi_{k',l_1}\varphi^*_{k',l_2}\varphi_{k,l_3}\varphi^*_{k,l_4}e^{i(\epsilon^F_{k}-\epsilon^F_{k'})(t-t')}
|\bra{k}\chi_q\ket{k'}|^2
(1-f_{k',t'})f_{k,t'}\bigg] +h.c.
\end{aligned}\end{align}
The time integration can be performed assuming that the interaction with the bath started in the distant past, and the system has reached a steady state with respect to its interaction with the bath, without retaining its initial information. This allows us to extend the lower limit of integration to negative infinity $\int_{t_0}^t d t'\to \int_{-\infty} ^t dt' e^{\eta t'}$ and express $f_{k,t'}$ into Floquet expansion:
\begin{align}\begin{aligned}
& \bra{k}I_e[t,\rho]\ket{k} = \sum_{l_1\ldots l_6} \sum_{k^\prime,q}  (N_q+1)e^{-i(l_1-l_2+l_3-l_4+l_5+l_6)\Omega t} \\
& \qquad \times\bigg[\frac{\varphi_{k,l_1}\varphi^*_{k,l_2}\varphi_{k^\prime,l_3}\varphi^*_{k^\prime,l_4}|\bra{k^\prime}\chi_q\ket{k}|^2\bar f_{k l_5}f_{k^\prime l_6}}{i(\omega_q+(l_2-l_3-l_5-l_6)\Omega+\epsilon^F_{k}-\epsilon^F_{k^\prime})+\eta}-\frac{\varphi_{k^\prime,l_1}\varphi^*_{k^\prime,l_2}\varphi_{k,l_3}\varphi^*_{k,l_4}|\bra{k}\chi_q\ket{k^\prime}|^2\bar f_{k^\prime l_5}f_{k l_6}(t')}{i(\omega_q-\epsilon^F_{k}+\epsilon^F_{k^\prime}+(l_2-l_3-l_5-l_6)\Omega)+\eta}\bigg] +h.c.,
\end{aligned}\end{align}
where we used the compact notation $\bar f_{k,t}= 1-f_{k,t}$, with the Floquet modes $\bar f_{l,k} = \delta_{l0}- f_{l,k}$. By projecting the Boltzmann equation onto Floquet mode $l$, we can compactly write the complete Floquet-Boltzmann equation as:
\begin{align}\begin{aligned}
-il\Omega f_{kl} =  \sum_{s = e,a}\sum_{l_1\ldots l_6} \sum_{k^\prime,q}  N^s_q\delta^{l-l_5-l_6}_{l_1-l_2+l_3-l_4}\bigg[\frac{\varphi_{k,l_1}\varphi^*_{k,l_2}\varphi_{k^\prime,l_3}\varphi^*_{k^\prime,l_4}|\bra{k^\prime}\chi^s_q\ket{k}|^2\bar f_{k l_5}f_{k^\prime l_6}}{i({\rm sgn_s}\omega_q+(l_2-l_3-l_5-l_6)\Omega+\epsilon^F_{k}-\epsilon^F_{k^\prime})+\eta}-(k\leftrightarrow k^\prime)\bigg]+\\\sum_{s = e,a}\sum_{l_1\ldots l_6} \sum_{k^\prime,q}  N^s_q\delta^{l-l_5-l_6}_{l_1-l_2+l_3-l_4}\bigg[\frac{\varphi_{k,l_1}\varphi^*_{k,l_2}\varphi_{k^\prime,l_3}\varphi^*_{k^\prime,l_4}|\bra{k^\prime}\chi^s_q\ket{k}|^2\bar f_{k l_5}f_{k^\prime l_6}}{-i({\rm sgn_s}\omega_q+(l_2-l_3+l)\Omega+\epsilon^F_{k}-\epsilon^F_{k^\prime})+\eta}-(k\leftrightarrow k^\prime)\bigg].
\label{FLoqutBoltzmann}\end{aligned}\end{align}
Here $ N^e_q =  N_q+1, N^a_q = N_q$, the sign in the denominator ${\rm sgn_e}=1,{\rm sgn_a}=-1$ and $\delta_a^b\equiv\delta_{ab}$ is the Kronecker delta. Importantly, the right-hand side of Eq.~\eqref{FLoqutBoltzmann} is of order $\hat\chi^2$.
This implies that in the limit of zero coupling to the bath $\hat\chi \to 0$, all oscillating components ($l\neq0$) of $f_{k,l}$ vanish, and the system's state is defined by the self-consistent solution of the Floquet-Boltzmann equation for the mode $l=0$. Consequently, the solution of the driven Boltzmann equation with weak coupling to the bath is time-independent. The corresponding static emission component of the collision integral is (here $f_{k,0}\equiv f_k$):
\begin{align}\begin{aligned}
& \bra{k}I_e[\rho]\ket{k} =\sum_{k'} \sum_{l_1l_2l_3l_4} \sum_q (N_q+1)\delta_{l_1+l_3,l_2+l_4}
\\
& \qquad \times\left[\frac{\varphi_{k,l_1}\varphi^*_{k,l_2}\varphi_{k',l_3}\varphi^*_{k',l_4}|\bra{k'}\chi_q\ket{k}|^2(1-f_{k})f_{k'}}{i(\omega_q+\epsilon^F_{k}-\epsilon^F_{k'}+l_2\Omega-l_3\Omega)+\eta}-\frac{\varphi_{k',l_1}\varphi^*_{k',l_2}\varphi_{k,l_3}\varphi^*_{k,l_4}|\bra{k}\chi_q\ket{k'}|^2(1-f_{k'})f_{k}}{i(\omega_q-\epsilon^F_{k}+\epsilon^F_{k'}+l_2\Omega-l_3\Omega)+\eta}\right] +h.c.
\label{ColIntstat}\end{aligned}\end{align}
Focusing on the Hermitian conjugate term, after relabelling ($l_1,l_3\leftrightarrow l_2,l_4$) and using the Kronecker delta condition $l_4-l_1 = l_3-l_2$, we observe that denominators combine into delta functions as $\frac{1}{ix+\eta} +\frac{1}{-ix+\eta}=\frac{2\eta}{x^2+\eta^2}\to 2\pi\delta(x)$. Thus, by introducing the scattering matrix elements:
\begin{align}\begin{aligned}
W^e_{k\to k_1} =2\pi \sum_{l_1l_2l_3l_4} \sum_q (N_q+1)\delta_{l_1+l_3,l_2+l_4}\delta(\omega_q+\epsilon^F_{k_1}-\epsilon^F_{k}+l_2\Omega-l_3\Omega)
{\rm Re}[\varphi_{k_1,l_1}\varphi^*_{k_1,l_2}\varphi_{k,l_3}\varphi^*_{k,l_4}]|\bra{k}\chi_q\ket{k_1}|^2.
\label{S-We-0}\end{aligned}\end{align}
the collision integral in Eq.~\eqref{ColIntstat}, together with the absorption component, can be written as:
\begin{align}\begin{aligned}
\bra{k}{\cal I}[\rho]\ket{k} = \sum_{s=e,a}\sum_{k'}[W^s_{k'\to k}f_{k'}(1-f_{k})-W^s_{k\to k'}f_{k}(1-f_{k'})] ,
\end{aligned}\end{align}
which provides the Floquet-Boltzmann equation for a single-band Floquet system.

\section{Analytical analysis of non-analyticities}
\label{S-Analytical-nonanalyticities}

This section presents derivations for the scattering matrix and formal expressions for the steady-state occupation and its derivatives. Based on these derivations, we demonstrate the prevalence of non-analyticities in the steady-state occupation.

\subsection{Scattering matrix element \texorpdfstring{$W(q \to p)$}{W(q to p)}}
\label{S-ScatteringMatrix}

We begin with the expression for the scattering matrix $W_e(q \to p)$ [see Eq.~\eqref{S-We-0}]:
\begin{align}
W_e (q \to  p) & =
2\pi \sum_{l_1l_2l_3l_4} 
{\rm Re}[\varphi_{q,l_1}\varphi^*_{q,l_2}\varphi_{p,l_3}\varphi^*_{p,l_4}]
\delta_{l_1+l_3,l_2+l_4}
\sum_\lambda [N_{b} (\lambda)+1]\delta(\omega_\lambda+\epsilon^F_{q}-\epsilon^F_{p}+l_2\Omega-l_3\Omega)
|\bra{q}\chi_\lambda\ket{p}|^2
\nonumber\\
& =
2\pi \sum_{l} 
\sum_{l_1}
|\varphi_{l_1,p} \varphi_{l_1+l, q}^* |^2
\sum_\lambda [N_{b} (\lambda)+1]\delta(\omega_\lambda+\epsilon^F_{q}-\epsilon^F_{p}+ l \Omega)
|\bra{q}\chi_\lambda\ket{p}|^2
,
\label{S-We-1}
\end{align}
in which we let $l = l_2 - l_3$.
We introduce the density of states $\nu_B(\epsilon)$ for the bosonic bath:
\begin{align}
\nu_B(\epsilon)=\sum_\lambda \delta(\epsilon-\omega_\lambda).
\label{S-Nu}
\end{align}
For an arbitrary function $f(\omega_\lambda)$, we can use the following property:
\begin{align}
\sum_\lambda f(\omega_\lambda)\delta(\omega_\lambda - x) = f(x)\sum_\lambda \delta(\omega_\lambda - x) = f(x)\nu_B(x) .
\label{S-delta-property}
\end{align}
This property can be verified by multiplying both sides by any test function $g(x)$ and integrating over $x$. Using Eqs.~(\ref{S-Nu}) and (\ref{S-delta-property}), we can reformulate the sum over $\lambda$ in terms of $\nu_B$ in Eq.~\eqref{S-We-1}:
\begin{align}
& W_e(q \to p) = 
\sum_{l \in \mathbb{Z}}
\Gamma_l (q, p)
\times
\big[ N_b ( \epsilon_q^F - \epsilon_p^F + l \Omega ) + 1 \big]
\times \nu_B ( \epsilon_q^F - \epsilon_p^F + l \Omega ) ,
\\
& W_a(q \to p) =
\sum_{l \in \mathbb{Z}}
\Gamma_l (q, p)
\times 
N_b ( \epsilon_p^F - \epsilon_q^F - l \Omega )
\times 
\nu_B ( \epsilon_p^F - \epsilon_q^F - l \Omega ) .
\end{align}
Here, $\Gamma_l (q, p)$ is defined as:
\begin{align}
& \Gamma_l (q, p) \equiv 
2\pi | M_l (q, p)|^2
\sum_{l_1 \in \mathbb{Z}}
|\varphi_{l_1,p} \varphi_{l_1+l, q}^* |^2 
=
2\pi | M_l (q, p)|^2 \Phi_{q,p}^{(l)} 
= \Gamma_l (p, q) ,
\\
& | M_l (q, p)|^2
= |\bra{q}
\chi (|\epsilon_q^F - \epsilon_p^F + l \Omega|) \ket{p}|^2 .
\label{S-Gammal}
\end{align}
We note that $W_a(q \to p)$ is obtained from $W_e(q \to p)$ by replacing $N_{b} (\lambda)+1 \to  N_{b} (\lambda)$, $\omega_\lambda \to -\omega_\lambda$, and $\chi_\lambda \to \chi_\lambda^\dagger$.
Using the symmetry $\Gamma_l (q, p) = \Gamma_l (p, q)$ and observing the pattern of appearances of $\epsilon_p^F - \epsilon_q^F - l \Omega$ in $W_e(q \to p)$ and $W_a(q \to p)$, we obtain the total scattering matrix element:
\begin{align}
& W (q \to p) 
= W_e(q \to p) +  W_a(q \to p)
=
\sum_{l \in \mathbb{Z}}
\Gamma_l (q, p)
S ( \varepsilon_q - \varepsilon_p + l \Omega ),
\label{S-Wqp}
\end{align}
where we define $\varepsilon_p \equiv \epsilon_q^F$ for conciseness, and the function $S(x)$ as:
\begin{align}
S (x) \equiv [ N_b (|x|) + \Theta(x) ] \,
\nu_B (|x|) .
\end{align}
In the main text, we employ a simplified bath consisting of dispersionless ``Einstein'' phonons for clarity. While this simplification proves instructive, our general formalism for the scattering matrix element given by Eq.~(\ref{S-Wqp}) remains applicable to more realistic bosonic baths. For instance, when considering a physical phonon bath, one can incorporate the appropriate boson dispersion relation and modify the fermion-boson coupling matrix $\hat{\chi}_\lambda$ accordingly.

\subsection{Formal expressions for the steady state occupation and its derivatives}
\label{Sub-formal-generic}

In this subsection, we show formal expressions for the steady-state occupation and its derivatives.
We start from the steady-state condition [Eq.~\eqref{Floquet-Boltzmann} in the main text]:
\begin{align}
& \sum_{q}
( 
f_q W_{q \to p} \bar{f}_p
-
f_p W_{p\to q }\bar{f}_q
)=0,
\quad
\bar{f}_p \equiv 1 - f_p,
\end{align}
where $f_p$ is the occupation at momentum $p$, $W_{q \to p}$ is the scattering rate from momentum $q$ to $p$, and $\bar{f}_p$ is the vacancy at momentum $p$.

We relabel the momentum $p$ by $\{\varepsilon_p, \eta_p\}$, where $\varepsilon_p$ is the quasi-energy and $\eta_p$ is a generic parameter, which parametrizes the equi-energy surface at $\varepsilon_p$ [e.g., $(\theta,\phi)$ angles in 3D]. This yields:
\begin{align}
\int_{\varepsilon_b}^{\varepsilon_t} d \varepsilon_q
\oint d \eta_q 
|J_{\varepsilon_q, \eta_q}|
( 
f_{\varepsilon_q, \eta_q} W_{ {\varepsilon_q, \eta_q} \to {\varepsilon_p, \eta_p}} \bar{f}_{\varepsilon_p, \eta_p}
-
f_{\varepsilon_p, \eta_p} W_{ {\varepsilon_p, \eta_p} \to {\varepsilon_q, \eta_q} } \bar{f}_{\varepsilon_q, \eta_q}
)=0,
\label{S-fbe-energy}
\end{align}
where $J_{\varepsilon_q, \eta_q}$ is the Jacobian for the transformation $p \to \{\varepsilon_p, \eta_p\}$.

We then take the $n$-th derivative with respect to $\varepsilon_p$ on both sides, apply the general Leibniz rule, and obtain:
\begin{align}
f_{\varepsilon_p, \eta_p}^{[n]}
= 
\frac{1}{ R_{\varepsilon_p, \eta_p} }
\int_{\varepsilon_b}^{\varepsilon_t} d \varepsilon_q
\oint d \eta_q 
|J_{\varepsilon_q, \eta_q}|
\sum_{k=0}^{n-1}
\binom{n}{k}
\Big(
f_{\varepsilon_q, \eta_q}  W_{{\varepsilon_q, \eta_q} \to {\varepsilon_p, \eta_p} }^{[n-k]} \bar{f}_{\varepsilon_p, \eta_p}^{[k]}
- f_{\varepsilon_p, \eta_p}^{[k]} W_{{\varepsilon_p, \eta_p} \to {\varepsilon_q, \eta_q}}^{[n-k]}   \bar{f}_{\varepsilon_q, \eta_q}
\Big) ,
\quad
(n \geq 1) 
\end{align}
where the $k$-th derivative of a function $g$ with respect to $\epsilon_p$ is denoted by
\begin{equation}
g^{[k]}(\varepsilon_p) \equiv \frac{d^k g(\varepsilon_p)}{d\varepsilon_p^k} ,
\end{equation}
$\binom{n}{k} = \frac{n !}{k ! (n-k) !}$ is the binomial coefficient, and $R_{\varepsilon_p, \eta_p}$ represents the maximally allowed scattering rate at momentum $p$:
\begin{align}
R_{\varepsilon_p, \eta_p} = \int_{\varepsilon_b}^{\varepsilon_t} d \varepsilon_q
\oint d \eta_q 
|J_{\varepsilon_q, \eta_q}|
\big(
f_{\varepsilon_q, \eta_q} W_{ {\varepsilon_q, \eta_q} \to {\varepsilon_p, \eta_p} } + W_{ {\varepsilon_p, \eta_p} \to {\varepsilon_q, \eta_q} } \bar{f}_{\varepsilon_q, \eta_q} 
\big)
= 
\sum_q
\big(
f_{q} W_{ q \to p } + W_{ p \to q } \bar{f}_q 
\big)
.
\end{align}

We further apply the general Leibniz rule on $W_{ {\varepsilon_q, \eta_q} \to {\varepsilon_p, \eta_p} }^{[n-k]}$ and $W_{{\varepsilon_p, \eta_p} \to {\varepsilon_q, \eta_q}}^{[n-k]}$ [see Eq.~\eqref{S-Wqp}]:
\begin{align}
\begin{aligned}
W_{ {\varepsilon_q, \eta_q} \to {\varepsilon_p, \eta_p} }^{[n-k]} = 
\sum_{l \in \mathbb{Z}}
\sum_{j=0}^{n-k}
\binom{n-k}{j} &
\Big[\Gamma_l (\varepsilon_q, \eta_q; \varepsilon_p, \eta_p)
\Big]^{[n-k-j]}
S^{[j]} (\varepsilon_q - \varepsilon_p + l \Omega) ,
\\
W_{ {\varepsilon_p, \eta_p} \to {\varepsilon_q, \eta_q} }^{[n-k]} = 
\sum_{l \in \mathbb{Z}}
\sum_{j=0}^{n-k}
\binom{n-k}{j} &
\Big[\Gamma_l ( \varepsilon_q, \eta_q ; \varepsilon_p, \eta_p)
\Big]^{[n-k-j]}
S^{[j]} (\varepsilon_p - \varepsilon_p + l \Omega) .
\end{aligned}
\end{align}
Combining these results, we obtain:
\begin{align}
\begin{aligned}
R_{\varepsilon_p, \eta_p} \cdot f^{[n]}_{\varepsilon_p, \eta_p}
= 
\sum_{l \in \mathbb{Z}}
\sum_{k=0}^{n-1}
\sum_{j=0}^{n-k}
\binom{n}{k}
& \binom{n-k}{j}
\int_{\varepsilon_b}^{\varepsilon_t} d \varepsilon_q
\oint d \eta_q 
|J_{\varepsilon_q, \eta_q}|
\Big[\Gamma_l (\varepsilon_q, \eta_q; \varepsilon_p, \eta_p)
\Big]^{[n-k-j]}
\\
&
\times \Big(
f_{\varepsilon_q, \eta_q}  
S^{[j]} (\varepsilon_q - \varepsilon_p + l \Omega) \bar{f}_{\varepsilon_p, \eta_p}^{[k]}
-
\bar{f}_{\varepsilon_q, \eta_q}
S^{[j]} (\varepsilon_p - \varepsilon_q + l \Omega)
f_{\varepsilon_p, \eta_p}^{[k]}
\Big).
\label{RdotFn}
\end{aligned}
\end{align}

Finally, we can compactify the expression further and obtain:
\begin{align}
\begin{aligned}
f^{[n]}_{\varepsilon_p, \eta_p}
=
\sum_{l \in \mathbb{Z}}
\sum_{k=0}^{n-1}
\sum_{j=0}^{n-k}
\int_{\varepsilon_b}^{\varepsilon_t} d \varepsilon_q
& \Big(
B_{\varepsilon_p, \eta_p}^{lnkj} (\varepsilon_q)
S^{[j]} (\varepsilon_q - \varepsilon_p + l \Omega)
\bar{f}_{\varepsilon_p, \eta_p}^{[k]}
-
\bar{B}_{\varepsilon_p, \eta_p}^{lnkj} (\varepsilon_q)
S^{[j]} (\varepsilon_p - \varepsilon_q + l \Omega)
f_{\varepsilon_p, \eta_p}^{[k]}
\Big),
\label{fpn-e}
\end{aligned}
\end{align}
where the auxiliary functions $B_{\varepsilon_p, \eta_p}^{lnkj} (\varepsilon_q)$ and $\bar{B}_{\varepsilon_p, \eta_p}^{lnkj} (\varepsilon_q)$ are defined as:
\begin{align}
\begin{aligned}
B_{\varepsilon_p, \eta_p}^{lnkj} (\varepsilon_q)
& =
\frac{1}{R_{\varepsilon_p, \eta_p}}
\binom{n}{k}
\binom{n-k}{j}
\oint d \eta_q 
|J_{\varepsilon_q, \eta_q}| 
\big[\Gamma_l (\varepsilon_q, \eta_q; \varepsilon_p, \eta_p)
\big]^{[n-k-j]}
\times
f_{\varepsilon_q, \eta_q}  ,
\\
\bar{B}_{\varepsilon_p, \eta_p}^{lnkj} (\varepsilon_q)
& =
\frac{1}{R_{\varepsilon_p, \eta_p}}
\binom{n}{k}
\binom{n-k}{j}
\oint d \eta_q 
|J_{\varepsilon_q, \eta_q}| 
\big[\Gamma_l (\varepsilon_q, \eta_q; \varepsilon_p, \eta_p)
\big]^{[n-k-j]}
\times
\bar{f}_{\varepsilon_q, \eta_q}  .
\label{B-aux}
\end{aligned}
\end{align}

The calculation of derivatives is more transparent in momentum space:
\begin{align}
f_p^{\{n\}}
= 
\sum_{q}
\sum_{k=0}^{n-1}
\binom{n}{k}
\frac{
f_q  W_{q \to p}^{\{n-k\}} \bar{f}_{p}^{\{k\}}
- f_{p}^{\{k\}} W_{p\to q}^{\{n-k\}}   \bar{f}_{q}}
{R_p} ,
\quad
R_p
= 
\sum_q
\big(
f_{q} W_{ q \to p } + W_{ p \to q } \bar{f}_q 
\big)
\label{fpn-m}
\end{align}
where the superscript $\{k\}$ on a function denotes its $k$-th derivative with respect to the momentum $p$.
The momentum-space formulation of Eq.~\eqref{fpn-m} offers a compact representation; however, it obscures the essential role played by $S_x$. In contrast, Eq.~\eqref{fpn-e} underscores that non-analyticities arise exclusively from those present in $S_x$, which exhibits a sole dependence on the quasi-energy $\varepsilon_p$ and remains independent of the orientation $\eta_p$ characterizing the equi-energy surface.

\subsection{Generic analysis of non-analyticities}
\label{S-Generic-non-analyticities}

This subsection provides explicit calculations supporting the generic analysis of non-analyticities discussed in the main text. Consider an integral of the form:
\begin{align}
\begin{aligned}
R(\varepsilon) = \int_{\varepsilon_b}^{\varepsilon_t} d \varepsilon'  r(\varepsilon, \varepsilon') \delta (\varepsilon' - \varepsilon + \Lambda),
\label{S-R-definition}
\end{aligned}
\end{align}
where $r(\varepsilon, \varepsilon')$ is a continuous function. Evaluating $R(\varepsilon)$ in the vicinity of $\varepsilon = \varepsilon_{b,t} + \Lambda$ yields:
\begin{align}
\begin{aligned}
& R(\varepsilon = \varepsilon_{b} + \Lambda + 0^+)
=
\int_{\varepsilon_b}^{\varepsilon_t} d \varepsilon' 
r(\varepsilon_{b} + \Lambda + 0^+, \varepsilon') 
\delta (\varepsilon' - \varepsilon_b - 0^+ ) 
=
r(\varepsilon_{b} + \Lambda + 0^+, \varepsilon_b + 0^+) ,
\\
& R(\varepsilon = \varepsilon_{b} + \Lambda - 0^+)
=
\int_{\varepsilon_b}^{\varepsilon_t} d \varepsilon' 
r(\varepsilon_{b} + \Lambda - 0^+, \varepsilon') 
\delta (\varepsilon' - \varepsilon_b + 0^+ ) 
= 0,
\end{aligned}
\end{align}
and
\begin{align}
\begin{aligned}
& R(\varepsilon = \varepsilon_{t} + \Lambda + 0^+)
=
\int_{\varepsilon_b}^{\varepsilon_t} d \varepsilon' 
r(\varepsilon_{t} + \Lambda + 0^+, \varepsilon') 
\delta (\varepsilon' - \varepsilon_t - 0^+ ) 
=
0 ,
\\
& R(\varepsilon = \varepsilon_{t} + \Lambda - 0^+)
=
\int_{\varepsilon_b}^{\varepsilon_t} d \varepsilon' 
r(\varepsilon_{t} + \Lambda - 0^+, \varepsilon') 
\delta (\varepsilon' - \varepsilon_t + 0^+ )
= r(\varepsilon_{t} + \Lambda - 0^+, \varepsilon_{t} - 0^+).
\end{aligned}
\end{align}
Therefore non-analyticities manifest in $R(\varepsilon)$ near $\varepsilon = \varepsilon_{b,t} + \Lambda$ when the following condition is satisfied:
\begin{align}
\begin{aligned}
R(\varepsilon_{b,t} + \Lambda + 0^+) - R(\varepsilon_{b,t} + \Lambda - 0^+) = \pm r(\varepsilon_{b,t} + \Lambda, \varepsilon_{b,t}) \neq 0.
\label{S-R-discontinuity}
\end{aligned}
\end{align}

The analysis above, combined with the expression for the $n$-th derivative of the occupation function in Eq.~\eqref{fpn-e}, clarifies mathematically the origins of apparent non-analyticities or discontinuities in $f^{[n]}_{\varepsilon_p, \eta_p}$. Equation~(\ref{fpn-e}), repeated here for convenience,
\begin{align}
\begin{aligned}
f^{[n]}_{\varepsilon_p, \eta_p}
=
\sum_{l \in \mathbb{Z}}
\sum_{k=0}^{n-1}
\sum_{j=0}^{n-k}
\int_{\varepsilon_b}^{\varepsilon_t} d \varepsilon_q
\Big(
B_{\varepsilon_p, \eta_p}^{lnkj} (\varepsilon_q)
S^{[j]} (\varepsilon_q - \varepsilon_p + l \Omega)
\bar{f}_{\varepsilon_p, \eta_p}^{[k]}
-
\bar{B}_{\varepsilon_p, \eta_p}^{lnkj} (\varepsilon_q)
S^{[j]} (\varepsilon_p - \varepsilon_q + l \Omega)
f_{\varepsilon_p, \eta_p}^{[k]}
\Big),
\nonumber
\end{aligned}
\end{align}
has a structure similar to that of $R(\varepsilon)$ in Eq.~\eqref{S-R-definition}. If either $S^{[j]} (\varepsilon_q - \varepsilon_p + l \Omega)$, the $j$-th derivative of $S$ with respect to $\varepsilon_p$, or $f_{\varepsilon_p, \eta_p}^{[k]}$, the $k$-th derivative of $f_{\varepsilon_p, \eta_p}$ with respect to $\varepsilon_p$, contains a Dirac delta function, it can lead to discontinuities in $f^{[n]}_{\varepsilon_p, \eta_p}$. This conclusion follows the same reasoning as the one used to explain the discontinuity in $R(\varepsilon)$ through Eqs.~(\ref{S-R-definition}) to (\ref{S-R-discontinuity}). Therefore, the presence of Dirac delta functions in the derivatives of $S$ or $f_{\varepsilon_p, \eta_p}$ is responsible for the non-analytic or discontinuous behavior of $f^{[n]}_{\varepsilon_p, \eta_p}$.

\section{Analytical analysis of \texorpdfstring{$S$}{S} function for different types of bath}
\label{SAnalysis}

In Section~\ref{S-Analytical-nonanalyticities}, we demonstrated that the behavior of $S$ function can lead to non-analyticities in the system's occupation. 
To support the analysis in the main text, we now perform a detailed analytical examination of the $S$ function for the Ohmic and gapped baths.

\subsection{Ohmic bath with a finite temperature}

We first consider the following $S$ function with a finite bath temperature:
\begin{align}
\begin{aligned}
S(\omega) = [N_b (|\omega|)+\Theta(\omega)] \nu_B(|\omega|),
\quad
\nu_B(\omega) = (c_1 \omega + c_2 \omega^2 ) \Theta(\omega),
\end{aligned}
\end{align}
where $N_b(\omega)$ is the Bose-Einstein distribution and $\Theta(\omega)$ is the Heaviside step function.
More explicitly,
\begin{align}
\begin{aligned}
S(\omega) = c_1 y (\omega)
+ c_2 y (\omega) |x|,
\quad
y(\omega) =  \frac{ \omega e^{\beta \omega}}{e^{\beta \omega}-1} \text{~is a smooth function},
\quad
y(0) = 1/\beta .
\end{aligned}
\end{align}
By taking derivatives of $S(\omega)$ with respect to $\omega$, we obtain
\begin{align}
\begin{aligned}
S^{[1]}(\omega) = c_1 y^{[1]}(\omega)
+
c_2 y^{[1]}(\omega)\, |\omega| + c_2 y(\omega)\, {\rm sgn}(\omega),
\end{aligned}
\end{align}
\begin{align}
\begin{aligned}
S^{[2]}(\omega)
= c_1 y^{[2]}(\omega)
+
c_2 y^{[2]}(\omega)\, |\omega| 
+
2 c_2 y^{[1]}(\omega)\, {\rm sgn}(\omega) 
+
\frac{2 c_2}{\beta}\, {\color{blue}\delta (\omega)} ,
\end{aligned}
\end{align}
during which we used the identity $y(x) \delta(x) = y(0) \delta(x)$. Consequently, we observe that the second-order derivative $S^{[2]}(\omega)$ is the lowest order derivative containing a Dirac delta function at finite bath temperature. According to Eq.~\eqref{fpn-e} and following the reasoning used to explain the discontinuity in $R(\varepsilon)$ through Eqs.~\eqref{S-R-definition} to \eqref{S-R-discontinuity}, the Dirac delta function in $S^{[2]}$ leads to discontinuities at $l \Omega$ in $f^{[2]} (\epsilon)$ of the system's occupation function at finite bath temperature.

As a concrete example,
following Eq.~\eqref{S-Fn-1D2D} in the Appendix~\ref{Floquet-Boltzmann-in-1Dand2D}, for parabolic bands in 1D and 2D, we have
\begin{align}
\begin{aligned}
f^{[2]}(l \Omega^+) - 
& f^{[2]}(l \Omega^-) 
\propto \frac{2 c_2}{\beta} \,
\varrho^{d} (0^+)
\Gamma_{l}^{d} (0^+, l \Omega^+)
 \big[
f (0^+)
\bar{f} (l \Omega^+)
-
\bar{f} (0^+)
f (l \Omega^+)
\big] .
\label{S-f3-parabolic}
\end{aligned}
\end{align}
In the above equation, $l \Omega^\pm \equiv l \Omega \pm 0^+$, where $0^+$ is a positive infinitesimal, $\Gamma_l^{d} (\epsilon_q, \epsilon_p)$ is the scattering amplitude corresponding to the scattering matrix $W_{\epsilon_p \to \epsilon_q}^{d} = \sum_l \Gamma_l^{d} (\epsilon_q, \epsilon_p) S ( \epsilon_q - \epsilon_p + l \Omega )$, and $\varrho^{d} (\epsilon) = \epsilon^{(d-2)/2}$ is the density of states, both for the $d$-dimensional ($d = 1,2$) parabolic model. 
Since $\epsilon = k^2/2m$ for parabolic bands, the discontinuities in the above second energy derivative $f^{[2]}(\epsilon) = d^2f(\epsilon)/d\epsilon^2$ can be mapped to discontinuities in $d^2f(k)/dk^2$ through the chain rule of differentiation.

\subsection{Ohmic bath with zero temperature}
We now consider the following $S$ function with zero bath temperature:
\begin{align}
\begin{aligned}
S(\omega) = \Theta(\omega) \nu_B(|\omega|)
= c_1 \omega \Theta(\omega)
+ c_2 \omega^2  \Theta(\omega),
\quad
\nu_B(\omega) = (c_1 \omega + c_2 \omega^2 ) \Theta(\omega),
\end{aligned}
\end{align}
where we take $N_b(\omega) = 0$ in this case. By taking derivatives of $S(\omega)$ with respect to $\omega$, we obtain
\begin{align}
\begin{aligned}
S^{[1]}(\omega) = c_1 \Theta(\omega) + 2 c_2 \omega \Theta(\omega)
\end{aligned}
\end{align}
\begin{align}
\begin{aligned}
S^{[2]}(\omega) = c_1 {\color{blue}\delta(\omega)} + 2 c_2  \Theta(\omega)
\end{aligned}
\end{align}
during which we used the identity $x\delta(x) = 0$.
We observe that the second-order derivative $S^{[2]}(\omega)$ is the lowest order derivative containing a Dirac delta function at zero temperature.
According to Eq.~\eqref{fpn-e} and following the reasoning used to explain the discontinuity in $R(\varepsilon)$ through Eqs.~\eqref{S-R-definition} to \eqref{S-R-discontinuity}, this leads to discontinuities at $l \Omega$ in $f^{[2]} (\epsilon)$ of the system's occupation function at finite bath temperature.

As a concrete example,
following Eq.~\eqref{S-Fn-1D2D} in the Appendix~\ref{Floquet-Boltzmann-in-1Dand2D}, for parabolic bands in 1D and 2D, we have in this case
\begin{align}
\begin{aligned}
f^{[2]}(l \Omega^+) - 
& f^{[2]}(l \Omega^-) 
\propto c_1 \,
\varrho^{d} (0^+)
\Gamma_{l}^{d} (0^+, l \Omega^+)
 \big[
f (0^+)
\bar{f} (l \Omega^+)
-
\bar{f} (0^+)
f (l \Omega^+)
\big] ,
\label{S-f2-parabolic}
\end{aligned}
\end{align}
where $l \Omega^\pm \equiv l \Omega \pm 0^+$, $\Gamma_l^{d} (\epsilon_q, \epsilon_p)$ is the scattering amplitude and $\varrho^{d} (\epsilon) = \epsilon^{(d-2)/2}$ is the density of states, both for the $d$-dimensional ($d = 1,2$) parabolic model.
The discontinuities in the above second energy derivative $f^{[2]}(\epsilon) = d^2f(\epsilon)/d\epsilon^2$ can be mapped to discontinuities in $d^2f(k)/dk^2$ through the chain rule of differentiation.

\subsection{Gapped bath with a finite temperature}

Here we consider the following $S$ function with a finite bath temperature:
\begin{align}
\begin{aligned}
S(\omega) = [N_b (|\omega|)+\Theta(\omega)] \nu_B(|\omega|),
\quad
\nu_B (\omega) = \Theta(\omega - \Delta),
\quad
\Delta > 0.
\end{aligned}
\end{align}
The first-order derivative of $S(\omega)$ with respect to $\omega$ already contains Dirac delta functions:
\begin{align}
\begin{aligned}
S^{[1]}(\omega) = 
\begin{cases}
\partial_\omega N_b (\omega)\, \Theta(+\omega - \Delta) + [N_b (|\omega|)+1] {\color{blue}\delta(\omega - \Delta)} , & \omega >0
\\
\partial_\omega N_b (-\omega)\, \Theta(-\omega - \Delta) -N_b (|\omega|){\color{blue}\delta(\omega + \Delta)} , & \omega <0
\end{cases}
\end{aligned}
\end{align}
According to Eq.~\eqref{fpn-e} and following the reasoning used to explain the discontinuity in $R(\varepsilon)$ through Eqs.~\eqref{S-R-definition} to \eqref{S-R-discontinuity}, Dirac delta functions in $S^{[1]}(\omega)$ lead to discontinuities at $l \Omega \pm \Delta$ in $f^{[1]} (\epsilon)$ of the system's occupation function at finite bath temperature.
These discontinuities in $f^{[1]} (\epsilon)$ will propagate via the same mechanism to $l \Omega \pm n \Delta$ in higher order derivatives $f^{[n]} (\epsilon)$.

As a concrete example,
following Eq.~\eqref{S-Fn-1D2D} in the Appendix~\ref{Floquet-Boltzmann-in-1Dand2D}, for parabolic bands in 1D and 2D, we have in this case
\begin{align}
\begin{aligned}
& f^{[1]}_{l \Omega^+ \pm \Delta} - f^{[1]}_{l \Omega^- \pm \Delta}
\propto
\pm
\varrho^{d} (0^+)
\Gamma_{l}^{d} (0^+, l \Omega^+ \pm \Delta)
\Big(  f_{0^+} \bar{f}_{l \Omega^+ \pm \Delta}
N_b (\Delta) - f_{l \Omega^+ \pm \Delta} \bar{f}_{0^+}[ N_b (\Delta) + 1 ]  \Big) ,
\label{S-f1-parabolic}
\end{aligned}
\end{align}
in which $l \Omega^\pm \equiv l \Omega \pm 0^+$, $\Gamma_l^{d} (\epsilon_q, \epsilon_p)$ is the scattering amplitude and $\varrho^{d} (\epsilon) = \epsilon^{(d-2)/2}$ is the density of states, both for the $d$-dimensional ($d = 1,2$) parabolic model. The discontinuities in the above first energy derivative $f^{[1]}(\epsilon) = d f(\epsilon)/d\epsilon$ can be mapped to discontinuities in $df(k)/dk$ through the chain rule of differentiation.

\section{Dimensionless parabolic models in 1D and 2D}
\label{Dimensionless-Models}

In this section, we non-dimensionalize the parabolic models in one and two dimensions introduced in the main text for the purpose of numerical calculations. We define the relevant dimensionless quantities and relate the particle density to the equilibrium Fermi wave vector and Fermi energy at zero temperature.

For clarity and readability, we omit the bar symbols over dimensionless quantities in the subsequent supplementary materials whenever the context makes it clear that the quantities are dimensionless.

\subsection{Dimensionless parabolic model in 1D}

We begin with the 1D parabolic model driven by an oscillating electric field:
\begin{align}
\begin{aligned}
\epsilon_k = \frac{k^2}{2m}
\to
\epsilon_k (t) = 
\frac{[k - A_0 \sin(\Omega t + \phi_0 )]^2}{2m}.
\end{aligned} 
\end{align}
To obtain a dimensionless model, we introduce the characteristic energy $\epsilon_c$, and the following dimensionless quantities (note that in the main text we choose $\epsilon_c = \Omega$):
\begin{align}
\begin{aligned} 
& \bar{k}= \frac{k} { k_c } = \frac{k} { \sqrt{2m \epsilon_c} } ,
\quad
\bar{\epsilon}_k= \frac{ \epsilon_k }{ \epsilon_c } =\bar{k}^2 ,
\quad
\bar{A}_0= \frac{A_0} { k_c } =\frac{A_0 } { \sqrt{2m \epsilon_c} },
\quad
\bar{\Omega}= \frac{ \Omega }{ \epsilon_c } ,
\quad
\bar{t} = \epsilon_c \, t .
\end{aligned}
\end{align}
The dimensionless parameter $\bar{A}_0$ characterizes how far the drive pushes an electron back and forth in the parabolic band relative to $k_c = \sqrt{2m \epsilon_c}$. Using these dimensionless quantities, the 1D parabolic model becomes:
\begin{align}
\begin{aligned}
\bar{\epsilon}_k=\bar{k}^2 
\to
\bar{\epsilon}_k ( { \bar{t} } ) = [ \bar{k}-\bar{A}_0 
\sin (  \bar{\Omega} \bar{t} + \phi_0 ) ]^2 .
\label{S-Dless-1D}
\end{aligned}
\end{align}

Assuming the fermionic system is initially at zero temperature and equilibrium, the particle number is conserved due to the absence of particle exchange with the bosonic bath. 
We can relate the particle density to the initial equilibrium dimensionless Fermi wave vector $\bar{k}_{F, \text{initial}}$ and Fermi energy $\bar{k}_{F,\text{initial}}^2 = \bar{\epsilon}_{F,\text{initial}} = \bar{\mu}_0  $, where $\bar{\mu}_0 \equiv \mu_0 / \epsilon_c$:
\begin{align}
\bar{n}_0
=
\int_0^{+\infty} d \bar{k} \Theta\left(\bar{\mu}_0-\bar{k}^2\right)
=
\sqrt{\bar{\mu}_0}
=
\bar{k}_{F,\text{initial}} .
\end{align}
Thus, the particle density $n_0$ is simply related to the dimensionless Fermi wave vector and energy as $n_0=\sqrt{\bar{\mu}_0}=\sqrt{\bar{\epsilon}_F}=\bar{k}_F$ in 1D.

The expressions for the Floquet energy harmonics and Floquet wavefunction harmonics for the dimensionless parabolic model in 1D are provided below (with the bar symbol omitted). 
Floquet energy harmonics:
\begin{align}
\begin{aligned}
& \epsilon_{l, k}
= 
\int_0^T \frac{d t}{T} 
\epsilon_k (t) e^{+ i l \Omega t } 
= [ \epsilon_{-l, k} ]^* ,
\label{S-fe1D}
\end{aligned}
\end{align}
\begin{align}
\begin{aligned}
\varepsilon_k \equiv \epsilon_{l = 0, k} = k^2 + \frac{A_0^2}{2},
\quad
\epsilon_{+1, k} = - i A_0 k e^{- i \phi_0},
\quad
\epsilon_{+2, k} = -\frac{A_0^2}{4} e^{- 2 i \phi_0} ,
\quad
\epsilon_{| l |>2, k} = 0 .
\label{S-fe1D-1}
\end{aligned}
\end{align}
Floquet wavefunction harmonics:
\begin{align}
\begin{aligned}
\varphi_{l, k}
& = \int_0^T \frac{d t}{T}
\left[
\exp (+ i l \Omega t ) 
\exp 
\left(
-\frac{i}{\hbar} \int_0^t d t'
\big[ \epsilon_k (t')-\varepsilon_{k}\big]
\right) 
\right] .
\end{aligned}
\end{align}
While Floquet wavefunction harmonics lack a closed-form expression, we provide perturbative expressions for $\Gamma_l^{\rm 1D}(q,p)$ [see Eq.~\eqref{S-Gammal}] by calculating the factor $\Phi^{(l)}_{q,p} = \sum_{l_1 \in {\mathbf Z}} |\varphi_{l_1,p}\varphi_{l_1+l,q}^*|^2$ up to $O(A_0^6)$, which is necessary to obtain the scattering matrix $W(q \to p)$:
\begin{align}
\begin{aligned}
& \Gamma_{0}^{\rm 1D} (q, p) = 
2 \pi
| M_0 (q, p)|^2 \left[ 
1
-\frac{2(p-q)^2}{\Omega^2} A_0^2 
+\frac{3(p-q)^4}{2 \Omega^4} A_0^4
+ O(A_0^6)
\right] ,
\\
& 
\Gamma_{\pm 1}^{\rm 1D} (q, p) = 
2 \pi
| M_{\pm 1}(q, p)|^2 \left[ 
\frac{(p-q)^2}{\Omega^2} A_0^2 
- \frac{(p-q)^4}{\Omega^4} A_0^4
+ O(A_0^6)
\right] ,
\\
& 
\Gamma_{\pm2}^{\rm 1D} (q, p) = 
2 \pi
| M_{\pm 2}(q, p)|^2 \left[ 
\frac{(p-q)^4}{4 \Omega^4} A_0^4
+ O(A_0^6)
\right] ,
\\
&
\Gamma_{\pm3}^{\rm 1D} (q, p) \propto  
O(A_0^6) .
\label{S-Gammas-1D}
\end{aligned}
\end{align}

\subsection{Dimensionless parabolic model in 2D driven by a circularly polarized electric field}

The 2D parabolic model driven by a circularly polarized electric field is:
\begin{align}
\begin{aligned}
\epsilon_{\bf k} = \frac{{\bf k}^2}{2m}
\to
\epsilon_{\bf k}^{\pm} (t) = 
\frac{[k_x - A_0 \sin(\Omega t + \phi_0 )]^2}{2m} 
+ 
\frac{[k_y - A_0 \sin(\Omega t + \phi_0 \pm \pi/2)]^2}{2m} .
\end{aligned} 
\end{align}
where the superscript $\pm$ denotes the helicity of the circularly polarized electric field: $(+)$ for left-handed (counter-clockwise) and $(-)$ for right-handed (clockwise) polarization.
Similar to the previous subsection, we introduce the characteristic energy $\epsilon_c$ and the following dimensionless quantities (note that in the main text we choose $\epsilon_c = \Omega$):
\begin{align}
\begin{aligned} 
& \bar{k}_{x,y}= \frac{k_{ x,y }}{{ k_c }}=\frac{k_{ x,y } }{ \sqrt{2m \epsilon_c} } ,
\quad
\bar{\epsilon}_{\bf k}= \frac{ \epsilon_{\bf k} }{ \epsilon_c } =\bar{k}_x^2 +  \bar{k}_y^2,
\quad
\bar{A}_0=\frac{A_0} { k_c }= \frac{A_0 } { \sqrt{2m \epsilon_c} },
\quad
\bar{\Omega}= \frac{\Omega }{ \epsilon_c } ,
\quad
\bar{t} = \epsilon_c \, t .
\end{aligned}
\end{align}
The dimensionless 2D parabolic model is then:
\begin{align}
\begin{aligned}
\bar{\epsilon}_{\bf k}= \bar{\bf k}^2 = \bar{k}_x^2 +  \bar{k}_y^2
\to
\bar{\epsilon}_{\bf k}^{\pm} (t) = [ \bar{k}_x - \bar{A}_0 
\sin (  \bar{\Omega} \bar{t} + \phi_0 ) ]^2 
+ 
[ \bar{k}_y - \bar{A}_0 
\sin (  \bar{\Omega} \bar{t} + \phi_0 \pm \pi/2 ) ]^2 .
\label{S-Dless-2D}
\end{aligned}
\end{align}
In 2D, the relation between the particle density and the initial dimensionless Fermi wave vector and energy is:
\begin{align}
\bar{n}_0=
\frac{\pi k_{F,\text{initial}}^2}{\pi k_c^2}
=
\bar{k}_{F,\text{initial}}^2
=
\bar{\epsilon}_{F,\text{initial}}
=
\bar{\mu}_0 .
\end{align}
The expressions for the Floquet energy harmonics and Floquet wavefunction harmonics for the dimensionless parabolic model in 2D are provided below (with the bar symbol omitted).

Floquet energy harmonics:
\begin{align}
\begin{aligned}
& \epsilon_{+ l, \bf k}^{\pm}
= 
\int_0^T \frac{d t}{T} 
\epsilon_{\bf k} (t) \exp (+ i l \Omega t ) 
= [ \epsilon_{-l, \bf k}^{\pm} ]^*,
\end{aligned}
\end{align}
\begin{align}
\begin{aligned}
\varepsilon_{\bf k} \equiv \epsilon_{l = 0, \bf k}^{\pm} = {\bf k}^2 + A_0^2,
\quad
\epsilon_{+1, \bf k}^{\pm} = A_0 e^{ - i \phi_0} (-i k_x \mp k_y),
\quad
\epsilon_{| l |>1, \bf k}^{\pm} = 0 .
\label{S-fe2D-1}
\end{aligned}
\end{align}

Floquet wavefunction harmonics can be calculated using the Jacobi-Anger expansion and have closed forms:
\begin{align}
\begin{aligned}
\varphi_{l, \bf k}^{\pm} & = \int_0^T \frac{d t}{T}
\left[
\exp (+ i l \Omega t ) 
\exp 
\left(
-\frac{i}{\hbar} \int_0^t d t'
\big[ \epsilon_{\bf k} (t')-\epsilon _{\bf k}^{(0)}\big]
\right) 
\right] 
\\
& = \exp \left(
-\sum_{l_1 \neq 0}
\frac{ \epsilon_{\bf k}^{(l_1)} }{l_1 \Omega}
\right)
 \times 
 \int_0^T \frac{d t}{T} 
 \exp \left(
 \sum_{l_1 \neq 0}
 \frac{ \epsilon_{\bf k}^{(l_1)} e^{-i l_1 \Omega t}}{l_1 \Omega} + i l \Omega t\right) 
 \\
 & = 
 \exp \left(
-\sum_{l_1 \neq 0}
\frac{ \epsilon_{\bf k}^{(l_1)} }{l_1 \Omega}
\right) \times 
J_{\mp l}
\left(\frac{2 A_0 |{\bf k}|}{\Omega}\right)
\exp(\pm i l \theta_{\bf k} - i l \phi_0 ) ,
\quad
\tan \theta_{\bf k} = k_y / k_x .
\label{S-Fwh-2D}
\end{aligned}
\end{align}
Similar to the 1D case, we provide perturbative expressions for $\Gamma_l^{\rm 2D}({\bf q}, {\bf p})$ [see Eq.~\eqref{S-Gammal}] by calculating the factor $\Phi^{(l)}_{{\bf q},{\bf p}} =\sum_{l_1 \in {\mathbf Z}} |\varphi_{l_1, {\bf p} }\varphi_{l_1+l, {\bf q} }^*|^2$ up to $O(A_0^6)$, which is necessary to obtain the scattering matrix $W( {\bf q} \to {\bf p} )$:
\begin{align}
\begin{aligned}
& \Gamma_{0}^{\rm 2D} ( {\bf q}, {\bf p} ) = 
2 \pi
| M_0 ( {\bf q}, {\bf p} )|^2 \left[ 
1
-\frac{2 ({\bf p}^2+{\bf q}^2)}{\Omega^2} A_0^2
+\frac{3 ({\bf p}^4+4 {\bf p}^2 {\bf q}^2+{\bf p}^4)}{2 \Omega^4} A_0^4
+ O(A_0^6)
\right] ,
\\
& 
\Gamma_{\pm1}^{\rm 2D} ( {\bf q}, {\bf p} ) = 
2 \pi
| M_{\pm 1}( {\bf q}, {\bf p})|^2 \left[ 
\frac{({\bf p}^2+{\bf q}^2)}{\Omega^2}A_0^2
-\frac{({\bf p}^4+4 {\bf p}^2 {\bf q}^2+{\bf p}^4)}{\Omega^4} A_0^4
+ O(A_0^6)
\right] ,
\\
& 
\Gamma_{\pm2}^{\rm 2D} ( {\bf q}, {\bf p} ) = 
2 \pi
| M_{\pm 2}( {\bf q}, {\bf p})|^2 \left[ 
\frac{({\bf p}^4+4 {\bf p}^2 {\bf q}^2+{\bf p}^4)}{4 \Omega^4} A_0^4
+ O(A_0^6)
\right] ,
\\
&
\Gamma_{\pm3}^{\rm 2D} ( {\bf q}, {\bf p} ) \propto  
O(A_0^6) .
\label{S-Gammas-2D}
\end{aligned}
\end{align}
Here we note that $\Gamma_l^{\rm 2D}({\bf q}, {\bf p})$ does not depend on the helicity of the circularly polarized electric field.

\section{Floquet-Boltzmann equation in energy space for parabolic models with uniform bosonic baths}
\label{Floquet-Boltzmann-in-1Dand2D}

In this section, we derive Floquet-Boltzmann equations in quasi-energy space for the parabolic models in 1D and 2D. We consider these models coupled to bosonic baths, and for simplicity, we assume the bath has a momentum-independent coupling strength, which we refer to as the uniform coupling approximation, characterized by a constant form factor [see Eq.~\eqref{general-scattering} in the main text]:
\begin{align}
|\bra{q}\chi_\lambda\ket{p}|^2  \to |\chi_0|^2 ,
\end{align}
where $\chi_0$ is a constant. This approximation leads to a simplified expression for $\Gamma_l(q,p)$ [see Eq.~\eqref{S-Gammal}]:
\begin{align}
\Gamma_l (q, p)  
\to 
\Gamma_0 \sum_{l_1 \in {\mathbf Z}}
|\varphi_{l_1,p} \varphi_{l_1+l, q}^* |^2
=
\Gamma_0 \Phi^{(l)}_{q,p},
\quad
\Gamma_0 \equiv 2 \pi |\chi_0|^2 ,
\label{S-Gammal-1}
\end{align}
as well as a simplified scattering matrix [see Eq.~\eqref{S-Wqp}]:
\begin{align}
W (q \to p) 
= W_e(q \to p) +  W_a(q \to p)
=
\Gamma_0 \sum_{l \in {\mathbf Z}}
\Phi^{(l)}_{q,p}
S ( \varepsilon_q - \varepsilon_p + l \Omega ),
\label{S-WSimple}
\end{align}
which now only depends on the Floquet wavefunction harmonics $\varphi_{l, p}$ and Floquet quasi-energies $\varepsilon_p$.

Before presenting the detailed derivations below, we summarize the Floquet-Boltzmann equation in quasi-energy space for the parabolic models in 1D and 2D that we consider in the main text:
\begin{align}
0 = \int_{0}^{+\infty} 
f ( \epsilon_q )
W_{\epsilon_q \to \epsilon_p}^{d}
\bar{f} (\epsilon_p) 
\varrho^{d} (\epsilon_q)
d \epsilon_q  
-
\int_{0}^{+\infty} 
f ( \epsilon_p )
W_{\epsilon_p \to \epsilon_q}^{d}
\bar{f} (\epsilon_q) 
\varrho^{d} (\epsilon_q)
d \epsilon_q  ,
\quad
\bar{f}(\epsilon_p) \equiv 1 - f(\epsilon_p) ,
\label{S-FB1and2D}
\end{align}
where $f(\epsilon_p)$ is the occupation function labeled by the energy $\epsilon_p = p^2$, 
$W_{\epsilon_p \to \epsilon_q}^{d}$ and $\varrho^{d} (\epsilon_q) = \epsilon_q^{(d-2)/2}$ are the scattering matrix and the density of states  for the $d$-dimensional ($d = 1,2$) parabolic model in energy space, respectively.

Moreover, following the analysis in the subsection \ref{Sub-formal-generic}, their $n$-th derivative with respect to $\epsilon_p$ of the occupation reads
\begin{align}
\begin{aligned}
f^{[n]}_{\epsilon_p}
= 
\sum_{l \in \mathbb{Z}}
\sum_{k=0}^{n-1}
\sum_{j=0}^{n-k}
B_{\epsilon_p}^{(n j k)}
& \int_{0}^{+\infty} 
d \epsilon_q 
\varrho^{d} (\epsilon_q)
\big[ 
\Gamma_l^{d} (\epsilon_q, \epsilon_p)
\big]^{[n-k-j]}
\\
\times
\Big(
& f_{\epsilon_q}  
S^{[j]} (\epsilon_q - \epsilon_p + l \Omega) \bar{f}_{\epsilon_p}^{[k]}
-
\bar{f}_{\epsilon_q}
S^{[j]} (\epsilon_p - \epsilon_q + l \Omega)
f_{\epsilon_p}^{[k]}
\Big) ,
\label{S-Fn-1D2D}
\end{aligned}
\end{align}
with the coefficient $B_{\epsilon_p}^{(n j k)} = \binom{n}{k}
\binom{n-k}{j} / R_{\epsilon_p}$, and $R_{\epsilon_p}$ represents the maximally allowed scattering rate at momentum $p$:
\begin{align}
R_{\epsilon_p} = 
\int_{0}^{+\infty} 
d \epsilon_q 
\varrho^{d} (\epsilon_q)
\big(
f_{\epsilon_q} W_{ \epsilon_q, \epsilon_p }^{d} + W_{ \epsilon_p, \epsilon_q }^{d} \bar{f}_{\epsilon_q} 
\big)
\label{R-1D2D}
\end{align}

In the following sub-sections, we provide detailed derivations of the Floquet-Boltzmann equation for the 1D and 2D parabolic models coupled to simple bosonic baths under the uniform coupling approximation. The derivations lead to the specific scattering matrix $W_{\epsilon_p \to \epsilon_q}^{d} = \sum_l \Gamma_l^{d} (\epsilon_q, \epsilon_p) S ( \epsilon_q - \epsilon_p + l \Omega )$ and density of states $\varrho^{d} (\epsilon_q)$ used in Eqs.~(\ref{S-FB1and2D}), (\ref{S-Fn-1D2D}), and (\ref{R-1D2D}).

\subsection{Floquet-Boltzmann equation for the 1D parabolic model}

For the 1D parabolic model coupled to the bosonic bath described above, the Floquet-Boltzmann equation [see Eq.~\eqref{Floquet-Boltzmann} in the main text] takes the form:
\begin{align}
0 = 
\int_{-\infty}^{+\infty} 
f_qW_{q \to p} \bar{f}_p  \frac{d q}{2 \pi}
-
\int_{-\infty}^{+\infty} f_p W_{p \to q} \bar{f_q} \,\frac{ d q}{2 \pi}, 
\label{S-FB1D-1}
\end{align}
with the scattering matrix given by Eq.~\eqref{S-WSimple}. 
Given the momentum-independent difference between the original dispersion $\epsilon_k = k^2$ and the quasi-energy $\varepsilon_k = k^2 + A_0^2/2$ for the 1D parabolic model, we can recast the momentum-space equation in terms of the energy variable $\epsilon_k$:
\begin{align}
0 = \int_{0}^{+\infty} 
f ( \epsilon_q)
W_{\epsilon_q \to \epsilon_p}^{\rm 1D}
\bar{f} (\epsilon_p) \frac{d \epsilon_q}{\sqrt{\epsilon_q}}  
-
\int_{0}^{+\infty} f ( \epsilon_p )
W_{\epsilon_p \to \epsilon_q}^{\rm 1D}
\bar{f} (\epsilon_q) \frac{d \epsilon_q}{\sqrt{\epsilon_q}},
\label{S-FB1D-2}
\end{align}
in which we define the summed scattering matrix as
\begin{align}
W_{\epsilon_q \to \epsilon_p}^{\rm 1D}
\equiv
W_{ - \sqrt{\epsilon_q} \to +\sqrt{\epsilon_p} }
+
W_{ + \sqrt{\epsilon_q} \to +\sqrt{\epsilon_p} } ,
\quad
W_{\epsilon_p \to \epsilon_q}^{\rm 1D}
\equiv
W_{ + \sqrt{\epsilon_p} \to - \sqrt{\epsilon_q} }
+
W_{ + \sqrt{\epsilon_p} \to +\sqrt{\epsilon_q} } 
\end{align}
and use the symmetry $f ( \epsilon_p) = f (+p) = f(-p)$ to obtain the above equation. 
Eq.~\eqref{S-FB1D-2} is a specialized case of the general Eq.~\eqref{S-fbe-energy}.

Using Eqs.~(\ref{S-Gammas-1D}) and (\ref{S-WSimple}), we write the summed scattering matrix up to $O(A_0^6)$:
\begin{align}
\begin{aligned}
W_{\epsilon_q \to \epsilon_p}^{\rm 1D}
& =
\sum_l \Gamma_l^{\rm 1D} (\epsilon_q, \epsilon_p)
S (\epsilon_q - \epsilon_p + l \Omega)
\\
& = 
\Gamma_0
\Big[2
-
\frac{4 (\epsilon_p+\epsilon_q) }{\Omega^2} A_0^2
+
\frac{3(\epsilon_p^2+6 \epsilon_p \epsilon_q+\epsilon_q^2) }{\Omega^4} A_0^4 \Big] S (\epsilon_q - \epsilon_p)  \\
&\quad
+  \sum_{\eta = \pm 1}
\Gamma_0
\Big[
\frac{2 (\epsilon_p+\epsilon_q) }{\Omega^2} A_0^2
-
\frac{2(\epsilon_p^2+6 \epsilon_p \epsilon_q+\epsilon_q^2) }{\Omega^4} A_0^4 \Big] S (\epsilon_q - \epsilon_p + \eta \Omega)  \\
&\quad
+ 
\sum_{\eta = \pm 1}
\Gamma_0
\Big[
\frac{(\epsilon_p^2+6 \epsilon_p \epsilon_q+\epsilon_q^2) }{2 \Omega^4} A_0^4 \Big] S (\epsilon_q - \epsilon_p + 2 \eta \Omega)
+ O(A_0^6) ,
\label{S-W-1D}
\end{aligned}
\end{align}
while $W_{\epsilon_p \to \epsilon_q}^{\rm 1D}$ is obtained by swapping $\epsilon_q \leftrightarrow \epsilon_p$ in the above expression.

Here we also show the formal expressions of derivatives of $f(\epsilon_p)$ following the analysis in the subsection \ref{Sub-formal-generic}.
We take the $n$-th derivative with respect to $\epsilon_p$ on both sides of Eq.~\eqref{S-FB1D-2}, apply the general Leibniz rule, and obtain:
\begin{align}
f_{\epsilon_p}^{[n]}
= 
\frac{1}{ R_{\epsilon_p} }
\sum_{k=0}^{n-1}
\binom{n}{k}
\int_{0}^{+\infty} \frac{d \epsilon_q}{\sqrt{\epsilon_q}} 
\Big(
f_{\epsilon_q}  
\big[ W_{\epsilon_q \to \epsilon_p}^{\rm 1D} \big]^{[n-k]}
\bar{f}_{\epsilon_p}^{[k]}
 - 
 f_{\epsilon_p}^{[k]} 
 \big[ W_{\epsilon_p \to \epsilon_q}^{\rm 1D} \big]^{[n-k]}   \bar{f}_{\epsilon_q}
 \Big) ,
\end{align}
where the superscript $[k]$ on a function denotes its $k$-th derivative with respect to the energy $\epsilon_p$, $\binom{n}{k} = \frac{n !}{k ! (n-k) !}$ is the binomial coefficient, and $R_{\epsilon_p}$ represents the maximally allowed scattering rate at momentum $p$:
\begin{align}
R_{\epsilon_p} = 
\int_{0}^{+\infty} 
\frac{d \epsilon_q}{\sqrt{\epsilon_q}} 
\big(
f_{\epsilon_q} W_{ \epsilon_q, \epsilon_p }^{\rm 1D} + W_{ \epsilon_p, \epsilon_q }^{\rm 1D} \bar{f}_{\epsilon_q} 
\big)
\end{align}

We further apply the general Leibniz rule on $W_{ \epsilon_q \to \epsilon_p }^{[n-k]}$ and $W_{ \epsilon_p \to \epsilon_q }^{[n-k]}$ [see Eq.~\eqref{S-W-1D}]:
\begin{align}
\begin{aligned}
\big[ W_{ \epsilon_q \to \epsilon_p }^{\rm 1D} \big]^{[n-k]} = 
\sum_{l \in \mathbb{Z}}
\sum_{j=0}^{n-k}
\binom{n-k}{j} &
\big[\Gamma_l^{\rm 1D} (\varepsilon_q, \epsilon_p)
\big]^{[n-k-j]}
S^{[j]} (\epsilon_q - \epsilon_p + l \Omega) ,
\\
\big[ W_{ \epsilon_p \to \epsilon_q }^{\rm 1D} \big]^{[n-k]} = 
\sum_{l \in \mathbb{Z}}
\sum_{j=0}^{n-k}
\binom{n-k}{j} &
\big[\Gamma_l^{\rm 1D} ( \epsilon_q, \epsilon_p)
\big]^{[n-k-j]}
S^{[j]} (\epsilon_p - \epsilon_p + l \Omega) .
\end{aligned}
\end{align}
Combining these results, we obtain:
\begin{align}
\begin{aligned}
f^{[n]}_{\epsilon_p}
& = 
\frac{1}{ R_{\epsilon_p} }
\sum_{l \in \mathbb{Z}}
\sum_{k=0}^{n-1}
\sum_{j=0}^{n-k}
\binom{n}{k}
\binom{n-k}{j}
\int_{0}^{+\infty} 
\frac{d \epsilon_q}{\sqrt{\epsilon_q}} 
\big[ 
\Gamma_l^{\rm 1D} (\epsilon_q, \epsilon_p)
\big]^{[n-k-j]}
\Big(
f_{\epsilon_q}  
S^{[j]}_{\epsilon_q - \epsilon_p + l \Omega} \bar{f}_{\epsilon_p}^{[k]}
-
\bar{f}_{\epsilon_q}
S^{[j]}_{\epsilon_p - \epsilon_q + l \Omega}
f_{\epsilon_p}^{[k]}
\Big)
\\
& 
=
\sum_{l \in \mathbb{Z}}
\sum_{k=0}^{n-1}
\sum_{j=0}^{n-k}
B_{\epsilon_p}^{(n j k)}
\int_{0}^{+\infty} 
\frac{d \epsilon_q}{\sqrt{\epsilon_q}} 
\big[ 
\Gamma_l^{\rm 1D} (\epsilon_q, \epsilon_p)
\big]^{[n-k-j]}
\Big(
f_{\epsilon_q}  
S^{[j]} (\epsilon_q - \epsilon_p + l \Omega) \bar{f}_{\epsilon_p}^{[k]}
-
\bar{f}_{\epsilon_q}
S^{[j]} (\epsilon_p - \epsilon_q + l \Omega)
f_{\epsilon_p}^{[k]}
\Big) ,
\end{aligned}
\end{align}
with the coefficient $B_{\epsilon_p}^{(n j k)} = \binom{n}{k}
\binom{n-k}{j} / R_{\epsilon_p} $.

\begin{figure}
\includegraphics[width=0.48\textwidth]{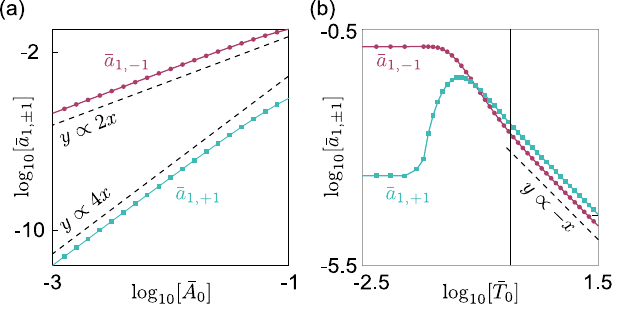}
\caption{
Dependence of $\bar{a}_{n = 1, s = + 1} = \Omega a_{n = 1, s = + 1}$ (green lines) and $\bar{a}_{n = 1, s = - 1} = \Omega a_{n = 1, s = - 1}$ (purple lines) as functions of (a) driving amplitude $\bar{A}_0= A_0 / \sqrt{2m \Omega} $ and (b) bath temperature $\bar{T}_0 = k_B T_0 / \Omega$ for the 2D parabolic model [see Eq.~\eqref{ParabolicModels}] coupled to the gapped bosonic bath [see Eq.~\eqref{gap-bath}].
Parameters used both panels: $\Delta / \Omega = 3/10$,  particle density $n_0 / (2m\Omega) =  3/13$.
For panel (a), bath temperature $T_0 = 0$. 
For panel (b), driving amplitude $A_0/\sqrt{6 m \Omega} = 1/5$.
}
\label{fig-s1}
\end{figure}

\subsection{Floquet-Boltzmann equation for the 2D parabolic model driven by a circularly polarized electric field}

For the 2D parabolic model driven by a circularly polarized electric field, the Floquet-Boltzmann equation takes the form:
\begin{align}
0 = 
\int_{-\infty}^{+\infty} 
f_{\bf q} W_{{\bf q} \to {\bf p}} \bar{f}_{\bf p}  \frac{d {\bf q}}{(2 \pi)^2}
-
\int_{-\infty}^{+\infty} f_{\bf p} W_{{\bf p} \to {\bf q}} \bar{f_{\bf q}} \frac{ d {\bf q}}{(2 \pi)^2}, 
\label{S-FB2D-1}
\end{align}
with the scattering matrix given by Eq.~\eqref{S-WSimple}. 
Given the momentum-independent difference between the original dispersion $\epsilon_k \equiv  \epsilon_{\bf k} = {\bf k}^2$ and the quasi-energy $\varepsilon_k \equiv \varepsilon_{\bf k} = {\bf k}^2 + A_0^2$ for the 2D parabolic model, we can recast the momentum-space equation in terms of the energy variable $\epsilon_k$:
\begin{align}
0 = 
\int_{0}^{+\infty} 
f ( \epsilon_q)
W_{ \epsilon_q \to \epsilon_p }^{\rm 2D}
\bar{f} (\epsilon_p) {d \epsilon_q}
-
\int_{0}^{+\infty} f ( \epsilon_p )
W_{ \epsilon_p \to \epsilon_q }^{\rm 2D}
\bar{f} (\epsilon_q) {d \epsilon_q},
\label{S-FB2D-2}
\end{align}
in which we define the scattering matrix as
\begin{align}
W_{ \epsilon_q \to \epsilon_p }^{\rm 2D}
\equiv
W_{ \sqrt{\epsilon_q} \to \sqrt{\epsilon_p} } ,
\quad
W_{ \epsilon_p \to \epsilon_q }^{\rm 2D}
\equiv
W_{ \sqrt{\epsilon_p} \to \sqrt{\epsilon_q} } 
\end{align}
by using the fact that $W_{ {\bf q} \to {\bf p} }$ is a function of $\{ \epsilon_p = {\bf p}^2, \epsilon_q = {\bf q}^2 \}$ and does not depend on the orientations of ${\bf p}$ and ${\bf q}$ [see Eq.~\eqref{S-Gammas-2D}] and $f({\bf p}) = f(|{\bf p}|)$,  for the 2D parabolic model driven by a circularly polarized electric field.

Using Eqs.~(\ref{S-Gammas-2D}) and (\ref{S-WSimple}), we write the scattering matrix up to $O(A_0^6)$:
\begin{align}
\begin{aligned}
W_{ \epsilon_q \to \epsilon_p }^{\rm 2D}
& =
\sum_l \Gamma_l^{\rm 2D} (\epsilon_q, \epsilon_p)
S (\epsilon_q - \epsilon_p + l \Omega)
\\
& = 
\Gamma_0
\Big[1
-
\frac{2 (\epsilon_p+\epsilon_q) }{\Omega^2} A_0^2
+
\frac{3(\epsilon_p^2+4 \epsilon_p \epsilon_q+\epsilon_q^2) }{2\Omega^4} A_0^4 \Big] S (\epsilon_q - \epsilon_p)  \\
&\quad
+  \sum_{\eta = \pm 1}
\Gamma_0
\Big[
\frac{(\epsilon_p+\epsilon_q) }{\Omega^2} A_0^2
-
\frac{(\epsilon_p^2+4 \epsilon_p \epsilon_q+\epsilon_q^2) }{\Omega^4} A_0^4 \Big] S (\epsilon_q - \epsilon_p + \eta \Omega)  \\
&\quad
+ 
\sum_{\eta = \pm 1}
\Gamma_0
\Big[
\frac{(\epsilon_p^2+4 \epsilon_p \epsilon_q+\epsilon_q^2) }{4 \Omega^4} A_0^4 \Big] S (\epsilon_q - \epsilon_p + 2 \eta \Omega)
+ O(A_0^6) ,
\label{S-W-2D}
\end{aligned}
\end{align}
while $W_{ \epsilon_p \to \epsilon_q }^{\rm 2D}$ is obtained by swapping $\epsilon_q \leftrightarrow \epsilon_p$ in the above expression.

Here we also show the formal expressions of derivatives of $f(\epsilon_p)$ following the analysis in the subsection \ref{Sub-formal-generic}.
We take the $n$-th derivative with respect to $\epsilon_p$ on both sides of Eq.~\eqref{S-FB2D-2}, apply the general Leibniz rule, and obtain:
\begin{align}
f_{\epsilon_p}^{[n]}
= 
\frac{1}{ R_{\epsilon_p} }
\sum_{k=0}^{n-1}
\binom{n}{k}
\int_{0}^{+\infty} d \epsilon_q 
\Big(
f_{\epsilon_q}  
\big[ W_{\epsilon_q \to \epsilon_p}^{\rm 2D} \big]^{[n-k]}
\bar{f}_{\epsilon_p}^{[k]}
 - 
 f_{\epsilon_p}^{[k]} 
 \big[ W_{\epsilon_p \to \epsilon_q}^{\rm 2D} \big]^{[n-k]}   \bar{f}_{\epsilon_q}
 \Big) ,
\end{align}
where the superscript $[k]$ on a function denotes its $k$-th derivative with respect to the energy $\epsilon_p$, $\binom{n}{k} = \frac{n !}{k ! (n-k) !}$ is the binomial coefficient, and $R_{\epsilon_p}$ represents the maximally allowed scattering rate at momentum $p$:
\begin{align}
R_{\epsilon_p} = 
\int_{0}^{+\infty} 
d \epsilon_q 
\big(
f_{\epsilon_q} W_{ \epsilon_q, \epsilon_p }^{\rm 2D} + W_{ \epsilon_p, \epsilon_q }^{\rm 2D} \bar{f}_{\epsilon_q} 
\big)
\end{align}

We further apply the general Leibniz rule on $W_{ \epsilon_q \to \epsilon_p }^{[n-k]}$ and $W_{ \epsilon_p \to \epsilon_q }^{[n-k]}$ [see Eq.~\eqref{S-W-2D}]:
\begin{align}
\begin{aligned}
\big[ W_{ \epsilon_q \to \epsilon_p }^{\rm 2D} \big]^{[n-k]} = 
\sum_{l \in \mathbb{Z}}
\sum_{j=0}^{n-k}
\binom{n-k}{j} &
\big[\Gamma_l^{\rm 2D} (\varepsilon_q, \epsilon_p)
\big]^{[n-k-j]}
S^{[j]} (\epsilon_q - \epsilon_p + l \Omega) ,
\\
\big[ W_{ \epsilon_p \to \epsilon_q }^{\rm 2D} \big]^{[n-k]} = 
\sum_{l \in \mathbb{Z}}
\sum_{j=0}^{n-k}
\binom{n-k}{j} &
\big[\Gamma_l^{\rm 2D} ( \epsilon_q, \epsilon_p)
\big]^{[n-k-j]}
S^{[j]} (\epsilon_p - \epsilon_p + l \Omega) .
\end{aligned}
\end{align}
Combining these results, we obtain:
\begin{align}
\begin{aligned}
f^{[n]}_{\epsilon_p}
& = 
\frac{1}{ R_{\epsilon_p} }
\sum_{l \in \mathbb{Z}}
\sum_{k=0}^{n-1}
\sum_{j=0}^{n-k}
\binom{n}{k}
\binom{n-k}{j}
\int_{0}^{+\infty} 
d \epsilon_q
\big[ 
\Gamma_l^{\rm 2D} (\epsilon_q, \epsilon_p)
\big]^{[n-k-j]}
\Big(
f_{\epsilon_q}  
S^{[j]}_{\epsilon_q - \epsilon_p + l \Omega} \bar{f}_{\epsilon_p}^{[k]}
-
\bar{f}_{\epsilon_q}
S^{[j]}_{\epsilon_p - \epsilon_q + l \Omega}
f_{\epsilon_p}^{[k]}
\Big)
\\
& 
= 
\sum_{l \in \mathbb{Z}}
\sum_{k=0}^{n-1}
\sum_{j=0}^{n-k}
B_{\epsilon_p}^{(n j k)}
\int_{0}^{+\infty} 
d \epsilon_q 
\big[ 
\Gamma_l^{\rm 2D} (\epsilon_q, \epsilon_p)
\big]^{[n-k-j]}
\Big(
f_{\epsilon_q}  
S^{[j]} (\epsilon_q - \epsilon_p + l \Omega) \bar{f}_{\epsilon_p}^{[k]}
-
\bar{f}_{\epsilon_q}
S^{[j]} (\epsilon_p - \epsilon_q + l \Omega)
f_{\epsilon_p}^{[k]}
\Big) ,
\end{aligned}
\end{align}
with the coefficient $B_{\epsilon_p}^{(n j k)} = \binom{n}{k}
\binom{n-k}{j} / R_{\epsilon_p}$.

\section{Equal time pair correlation function for the Floquet non-Fermi liquid}
\label{PairCorrelation}
In this section, we first derive the pair correlation function for a periodically driven Bloch band and then demonstrate that it exhibits long-range or power-law behavior for the Floquet non-Fermi liquid using specific examples.

The pair correlation function quantifies the probability density of finding a particle at position ${\bf r}_1$ given that another particle is located at position ${\bf r}_2$, making it a crucial characterization of the correlations in a fluid. The correlation function can be derived as follows~\cite{giuliani2008quantum}
\begin{align}\begin{aligned}
\label{gr1r2}
g({\bf r}_1, {\bf r}_2, t) = \frac{ \braket{ n({\bf r}_1, t) n({\bf r}_2, t) } }{
\braket{ n({\bf r}_1, t) } \braket{ n({\bf r}_2, t) } }
- \frac{ \delta ({\bf r}_2 - {\bf r}_1) }{ \braket{ n({\bf r}_1, t) } }
= 1 - \frac{
\braket{
c_{{\bf r}_1}^\dagger (t)
c_{{\bf r}_2} (t)
}
\braket{
c_{{\bf r}_2}^\dagger (t)
c_{{\bf r}_1} (t)
}
}{n_0^2} ,
\end{aligned}\end{align}
where Wick's theorem is applied, and the system is assumed to be translationally invariant, i.e., $\braket{ n({\bf r}_1, t_1) } = \braket{ n({\bf r}_2, t_2) } = n_0 $, which represents the averaged uniform density. 
For the single-band model under consideration, the creation/annihilation operators in Eq.~\eqref{gr1r2} can be expanded in terms of plane waves as follows:
\begin{align}\begin{aligned}
\label{c-expansion}
c_{\bf r}^\dagger(t)
= \frac{1}{\sqrt{V}} \sum_{\bf k}
e^{- i {\bf k} \cdot {\bf r}}
[\psi_{\bf k} (t)]^* c_{\bf k}^\dagger,
\end{aligned}\end{align}
where $\psi_{\bf k} (t)$ denotes the Floquet wave function [see Eq.~\eqref{S-FloquetDefinitions}].
In the Floquet non-Fermi liquid steady state, we have $\braket{ c_{{\bf k}'}^\dagger c_{\bf k} } = \delta_{{\bf k}' {\bf k}}f_{\bf k}$. By denoting ${\bf r} = {\bf r}_1 - {\bf r}_2$, we obtain the following expression for Eq.~\eqref{gr1r2}:
\begin{align}\begin{aligned}
g({\bf r}) = 1 - \frac{1}{n_0^2} 
\bigg|
\frac{1}{V}
\sum_{\bf k}
e^{-i {\bf k} \cdot {\bf r} }
f_{\bf k}
\bigg|^2 
=
1 -
\bigg|
\frac{  \sum_{\bf k}
e^{-i {\bf k} \cdot {\bf r} }
f_{\bf k}/ V   }
{   \sum_{\bf k}
f_{\bf k}/V   }
\bigg|^2
=
1 -
\bigg|
\frac{  \tilde{f} ({\bf r})  }
{    \tilde{f} ({\bf 0})   }
\bigg|^2
,
\label{S-gr}
\end{aligned}\end{align}
where $\tilde{f} ({\bf r})$ is defined as the $d$-dimensional Fourier transform of $f_{\bf k}$:
\begin{align}
\tilde{f}({\bf r}) =
\int \frac{d {\bf k}}{(2\pi)^d}
e^{-i {\bf k} \cdot {\bf r} }
f_{\bf k}  .
\label{f-tilde}
\end{align}

In a conventional equilibrium Fermi liquid at zero temperature, i.e., $T_0 = 0$, $f_{\bf k} \to \Theta (k_F - |{\bf k}|)$. The discontinuity in $f_{\bf k}$ at the Fermi surface causes long-range oscillatory behavior in $g({\bf r})$, which is determined by the Fermi wave vector $k_F$ and the system dimensionality~\cite{giuliani2008quantum}.
In contrast, the Floquet non-Fermi liquid exhibits power-law long-range oscillations originating from non-analyticities in its occupation function $f_{\bf k}$, even at finite temperatures. This is demonstrated next for 1D and 2D parabolic models.

\subsection{Parabolic model in 1D}

We now evaluate the integral $\int dk e^{-ikr} f_k / (2\pi)$ from Eq. (\ref{f-tilde}) for the dimensionless parabolic model in 1D [see Eq. (\ref{S-Dless-1D})]. Our focus is on the square-root-Theta type non-analyticities in the momentum distribution at specific momenta $q_{j}$ ($j = 1, 2, \ldots$), which arise from the coupling with gapped bosonic baths [see Eq.~\eqref{gap-bath} in the main text]. The momentum distribution can be decomposed as follows:
\begin{align}\begin{aligned}
f^{\{ 1 \}} (k) = f_{\rm regular}^{\{ 1 \}} (k)
+
\sum_j \frac{ a_j \Theta (k - q_j) }{(k - q_j)^{1/2}}
,
\label{S-f1-decompose-1D}
\end{aligned}\end{align}
where the superscript $\{ k \}$ denotes the $k$-th derivative with respect to momentum $p$, $f_{\text{regular}}^{\{1\}}(k)$ encompasses the analytic component and any non-analyticities weaker than $\Theta (k - q_j) (k - q_j)^{-1/2}$ with $\Theta(x)$ being the Heaviside step function.

Exploiting the symmetry property $f(-k) = f(k)$, which implies $f^{\{1\}}(-k) = -f^{\{1\}}(k)$, and applying integration by parts, we obtain:
\begin{align}\begin{aligned}
& \hspace{-7px}\frac{1}{2\pi}
\int_{-\infty}^{+\infty} d k \,
e^{-i k r } f_k
=
- \frac{1}{2\pi}
\int_{0}^{+\infty} d k \,
\frac{2 \sin (k r)}{r}
f^{\{ 1 \}} (k)
=
- \frac{1}{2\pi}
\int_{0}^{+\infty} d k \,
\frac{2 \sin (k r)}{r}
\Big[
f_{\rm regular}^{\{ 1 \}} (k)
+
\sum_j \frac{ a_j \Theta (k - q_j) }{(k - q_j)^{1/2}}
\Big]
\\
& = \text{regular part}
-
\frac{1}{r^{3/2}}
\sum_j \sqrt{ \frac{2 }{\pi} } a_j
\Big[
\cos(q_j r) S_F \Big( \sqrt{ (2 / \pi) \left(k-q_j\right) r} \Big)  
+
\sin(q_j r) C_F \Big( \sqrt{ (2 / \pi) \left(k-q_j\right) r} \Big)
\Big]_{q_j}^{+\infty}
\\
& = 
\text{regular part}
-
\frac{1}{r^{3/2}}
\sum_j \frac{a_j}{\sqrt{\pi}} \sin \left(q_j r + \frac{\pi}{4}\right),
\label{S-fr-1D}
\end{aligned}\end{align}
where $S_F(z)$ and $C_F(z)$ are Fresnel integrals, and their asymptotic properties $S_F(0) = C_F(0) = 0$ and $S_F(+\infty) = C_F(+\infty) = 1/2$ were used. The non-analyticities thus lead to power-law decay $\sim r^{-3/2}$ with oscillations at Floquet Fermi surfaces
$q_j$, contrasting the behavior of conventional Fermi liquids.

\begin{figure}
\includegraphics[width=0.48\textwidth]{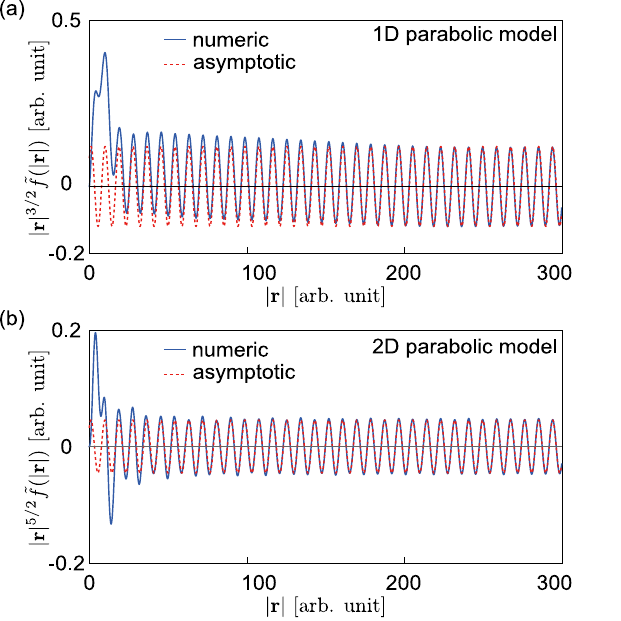}
\caption{
Re-scaled real space Fourier transform $\tilde{f}({\bf r})$ [Eq.~\eqref{f-tilde}] obtained from direct numerical solutions of $f({\bf k})$ (blue solid lines) and asymptotic analysis (red dashed lines) for (a) a one-dimensional parabolic model [Eq.~\eqref{S-fr-1D}] and (b) a two-dimensional parabolic model [Eq.~\eqref{S-fr-2D}] coupled with a gapped bosonic bath with $T_0 = 0$ and $\Delta = 0$ such that there is only one primary Fermi surface at $\Omega$, for visual clarity.
The asymptotic behavior of $\tilde{f}({\bf r})$ for both models can be described by a unified expression, as discussed in Eq.~\eqref{fr-asymptotic} of the main text.
The following parameters are used: $A_0/\sqrt{m \Omega} = 4/5$, particle density $n_0 / \sqrt{m \Omega} = 1/2$ (1D) or $n_0 /(m \Omega) = 1$ (2D).
}
\label{fig-s2}
\end{figure}

\subsection{Parabolic model in 2D}

Here we evaluate the integral $\int d {\bf k}  e^{-i{\bf k} \cdot {\bf r} } f_{\bf k} / (2 \pi)^2$ from Eq. (\ref{f-tilde}) for the dimensionless parabolic model in 2D [see Eq. (\ref{S-Dless-2D})].
Our focus is on the Theta-type non-analyticities in the momentum distribution at specific momenta $p_i$ ($i = 1, 2, \ldots$), which arise from the coupling with gapped bosonic baths [see Eq.~\eqref{gap-bath} in the main text]. The momentum distribution can be decomposed as follows:
\begin{align}\begin{aligned}
f^{\{ 1 \}} (k) = f_{\rm regular}^{\{ 1 \}} (k)
+
\sum_i a_i \Theta (k - p_i) 
,
\label{S-f1-decompose-2D}
\end{aligned}\end{align}
where the superscript $\{ k \}$ denotes the $k$-th derivative with respect to momentum $p$, $f_{\text{regular}}^{\{1\}}(k)$ encompasses the analytic component and any non-analyticities weaker than $\Theta (k - p_i)$.
Exploiting the symmetry property $f({\bf k} ) = f(k)$ with $k = |{\bf k}|$, and applying integration by parts, we obtain:
\begin{align}\begin{aligned}
& \frac{1}{(2\pi)^2}
\int d {\bf k} \,
e^{-i {\bf k} \cdot {\bf r} } f_{\bf k}
= 
\frac{1}{(2\pi)^2}
\int_{-\infty}^{+\infty} k d k 
\int_{0}^{2 \pi} d \theta
e^{- i k r \cos \theta}
f(k)
= 
- \frac{1}{2\pi}
\int_{0}^{+\infty} d k \,
\frac{ k J_1 (k r)}{r}
f^{\{ 1 \}} (k)
\\
&
= 
- \frac{1}{2\pi}
\int_{0}^{+\infty} d k \,
\frac{ k J_1 (k r)}{r}
\Big[
f_{\rm regular}^{\{ 1 \}} (k)
+
\sum_i a_i \Theta (k - p_i)
\Big]
\\
&
= 
- \frac{1}{2\pi}
\int_{0}^{+\infty} d k \,
\frac{ k J_1 (k r)}{r}
\Big[
\mathring{f}_{\rm regular}^{\{ 1 \}} (k)
+
\sum_i a_i \Theta (p_i - k)
\Big]
\\
& = \text{regular part}
- \frac{1}{2\pi}
\sum_i a_i
\frac{ k \pi \big[
J_1(k r) H_0(k r)
-
J_0(k r) H_1(k r)
\big]}
{ 2 r^2 } \bigg|_{0}^{p_i}
\\
& 
\xrightarrow{r \to +\infty}
\text{regular part}
-
\frac{1}{2\pi}
\frac{1}{r^{5/2}}
\sum_i a_i 
\sqrt{\frac{2}{\pi} p_i}
\sin \left( p_i r + \frac{\pi}{4} \right) .
\label{S-fr-2D}
\end{aligned}\end{align}
We note that in the above derivations, we define $\mathring{f}_{\text{regular}}^{\{ 1 \}} (k)$ such that
$\mathring{f}_{\rm regular}^{\{ 1 \}} (k)
+
\sum_i a_i \Theta (p_i - k) = f_{\rm regular}^{\{ 1 \}} (k)
+
\sum_i a_i \Theta (k - p_i) $.
Similar to $\lim_{k \to \infty} f^{\{1 \}}(k) = 0$, we have $\lim_{k \to \infty} \mathring{f}_{\rm regular}^{\{1 \}} (k) = 0$.
Consequently, the integral with respect to $\mathring{f}_{\rm regular}^{\{1\}} (k)$ is convergent.
The ``regular part'' refers to the part whose oscillation decays faster than $r^{-5/2}$ at large $r$, $J_{0,1}(x)$ and $H_{0,1} (x)$ are Bessel and Struve functions.
The non-analyticities produce power-law decay $\sim r^{-5/2}$ with oscillations at Floquet Fermi surfaces $p_i$, again distinct from conventional Fermi liquids.

\section{Density noise correlation function for the Floquet non-Fermi liquid}
\label{Appendix-DensityNoise}

In this section, we derive the density noise correlation function for a periodically driven Bloch band and provide explicit expressions for the parabolic models in 1D and 2D.

The spectrum of density fluctuations can be obtained from the correlation function
\begin{align}\begin{aligned}
C({\bf r}_1, t_1 ; {\bf r}_2, t_2)=
\braket{ n({\bf r}_1, t_1) n({\bf r}_2, t_2) }
-
\braket{ n({\bf r}_1, t_1) }
\braket{ n({\bf r}_2, t_2) }.
\end{aligned}\end{align}
Due to translational invariance, $\braket{ n({\bf r}_1, t_1) } = \braket{ n({\bf r}_2, t_2) } = n_0$. Thus, the non-trivial part is the correlation function:
\begin{align}\begin{aligned}
\braket{ n({\bf r}_1, t_1) n({\bf r}_2, t_2) }
=
\braket{
c_{{\bf r}_1}^\dagger (t_1) c_{{\bf r}_1} (t_1)
c_{{\bf r}_2}^\dagger (t_2) c_{{\bf r}_2} (t_2)
}.
\end{aligned}\end{align}
Using the relation in Eq.~\eqref{c-expansion} and performing contractions $\braket{ 
c_{ {\bf k}_1 }^\dagger 
c_{ {\bf k}_2 } 
c_{ {\bf k}_3 }^\dagger 
c_{ {\bf k}_4 } } = 
\braket{ c_{ {\bf k}_1 }^\dagger c_{ {\bf k}_2 } }
\braket{ c_{ {\bf k}_3 }^\dagger c_{ {\bf k}_4 } }
+
\braket{ c_{ {\bf k}_1 }^\dagger c_{ {\bf k}_4 } } 
\braket{ c_{ {\bf k}_2 } c_{ {\bf k}_3 }^\dagger }$ according to Wick's theorem, we obtain
\begin{align}\begin{aligned}
C({\bf r}_1, t_1 ; {\bf r}_2, t_2)=
\frac{1}{V^2}
\sum_{{\bf k}_1, {\bf k}_2 }
f_{ {\bf k}_1 } \bar{f}_{ {\bf k}_2 }
e^{- i ( {\bf k}_1 - {\bf k}_2 ) \cdot ( {\bf r}_1 - {\bf r}_2 )}
[ \psi_{ {\bf k}_1 } (t_1) ]^* 
\psi_{ {\bf k}_2 } (t_1)
[ \psi_{ {\bf k}_2 } (t_2) ]^*
\psi_{ {\bf k}_1 } (t_2).
\end{aligned}\end{align}
By Fourier transforming into momentum space, we have
\begin{align}\begin{aligned}
C({\bf q}; t_1, t_2)=
\frac{1}{V}
\sum_{\bf k}
f_{ {\bf k} + {\bf q} } \bar{f}_{\bf k}
[ \psi_{ {\bf k} + {\bf q} } (t_1) ]^* 
\psi_{ {\bf k} } (t_1)
[ \psi_{ {\bf k} } (t_2) ]^*
\psi_{ {\bf k} + {\bf q} } (t_2).
\end{aligned}\end{align}
Averaging over $t_2$, Fourier transforming $t = t_1 - t_2$ into frequency space, and performing Floquet expansions [see Eq.~\eqref{S-FloquetDefinitions}], we obtain
\begin{align}\begin{aligned}
\bar{C} ({\bf q}; \omega)
& \equiv
\int_0^T
\frac{d t_2}{T}
\left[
\int_{-\infty}^{+\infty}
\frac{d t}{2\pi}
C({\bf q}; t_1, t_2) e^{+ i \omega t}
\right] =
\frac{1}{V}
\sum_{ {\bf k}}
f_{ {\bf k} + {\bf q} } \bar{f}_{\bf k}
\sum_l
\Phi_l ({\bf k}, {\bf k}+ {\bf q})
\delta ( \varepsilon_{\bf k} - \varepsilon_{ {\bf k} + {\bf q} } - \omega + l \Omega ),
\label{S-DensityNoise}
\end{aligned}\end{align}
where the $l$-dependent amplitude factor reads
\begin{align}\begin{aligned}
\Phi_l ({\bf k}, {\bf k}+ {\bf q})
=
\Big|
\sum_{l_1}
\varphi_{ l, {\bf k} }
\varphi_{ l + l_1, {\bf k} + {\bf q} }^*
\Big|^2 .
\end{aligned}\end{align}
In equilibrium, $\Phi_l ({\bf k}, {\bf k}+ {\bf q}) = \delta_{l 0}$ and the density noise correlation function reduces to the dynamic structure factor:
\begin{align}\begin{aligned}
\bar{C} ({\bf q}; \omega) 
\to C_{\rm equi} ({\bf q}; \omega)
=
\frac{1}{V}
\sum_{ {\bf k}}
f_{ {\bf k} + {\bf q} } \bar{f}_{\bf k}
\delta ( \epsilon_{\bf k} - \epsilon_{ {\bf k} + {\bf q} } - \omega ).
\end{aligned}\end{align}
At zero temperature, the dynamic structure factor $C_{\rm equi} ({\bf q}; \omega)$ is non-zero only when ${\bf q}$ and $\omega$ lie within the particle-hole continuum. At finite temperatures, this region smoothly broadens due to thermal excitations. However, as we will demonstrate, this is not the case for our Floquet non-Fermi liquid, where the particle-hole continuum remains sharply defined even at finite temperatures.

\subsection{Parabolic model in 1D}

We now evaluate explicitly the density noise correlation function, Eq. (\ref{S-DensityNoise}) for the parabolic model in 1D [see Eq. (\ref{S-Dless-1D})]. 
Given the quasi-energy $\varepsilon_k = k^2 + A_0^2/2$ for the 1D parabolic model, we obtain
\begin{align}\begin{aligned}
\bar{C}_{\rm 1D} (q; \omega)
=
\int_{-\infty}^{+\infty} \frac{d k}{2 \pi}
f_{ k + q } \bar{f}_{k}
\sum_l
\Phi_l (k, k+q)
\delta \big( k^2 - (k+q)^2 - \omega + l \Omega \big).
\label{S-DensityNoise-1D-1}
\end{aligned}\end{align}
The $l$-dependent amplitude $\Phi_l (k, k+q) = |\sum_{l_1}
\varphi_{l, k}
\varphi_{l+l_1, k + q}^*
|^2 \to \Phi_l (q)$ is $k$-independent for the 1D parabolic model and can be read from $\Gamma_l^{\rm 1D} (k, k+q)$ [see Eq.~\eqref{S-Gammas-1D}]:
\begin{align}
\begin{aligned}
& \Phi_{l=0} (q) =
1
-\frac{2 q^2}{\Omega^2} A_0^2 
+\frac{3 q^4}{2 \Omega^4} A_0^4
+ O(A_0^6) ,
\\
& \Phi_{l=\pm 1} (q) = 
\frac{q^2}{\Omega^2} A_0^2 
- \frac{q^4}{\Omega^4} A_0^4
+ O(A_0^6) ,
\\
& \Phi_{l=\pm2} (q) =
\frac{q^4}{4 \Omega^4} A_0^4
+ O(A_0^6),
\\
& \Phi_{l=\pm3} (q) 
\propto  
O(A_0^6) .
\end{aligned}
\end{align}
Evaluating the integral in Eq. (\ref{S-DensityNoise-1D-1}) by eliminating the Dirac delta function,
\begin{align}
\begin{aligned}
\delta \big( k^2 - (k+q)^2 - \omega + l \Omega \big)
\to
\frac{1}{2|q|}
\delta \left( 
k - \frac{ - q^2 - \omega + l \Omega }{2 q}
\right)
\end{aligned}
\end{align}
we arrive at the expression
\begin{align}\begin{aligned}
\bar{C}_{\rm 1D} (q; \omega)
=
\frac{1}{4 \pi |q|}
\sum_l
\Phi_l (q)
f\left( 
\frac{-q^2-\omega+l \Omega}{2 q}+q
\right)
\bar{f}
\left(
\frac{-q^2-\omega+l \Omega}{2 q}
\right) .
\label{S-DensityNoise-1D-2}
\end{aligned}\end{align}

\subsection{Parabolic model in 2D}
We now explicitly evaluate the density noise correlation function, Eq. (\ref{S-DensityNoise}), for the parabolic model in 2D [see Eq. (\ref{S-Dless-2D})].
Given the quasi-energy $\varepsilon_{\bf k} = {\bf k}^2 + A_0^2$, the rotationally invariant occupation function $f(|{\bf k}|)$, and the closed form for its Floquet wave function harmonics [see Eq.~\eqref{S-Fwh-2D}], we have:

\begin{align}\begin{aligned}
\bar{C}_{\rm 2D}^{\pm} (q, \theta_q; \omega)
=
\frac{1}{(2 \pi)^2}
\int_0^{2 \pi} d \theta_k
\int_0^{+ \infty} k dk
& \,
f(|{\bf k}+{\bf q}|)
\bar{f}(k) 
\sum_l 
\left|
\sum_{l_1}
J_l \left( 
\frac{2 A_0 k}{\Omega}
\right)
J_{l+l_1} \left( 
\frac{2 A_0 |{\bf k}+{\bf q}|}{\Omega}
\right)
e^{i l_1 (\phi_0 \mp \theta_{{\bf k}+{\bf q}})}
\right|^2
\\
& \times 
\delta( k^2 - |{\bf k}+{\bf q}|^2 - \omega + l \Omega) ,
\label{S-DensityNoise-2D-1}
\end{aligned}\end{align}
where ${\bf k} \to (k, \theta_k)$, ${\bf q} \to (q, \theta_q)$, and $|{\bf k}+{\bf q}|=\sqrt{k^2+q^2+2 k q \cos(\theta_k-\theta_q)}$ are represented in polar coordinates. The superscript $\pm$ in $\bar{C}_{\rm 2D}^{\pm} (q, \theta_q; \omega)$ denotes left- or right-handed circularly polarized light. After carefully eliminating the Dirac delta function and integrating over $\theta_k$, we obtain:

\begin{align}\begin{aligned}
\bar{C}_{\rm 2D}^{\pm} (q, \theta_q; \omega)
& =
\frac{1}{(2\pi)^2}
\sum_l
\int_0^{+\infty}
k dk
\frac{ f \left(\sqrt{k^2-\omega+l \Omega}\right)
\bar{f}(k)
 }
 {\big|\sqrt{4 k^2 q^2-(q^2+\omega-l \Omega)^2}\big|}
 \left|J_{l}\left(\frac{2 A_0}{\Omega} k\right)\right|^2
\\
& \quad \times
2 \sum_{l_1} \sum_{l_2} J_{l+l_1}\left(\frac{2 A_0}{\Omega} \sqrt{k^2-\omega+l \Omega}\right) J_{l+l_2}\left(\frac{2 A_0}{\Omega} \sqrt{k^2-\omega+l \Omega}\right)
\\
& \quad \times
T_{\left(l_1-l_2\right)}\left(\frac{q^2-\omega+l \Omega}{2 q \sqrt{k^2-\omega+l \Omega}}\right) 
\exp [+i (l_1-l_2)(\phi_0 \mp \theta_q)] ,
\label{S-DensityNoise-2D-2}
\end{aligned}\end{align}
where $T_n (x)$ is the Chebyshev polynomial of the first kind.
We observe that $\bar{C}_{\rm 2D}^{\pm} (q, \theta_q; \omega)$ depends on the angle $\theta_q$, initial phase $\phi_0$, and the helicity of the drive.
However, integrating over $\theta_q$ leads to $\delta_{l_1 l_2}$ and yields a simpler final expression:
\begin{align}\begin{aligned}
\overline{\bar{C}_{\rm 2D} (q, \omega)}
\equiv
\int_0^{2 \pi} d \theta_q \bar{C}_{\rm 2D}^{\pm} (q, \theta_q; \omega)
& = 
\frac{2}{(2\pi)^2}
\sum_l
\int_0^{+\infty}
k dk
\frac{ f \left(\sqrt{k^2-\omega+l \Omega}\right)
\bar{f}(k)
 }
 {\big|\sqrt{4 k^2 q^2-(q^2+\omega-l \Omega)^2}\big|}
 \left|J_{l}\left(\frac{2 A_0}{\Omega} k\right)\right|^2 ,
\label{S-DensityNoise-2D-3}
\end{aligned}\end{align}
where we used the fact that $T_0(x)=1$ and $\sum_l |J_l (x)|^2 = 1$.

Interestingly, despite the simplification achieved in Eq. (\ref{S-DensityNoise-2D-3}), the underlying angular dependence of $\bar{C}_{\rm 2D}^{\pm} (q, \theta_q; \omega)$ reveals a subtle physical feature of the system. Even when the 2D system is driven by circularly polarized light, $\bar{C} ({\bf q}, \omega)$ is not fully isotropic in ${\bf q}$. This anisotropy manifests as an explicit dependence on the angle difference between the initial phase $\phi_0$ of the drive and the direction of ${\bf q}$, represented by $\theta_q$. 
This is because in the steady state the physical direction of flow of the fermion fluid rotates in time with the drive, so as to instantaneously remain orthogonal to the electric field. While invisible in the equal-time density correlations, this instantaneous directionality of the flow leads to an explicit angular dependence of the correlations measured at different times and different angles. But, despite the dependence of $\bar{C} ({\bf q}, \omega)$ on the direction of ${\bf q}$, its non-analyticities are located at frequencies that depend only on the magnitude of $|{\bf q}|$.

\section{Persistence of non-analyticities in the presence of electron-electron interactions}
\label{Persistence}

In this section, we demonstrate that the non-analyticities in the Floquet occupation function arising from electron-phonon coupling persist even when electron-electron interactions are included in the Floquet-Boltzmann description.
For the sake of brevity, we will use $\epsilon_p$ to denote Floquet energy in this section.

In the Floquet-Boltzmann framework, the steady state occupation function $f(\epsilon_p)$ is determined by the condition that the total collision integral vanishes:
\begin{equation}
I_{\text{tot}}[f(\epsilon_p)] =
I_{\text{e-ph}}[f(\epsilon_p)] + 
I_{\text{e-e}}[f(\epsilon_p)] = 0 ,
\label{S-Itot}
\end{equation}
where $I_{\text{e-ph}}$ is the electron-phonon collision integral and $I_{\text{e-e}}$ is the electron-electron collision integral.
As shown in the main text, $I_{\text{e-ph}}$ has the form [see Eq.~\eqref{Floquet-Boltzmann} in the main text]:
\begin{equation}
I_{\text{e-ph}}[f(\epsilon_p)] = 
\sum_q \Big(f(\epsilon_q) W_{q \to p} \bar{f}(\epsilon_p) - f(\epsilon_p) W_{p \to q} \bar{f}(\epsilon_q)\Big) ,
\end{equation}
with $\bar{f}(\epsilon_p) = 1 - f(\epsilon_p)$ and the scattering rate:
\begin{equation}
W_{q \to p} = \sum_l \Phi^{(l)}_{q,p} S(\epsilon_q - \epsilon_p + l\Omega) .
\end{equation}
As discussed in the main text and Appendix~\ref{S-Analytical-nonanalyticities} ``Analytical analysis of non-analyticities'' of the supplementary material, the non-analyticities in the occupation function originate from non-analyticities in the $S$ function. These non-analyticities appear at specific energies $\epsilon_* = n\Omega + s\omega_*$, where $\omega_*$ represents the energy at which the bath's density of states have a non-analyticity.

The electron-electron collision integral in the Floquet-Boltzmann framework has the general form (see e.g., Eq.~(41) in Ref.~\cite{genske2015floquet}):
\begin{equation}
\begin{aligned}
I_{\text{e-e}}[f(\epsilon_p)] = 
\sum_{k,q,q'} K(p,k,q,q') &\left[f(\epsilon_p) f(\epsilon_k) \bar{f}(\epsilon_q)\bar{f}(\epsilon_{q'}) - \bar{f}(\epsilon_p)\bar{f}(\epsilon_k) f(\epsilon_q) f(\epsilon_{q'})\right] ,
\end{aligned}
\end{equation}
where $K(p,k,q,q')$ incorporates the Floquet interaction matrix elements, and enforces energy conservation modulo $\Omega$ to account for Floquet-Umklapp processes.

Let us consider the energy $\epsilon_p = \epsilon_*$ where non-analyticities occur in the absence of the the electron-electron collision operator.
We will denote the $n$-th derivative of a function $g$ with respect to $\epsilon_p$ used in Appendix~\ref{S-Analytical-nonanalyticities}:
\begin{equation}
g^{[n]}(\epsilon_p) = \frac{d^n g(\epsilon_p)}{d\epsilon_p^n} .
\end{equation}
In order to examine the presence of a non-analyticities, we define the following difference for any function $F$ at a specific point $\epsilon_*$:
\begin{equation}
\Delta F(\epsilon_*) = F(\epsilon_* + 0^+) - F(\epsilon_* - 0^-) .
\label{S-DeltaF}
\end{equation}
If the above limit is finite, then the function F has a discontinuity at $\epsilon_*$ from above versus from below.

We now differentiate Eq.~\eqref{S-Itot} $n$ times with respect to $\epsilon_p$:
\begin{equation}
\frac{d^n}{d\epsilon_p^n} I_{\text{e-ph}}[f(\epsilon_p)] + 
\frac{d^n}{d\epsilon_p^n} I_{\text{e-e}}[f(\epsilon_p)] = 0 .
\end{equation}
And evaluate the limit of the above expression as defined in Eq.~\eqref{S-DeltaF} around $\epsilon_p = \epsilon_*$:
\begin{equation}
\Delta \left(\frac{d^n}{d\epsilon_p^n} I_{\text{e-ph}}[f(\epsilon_p)]\right)_{\epsilon_p=\epsilon_*} +
\Delta \left(\frac{d^n}{d\epsilon_p^n} I_{\text{e-e}}[f(\epsilon_p)]\right)_{\epsilon_p=\epsilon_*} = 0 .
\end{equation}
As shown in Appendix~\ref{S-Analytical-nonanalyticities}, particularly in relation to Eqs.~\eqref{RdotFn} and \eqref{S-R-discontinuity} [as well as Eqs.~\eqref{S-f3-parabolic}, \eqref{S-f2-parabolic}, \eqref{S-f1-parabolic} for more specific models], 
the above term coming from electron-phonon collisions can be expressed as:
\begin{equation}
\Delta \left(\frac{d^n}{d\epsilon_p^n} I_{\text{e-ph}}[f(\epsilon_p)]\right)_{\epsilon_p=\epsilon_*} = A_{n,\epsilon_*}[f,S] \cdot \Delta f^{[n]}(\epsilon_*) + B_{n,\epsilon_*}[f, f^{[1]},\cdots,f^{[n-1]}; S, S^{[1]}, \cdots,S^{[n]}]
\end{equation}
This decomposition has the following interpretation:
\begin{itemize}
\item $A_{n,\epsilon_*}[f,S]$ is a general way of writing the (functional) coefficient [see $R_{\varepsilon_p, \eta_p}$ in Eq.~\eqref{RdotFn}] that captures how changes in the $n$-th derivative of $f$ affect the $n$-th derivative of the collision integral. 

\item $B_{n,\epsilon_*}[f, f^{[1]},\cdots,f^{[n-1]}; S, S^{[1]},\cdots,S^{[n]}] \neq 0 $ emerges from differentiating the $S$ function in the collision integral, and corresponds to the right-hand side of Eq.~\eqref{RdotFn}. When the $S$ function is differentiated $n$ times, it can produce terms containing Dirac delta functions at $\epsilon_p = \epsilon_*$ and leads to $B_{n,\epsilon_*} \neq 0$ follwing the discussion in Appendix~\ref{S-Generic-non-analyticities}.
\end{itemize}
Similarly, for the electron-electron collision integral, we can express the jump in its $n$-th derivative as:
\begin{equation}
\Delta \left(\frac{d^n}{d\epsilon_p^n} I_{\text{e-e}}[f(\epsilon_p)]\right)_{\epsilon_p=\epsilon_*} = C_{n,\epsilon_*}[f,K] \cdot \Delta f^{[n]}(\epsilon_*) + D_{n,\epsilon_*}[f, f^{[1]},\cdots,f^{[n-1]}; K, K^{[1]},\cdots,K^{[n]}]
\end{equation}
where $C_{n,\epsilon_*}$ and $D_{n,\epsilon_*}$ have similar interpretations to $A_{n,\epsilon_*}$ and $B_{n,\epsilon_*}$, respectively, but arise from the structure of the electron-electron collision integral $I_{\text{e-e}}[f(\epsilon_p)]$.

Combining these equations, we get:
\begin{equation}
(A_{n,\epsilon_*} + C_{n,\epsilon_*}) \cdot \Delta f^{[n]}(\epsilon_*) + ( B_{n,\epsilon_*} + D_{n,\epsilon_*}) = 0
\label{S-general-jump}
\end{equation}

The crucial point in our analysis is that generically $(B_{n,\epsilon_*} + D_{n,\epsilon_*}) \neq 0$ at the energy $\epsilon_p = \epsilon_*$. This is because:

\begin{itemize}
\item $B_{n,\epsilon_*} \neq 0$ as it originates from the non-analyticity in the $S$ function at $\epsilon_p = \epsilon_*$ which contains the detailed information of the bosonic bath, such as its density of states.

\item $D_{n,\epsilon_*}$ arises from a completely different physical origin, the electron-electron collision integral, and has no particular reason to exactly cancel $B_{n,\epsilon_*}$. For $D_{n,\epsilon_*}$ to exactly cancel $B_{n,\epsilon_*}$, the electron-electron interactions would need to be fine-tuned in a way that depends on the precise details of the bosonic bath, which is not physical.
\end{itemize}
Therefore, the equation $(A_{n,\epsilon_*} + C_{n,\epsilon_*}) \cdot \Delta f^{[n]}(\epsilon_*) + (B_{n,\epsilon_*} + D_{n,\epsilon_*}) = 0$ implies $\Delta f^{[n]}(\epsilon_*) \neq 0$ [assuming $(A_{n,\epsilon_*} + C_{n,\epsilon_*})$ is finite, which is generally true]. This means that the non-analyticity in the $n$-th derivative of the occupation function persists even when electron-electron interactions are included in the Floquet-Boltzmann framework.

As a specific example, we can consider a 2D system coupled to a gapped bath with a gap at $\Delta$. In this case, $D_{n,\epsilon_*}$ would be expected to be zero because the electron-electron collision integral does not contain any information about the bath. Consequently, the Floquet interaction kernel $K(p,k,q,q')$ is smooth around $\epsilon_* = \Delta$, which leads to $D_{n,\epsilon_*} = 0$. This simplifies Eq.~\eqref{S-general-jump} to:
\begin{equation}
\Delta f^{[n]}(\epsilon_*) 
= 
- \frac{B_{n,\epsilon_*}}{A_{n,\epsilon_*} + C_{n,\epsilon_*}} .
\end{equation}
Therefore the non-analyticities in the steady state occupation function at energies $\epsilon_*$ persist even when electron-electron interactions are included as a collision integral. While the electron-electron collision integral may modify the magnitude of these non-analyticities, particularly as the relative strength of interactions changes, it cannot eliminate them entirely when their relative strength are finite, thus preserving the essential character of the Floquet non-Fermi liquid state.

\clearpage

\end{document}